\documentclass{article}

\usepackage[T1]{fontenc}
\usepackage[margin=1in]{geometry}
\usepackage{amsfonts,amsmath,amsthm,amssymb,mathtools}  %

\mathtoolsset{centercolon}
\usepackage{xfrac,nicefrac}
\usepackage{mathdots}
\usepackage{mleftright}  %
\let\left\mleft
\let\right\mright

\usepackage{xspace}
\xspaceaddexceptions{]\}}  %
\usepackage{regexpatch}

\usepackage{bm,bbm,dsfont}  %
\usepackage{caption}
\usepackage[normalem]{ulem}
\usepackage{enumitem}

\usepackage{graphicx}
\usepackage{float}
\usepackage{subcaption}  %
\usepackage{tcolorbox}
\usepackage{tikz}
\usetikzlibrary{decorations.pathreplacing}
\usetikzlibrary{calc}
\usetikzlibrary{positioning}
\usetikzlibrary{arrows.meta}

\usepackage[linesnumbered,boxed,ruled,vlined]{algorithm2e}

\SetCommentSty{mycommfont}

\usepackage{thmtools,thm-restate}
\usepackage[colorlinks,citecolor=blue,linkcolor=blue,urlcolor=red]{hyperref}

\theoremstyle{plain}

\newtheorem{theorem}{Theorem}[section]  %
\newtheorem{lemma}[theorem]{Lemma}

\newtheorem{proposition}[theorem]{Proposition}

\newtheorem{claim}[theorem]{Claim}

\theoremstyle{definition}  %

\usepackage[capitalise]{cleveref}
\crefname{algocf}{Algorithm}{Algorithms}
\Crefname{algocf}{Algorithm}{Algorithms}
\crefname{claim}{Claim}{Claims}
\Crefname{claim}{Claim}{Claims}

\usepackage{xurl}

\newfloat{Distribution}{htbp}{loa}
\crefname{Distribution}{Distribution}{Distributions}
\Crefname{Distribution}{Distribution}{Distributions}
\newfloat{Protocol}{htbp}{loa}
\crefname{Protocol}{Protocol}{Protocols}
\Crefname{Protocol}{Protocol}{Protocols}

\SetKwProg{Function}{Function}{:}{}
\let\Fn\Function
\SetKwProg{CodeBlock}{}{}{}
\SetKwProg{Repeat}{Repeat}{:}{}

\DeclarePairedDelimiter{\ceil}{\lceil}{\rceil}
\DeclarePairedDelimiter{\floor}{\lfloor}{\rfloor}

\DeclarePairedDelimiter{\bk}{(}{)}
\DeclarePairedDelimiter{\Bk}{[}{]}
\DeclarePairedDelimiter{\BK}{\{}{\}}

\DeclarePairedDelimiter{\abs}{\lvert}{\rvert}

\DeclarePairedDelimiterX\mysetbase[2]{\lbrace}{\rbrace}{#1\,\delimsize\vert\,#2}
\NewDocumentCommand{\myset}{sO{}m m}{%
  \IfBooleanTF{#1}%
    {\mysetbase*{#3}{#4}}%
    {\mysetbase[#2]{#3}{#4}}%
}

\DeclareMathOperator*{\E}{\mathbb{E}}

\let\Pr\PrAux
\DeclareMathOperator{\poly}{poly}

\DeclareMathOperator{\polylog}{polylog}
\DeclareMathOperator*{\argmax}{arg\,max}

\DeclareMathOperator{\supp}{supp}

\renewcommand{\tilde}{\widetilde}
\newcommand{\eqdef}{\eqqcolon}
\newcommand{\defeq}{\coloneqq}
\newcommand{\eps}{\varepsilon}

\newcommand{\N}{\mathbb{N}}

\renewcommand{\l}{\ell}
\renewcommand{\emptyset}{\varnothing}
\renewcommand{\epsilon}{\eps}
\newcommand{\Patrascu}{\textup{P{\v{a}}tra{\c{s}}cu}\xspace}

\newcommand{\defn}[1]{\textit{\boldmath\textbf{#1}}}

\newcommand{\numberthis}{\addtocounter{equation}{1}\tag{\theequation}}

\newcommand{\EquationOnSameLine}[1]{\hspace*{\fill}$\displaystyle #1$\hspace*{\fill}\mbox{}}

\makeatletter
\xpatchcmd\thmt@restatable{%
\csname #2\@xa\endcsname\ifx\@nx#1\@nx\else[{#1}]\fi
}{%
\ifthmt@thisistheone
\csname #2\@xa\endcsname\ifx\@nx#1\@nx\else[{#1}]\fi
\else
\csname #2\@xa\endcsname[{Restated}]
\fi}{}{}
\makeatother

\usepackage{xparse}
\usepackage{booktabs}

\makeatletter
\let\oldparagraph\paragraph
\renewcommand{\paragraph}[1]{%
  \oldparagraph{\boldmath #1}%
}
\makeatother

\NewDocumentCommand{\xhigh}{o}{%
  \IfValueTF{#1}
    {x_{#1, \textup{high}}} %
    {x_{\textup{high}}} %
}
\NewDocumentCommand{\xmid}{o}{%
  \IfValueTF{#1}
    {x_{#1, \textup{mid}}} %
    {x_{\textup{mid}}} %
}
\NewDocumentCommand{\xlow}{o}{%
  \IfValueTF{#1}
    {x_{#1, \textup{low}}} %
    {x_{\textup{low}}} %
}
\NewDocumentCommand{\xhighmid}{o}{%
  \IfValueTF{#1}
    {x_{#1, \textup{high-mid}}} %
    {x_{\textup{high-mid}}} %
}
\NewDocumentCommand{\xmidlow}{o}{%
  \IfValueTF{#1}
    {x_{#1, \textup{mid-low}}} %
    {x_{\textup{mid-low}}} %
}

\newcommand{\phighmid}{p_{\textup{high-mid}}}
\newcommand{\plow}{p_{\textup{low}}}

\newcommand{\nmax}{n_{\max}}
\newcommand{\Umax}{U_{\max}}
\newcommand{\Sleft}{S_{\textup{left}}}
\newcommand{\Sright}{S_{\textup{right}}}
\newcommand{\rank}{\mathrm{rank}}

\newcommand{\hmax}{\overline{h}}

\newcommand{\DistSet}[1][\tilde a, \tilde b, n, \hmax]{D^S_{#1}}
\newcommand{\DistSetSmall}[1][a, b, n, \hmax]{D^S_{#1}}
\newcommand{\DistWeight}[1][\tilde a, \tilde b, n, \hmax]{D^\mathrm{weight}_{#1}}
\newcommand{\DistWeightTd}[1][n, \hmax]{\tilde{D}^\mathrm{weight}_{#1}}
\newcommand{\DistRank}[1][n, V_L, V_R]{D^\mathrm{\,rank}_{#1}}
\newcommand{\DistRankTd}[1][n, \tilde{V}_L, \tilde{V}_R]{D^\mathrm{\,rank}_{#1}}
\newcommand{\DistSmall}[1][i, j, a', b', n, \hmax]{D^\mathrm{\,small}_{#1}}

\renewcommand{\phi}{\varphi}
\newcommand{\mIn}{m_{\textup{in}}}
\newcommand{\mOut}{m_{\textup{out}}}
\newcommand{\kIn}{k_{\textup{in}}}
\newcommand{\kOut}{k_{\textup{out}}}
\newcommand{\MIn}{M_{\textup{in}}}
\newcommand{\MOut}{M_{\textup{out}}}
\newcommand{\KIn}{K_{\textup{in}}}
\newcommand{\KOut}{K_{\textup{out}}}

\newcommand{\mLeft}{m_{\mathrm{left}}}
\newcommand{\kLeft}{k_{\mathrm{left}}}
\newcommand{\mRight}{m_{\mathrm{right}}}
\newcommand{\kRight}{k_{\mathrm{right}}}
\newcommand{\mCat}{m_{\mathrm{cat}}}
\newcommand{\kCat}{k_{\mathrm{cat}}}
\newcommand{\MCat}{M_{\mathrm{cat}}}
\newcommand{\KCat}{K_{\mathrm{cat}}}

\newcommand{\mRank}{m_{\mathrm{rank}}}
\newcommand{\kRank}{k_{\mathrm{rank}}}
\newcommand{\MRank}{M_{\mathrm{rank}}}
\newcommand{\KRank}{K_{\mathrm{rank}}}

\newcommand{\mPivot}{m_{\mathrm{pivot}}}
\newcommand{\kPivot}{k_{\mathrm{pivot}}}
\newcommand{\MPivot}{M_{\mathrm{pivot}}}
\newcommand{\KPivot}{K_{\mathrm{pivot}}}

\newcommand{\mWeight}{m_{\mathrm{weight}}}
\newcommand{\kWeight}{k_{\mathrm{weight}}}
\newcommand{\MWeight}{M_{\mathrm{weight}}}
\newcommand{\KWeight}{K_{\mathrm{weight}}}

\newcommand{\mSpill}{m_{\mathrm{spill}}}

\newcommand{\mAll}{m_{\mathrm{all}}}
\newcommand{\kAll}{k_{\mathrm{all}}}
\newcommand{\MAll}{M_{\mathrm{all}}}
\newcommand{\KAll}{K_{\mathrm{all}}}

\newcommand{\MLarge}{M_{\mathrm{large}}}
\newcommand{\KLarge}{K_{\mathrm{large}}}

\newcommand{\HIn}{H_{\textup{in}}}

\newcommand{\HMax}{H_{\max}}
\newcommand{\MMax}{M_{\max}}
\newcommand{\KMax}{K_{\max}}
\newcommand{\MFix}{M_{\textup{fix}}}
\newcommand{\mFix}{m_{\textup{fix}}}
\newcommand{\mRem}{m_{\textup{rem}}}
\newcommand{\NSpill}{N_{\textup{spill}}}
\newcommand{\NEnc}{N_{\textup{enc}}}
\newcommand{\KEnc}{K_{\textup{enc}}}
\newcommand{\kEnc}{k_{\textup{enc}}}
\newcommand{\mEnc}{m_{\textup{enc}}}

\newcommand{\kShort}{k_{\textup{short}}}
\newcommand{\NShort}{N_{\textup{short}}}
\newcommand{\KShort}{K_{\textup{short}}}
\newcommand{\mShort}{m_{\textup{short}}}
\newcommand{\HShort}{H_{\textup{short}}}

\newcommand{\mAdj}{m_{\textup{adj}}}
\newcommand{\kAdj}{k_{\textup{adj}}}

\title{Succinct Dynamic Rank/Select: \\
Bypassing the Tree-Structure Bottleneck}
\author{
  William Kuszmaul\thanks{\texttt{kuszmaul@cmu.edu}. Partially supported by NSF grant CNS-2504471.} \\
  CMU
  \and
  Jingxun Liang\thanks{\texttt{jingxunl@andrew.cmu.edu}.} \\
  CMU
  \and
  Renfei Zhou\thanks{\texttt{renfeiz@andrew.cmu.edu}. Partially supported by the Jane Street Graduate Research Fellowship and the MongoDB PhD Fellowship.} \\
  CMU
}
\date{}

\begin{document}

\maketitle

\begin{abstract}
We show how to construct a dynamic ordered dictionary, supporting insert/delete/rank/select on a set of $n$ elements from a universe of size $U$, that achieves the optimal amortized expected time complexity of $O(1 + \log n / \log \log U)$, while achieving a nearly optimal space consumption of 
$$\log \binom{U}{n} + n / 2^{(\log n)^{\Omega(1)}} + \polylog U$$
bits in the regime where $U = \poly(n)$. This resolves an open question by Pibiri and Venturini as to whether a redundancy (a.k.a.\ space overhead) of $o(n)$ bits is possible, and is the first dynamic solution to bypass the so-called tree-structure bottleneck, in which the bits needed to encode some dynamic tree structure are themselves enough to force a redundancy of $\tilde{\Omega}(n)$ bits.

Our main technical building block is a dynamic balanced binary search tree, which we call the \emph{compressed tabulation-weighted treap}, that itself achieves a surprising time/space tradeoff. The tree supports $\polylog n$-time operations and requires a static lookup table of size $\poly(n) + \polylog U$---but, in exchange for these, the tree is able to achieve a remarkable space guarantee. Its total space redundancy is $O(\log U)$ bits. In fact, if the tree is given $n$ and $U$ for free, then the redundancy further drops to $O(1)$ bits. 
\end{abstract}

\thispagestyle{empty}
\clearpage
\setcounter{page}{1}

\section{Introduction}\label{sec:intro}

In this paper, we revisit one of the most basic questions in space-efficient data structures: How space-efficiently can one implement a dictionary that efficiently supports insertions, deletions, and rank/select queries?

A dictionary that supports rank/select queries is known as a \defn{Fully Indexable Dictionary (FID)}. An FID is said to be \defn{dynamic} if it additionally supports insertions and deletions. In this paper, we will use $n$ to denote the number of elements in an FID, and $U$ to denote the size of the universe $[U]$ from which these elements are selected.

A standard way of building a dynamic FID is with a balanced binary search tree---this gives a time bound of $O(\log n)$ per operation. With more sophisticated data structures, one can do slightly better, achieving an expected time bound of $\Theta(1 + \log n / \log \log U)$ per operation \cite{fredman1993surpassing, ajtai1984hash, patrascu2014dynamic}. So long as $n \le U/2$,  this time bound is known to be optimal for any solution \cite{fredman1989cell}.

What is not known is whether one can construct a dynamic FID that is both time-optimal \emph{and} very space efficient. Any FID must take space at least $\lceil \log \binom{U}{n}\rceil$ bits. A solution that takes space
$$\Big\lceil \log \binom{U}{n}\Big\rceil + R$$
bits is said to have \defn{redundancy} $R$. The goal of a space-efficient solution is to achieve as small a redundancy as possible.

In the static setting \cite{elias1974efficient, fano1971number, patrascu2006timespace, gupta2007compressed, patrascu2010cellprobe, viola2023new, liang2025optimal}, where insertions and deletions are disallowed, it is known how to achieve very small redundancy without compromising query time. In this setting, with a natural universe size $U = \poly(n)$, the optimal time per query is $O(\log \log n)$ \cite{patrascu2006timespace}, and subject to this time bound it is known how to achieve redundancy $R = n/(\log n)^{\Omega(\log \log n)}$ \cite{liang2025optimal}. In fact, for any time bound $T \in \Omega(\log \log n) \cap O(\log n / \log \log n)$, it is known how to support $O(T)$-expected-time queries while achieving redundancy $R = n / (\log n / T)^{\Omega(T)}$. And, although this time vs redundancy tradeoff may seem odd at first glance, it is even known to be optimal for any $T \in (\log n)^{1 - \Omega(1)}$ \cite{viola2023new, patrascu2010cellprobe}.

In contrast, the dynamic setting \cite{pibiri2017dynamic,pibiri2020succinct} has proven much more difficult to get good bounds for. Here, the state of the art, due to Pibiri and Venturini \cite{pibiri2017dynamic,pibiri2020succinct}, is a time-optimal solution with redundancy $R = O(n)$. Whether or not one can achieve redundancy significantly smaller than $O(n)$---even with, say, $(\polylog n)$-time operations---remains a stubbornly open problem \cite{pibiri2017dynamic,pibiri2020succinct}.

\paragraph{The tree-structure bottleneck.} One of the major challenges in designing a space-efficient dynamic solution is the following basic issue. Every known time-optimal solution relies on some sort of dynamic tree structure to store a large portion of the keys (at least, say, $\tilde{\Omega}(n)$ keys). As elements are inserted/deleted, this tree has a dynamically changing structure that must be stored in the data structure. Even if one could eliminate all other sources of redundancy, it is difficult to see how one can avoid incurring redundancy of at least $\tilde{\Omega}(n)$ bits just to encode the dynamically-changing structure of the tree in such a way that it can be navigated by queries. We refer to this obstacle as the \defn{tree-structure bottleneck}.

The tree-structure bottleneck suggests that, even if one could get the redundancy down to $o(n)$, the best bound that one could hope for without significant new techniques would be a bound of $R = n / \polylog n$. This is the barrier that we seek to break in the current paper.

\paragraph{This paper: A very succinct time-optimal solution.} In this paper, we show how to bypass the tree-structure bottleneck and construct a time-optimal solution that, in most parameter regimes, offers a redundancy of $R = n / 2^{(\log n)^{\Omega(1)}}$ bits.

\begin{restatable}{theorem}{MainTheoremFID}
In the word RAM model with $w = \Theta(\log U)$-bit words, we can construct a dynamic FID with the following properties, where $n$ denotes the current size of the FID at any given moment: 
\begin{itemize}
    \item Insertions and deletions take amortized expected time $O(1 + \log n / \log w)$.
    \item Rank/select queries take expected time $O(1 + \log n / \log w)$.
    \item The total space usage in bits is 
    $$\log \binom{U}{n} + O\left(w n \Big/ 2^{(\log n / \log w)^{1/3}}\right) + \polylog U,$$
    and this bound is satisfied even when the data structure is being modified between states. 
    Note that, in the common case where $w \le 2^{(\log n)^{o(1)}}$, this space bound simplifies to $\log \binom{U}{n} + n \big/ 2^{(\log n)^{1/3 - o(1)}}$ bits.
\end{itemize}
    \label{thm:fid}
\end{restatable}

To interpret Theorem \ref{thm:fid}, it is helpful to consider the natural parameter regime of $U = \poly(n)$, and to consider the average redundancy \emph{per key}, given by $r = R / n$. Prior to our work, the state of the art for dynamic time-optimal solutions was a per-key redundancy of $r = O(1)$. Our solution decreases this to $r = 1 / 2^{(\log n)^{\Omega(1)}}$, which is not just sub-constant but is even \emph{inverse quasi-polynomial}.

\paragraph{The main technical building block: A very space-efficient search tree.}
At the heart of our construction is a building block that, even on its own, achieves a quite surprising time-space tradeoff. We show how to construct a balanced binary search tree that completely avoids the tree-structure bottleneck. This binary search tree has several deficits when compared to the data structure in Theorem \ref{thm:fid}: It supports operations with a sub-optimal time bound of $\polylog n$; requires free access to a static lookup table of size $\poly(\nmax) + \polylog U$, where $\nmax$ is a fixed upper bound on $n$; and uses a potentially large amount of temporary space \emph{during} operations. But, subject to these limitations, the tree structure is able to achieve a remarkable space guarantee: It incurs a total redundancy of $O(\log U)$ bits.\footnote{In fact, even this redundancy bound of $O(\log U)$ is in some sense an overcount. If we give the tree free access to the parameters $n, \nmax, U$, without requiring it to store the parameters explicitly, then the redundancy further reduces to $O(1)$ bits.}

\begin{proposition}[Summarized version of \cref{lem:treap}]
  \label{prop:mainsummary}
  Consider the word RAM model with word size $w = \Theta(\log U)$, and consider the problem of constructing a dynamic FID storing a set $S\subseteq [U]$ whose size $n$ always satisfies $n \le \nmax$. Then, there exists a solution that, assuming free access to a static lookup table of size $\poly(\nmax) + \polylog U$, supports $(\polylog n)$-expected-time operations and incurs a redundancy of $O(w)$.
\end{proposition}

We refer to the data structure in Proposition \ref{prop:mainsummary} as the \defn{compressed tabulation-weighted treap}. As the name suggests, the starting point for the data structure is a treap, which is a classical way of implementing a randomized balanced binary search tree \cite{seidel1996randomized}. What makes the treap special---compared to other solutions such as red-black trees or AVL trees---is that it is \defn{(strongly) history-independent} \cite{naor2001antipersistence, hartline2005characterizing}. This means that the structure of the tree, at any given moment, is completely determined by the \emph{current} set of elements, rather than being determined by the history of how the elements got there (e.g., by the order in which elements were inserted).

Notice that history independence offers a potential path around the tree-structure bottleneck. If the tree structure is completely determined by the elements being stored, then in principle it might be possible to encode both the tree structure and the set of elements with the \emph{same bits}. What makes such an approach difficult is that these bits must simultaneously serve two seemingly orthogonal roles: They must encode elements with essentially no redundancy, but they must also allow us to navigate the tree as though we had pointers between its nodes. Moreover, on top of this, the encoding must be dynamic, supporting insertions and deletions, while preserving near-zero redundancy after each operation.

The main insight in this paper is that such an encoding \emph{actually is possible}. First, in Section \ref{sec:treap}, we design a new version of the treap, which we call the \defn{tabulation-weighted treap}, whose structure is designed to be more amenable to compression. The basic idea behind the tabulation-weighted treap is to re-design the weight function used by the treap in a way so that the line between ``elements'' and ``tree structure'' gets blurred. Second, we show how to encode the tabulation-weighted treap as a recursive tree of random variables, where each random variable satisfies a set of properties that makes it compatible with time-efficient compression techniques---we capture this encoding formally in Section \ref{sec:information_model}, where we introduce an intermediate model called the \emph{information model} that allows us to focus on this recursive structure without worrying about the specifics of how to encode variables. Finally, we show how to translate this tree of random variables into an explicit time- and space-efficient encoding whose space bound is deterministic (Section \ref{sec:ram}). This allows us to prove Proposition \ref{prop:mainsummary} (Section \ref{sec:ram_summary}), which we are then able to use in a relatively straightforward way to prove Theorem \ref{thm:fid} (Section \ref{sec:mainthm}).

\paragraph{The surprising power of history independence.} In recent decades, there has been a great deal of work on how to build efficient history-independent data structures. Most of this work has taken one of two perspectives: (1) using history independence as a \emph{privacy property} \cite{naor2001antipersistence, bender2024historyindependent, bender2016antipersistence, buchbinder2006lower, naor2008historyindependent, blelloch2007strongly, tzouramanis2012historyindependence, bajaj2013hifs, bajaj2016practical, hartline2005characterizing, berger2022memoryless} so that, if the state of a data structure is leaked to an adversary, the adversary is not able to recover any nontrivial information about the history; or (2) using history independence as an \emph{algorithmic tool} \cite{amble1974ordered, celis1985robin, behnezhad2019fully, bansal2022balanced, bender2024online, kuszmaul2023strongly} to simplify the analysis of data structures and even to enable the introduction of new faster data structures. Our work brings to the forefront a third perspective: the use of history independence as a technique for building highly space-efficient data structures. 

Of course, the relationship between history independence and space efficiency is in some sense unavoidable: Any data structure that achieves $0$ redundancy must be history-independent, since otherwise it would implicitly store unnecessary information about the history of past operations. What our work suggests, however, is that this relationship should be viewed as more than just a casual observation. It suggests that, at a high level, the right way to design a very space-efficient dynamic data structure is to take a ``history-independence first'' perspective: to start by designing a history-independent data structure whose properties make it amenable to compression, and then to design an encoding of that data structure that allows one to get to near-zero redundancy. 

In addition to bringing history independence to the forefront, our work provides a framework for \emph{how} to convert a history-independent data structure into a very space-efficient one. In designing the tabulation-weighted treap, and showing how to encode it in our ``information model'', we are able to isolate the key structural properties that a history-independent data structure must have in order for it to be converted into a highly space-efficient data structure. We believe that this perspective, and more generally, the high-level template that our work offers, is one of the main contributions of the paper.

\section{Preliminaries and Formalization of Proposition \ref{prop:mainsummary}}

\paragraph{Preliminaries.} Formally, the data structures that we construct in this paper are \defn{Fully Indexable Dictionaries} (FIDs). These are data structures that support the following set of operations on a set $S$ of keys from a universe $[U]$: \texttt{Insert}$(x)$ adds a key $x \in [U] \setminus S$ to $S$; \texttt{Delete}$(x)$ removes a key $x \in S$ from $S$; \texttt{Rank}$(x)$ returns $\# \{y \in S : y < x\}$; and \texttt{Select}$(i)$ returns the $i$-th smallest key in $S$ (if such a key exists). When referring to an FID, we will typically use $n$ to refer to the current number of keys and $\nmax$ to refer to the maximum number of keys that the FID can support.

We are interested in constructing FIDs in the word RAM model with some machine word size $w$. We will assume constant-time arithmetic operations, including logarithms and exponentials.\footnote{The assumption that logarithms and exponentials take $O(1)$ time is simply for convenience. Whenever we need to compute a logarithm or an exponential, it will be to translate between the number of options for a random variable and the number of (fractional) bits needed to encode it, or to round a number to the nearest power of $(1 + 1 / \nmax^3)$. It suffices to perform such calculations with $O(\log \nmax)$ bits of precision (after the decimal point), which means that all of the calculations can (up to appropriate approximations) be performed in $O(1)$ time with the help of an additional lookup table of $\poly(\nmax, w)$ words. Such a lookup table can be incorporated into \cref{lem:treap} while only increasing the sizes of the lookup tables that it uses by at most a $\poly(w)$ factor. When using \cref{lem:treap} to prove \cref{thm:fid}, this increase to the lookup-table size would have no effect on the final bounds of Theorem \ref{thm:fid}.}
We view memory as an infinite tape, and consider the data structure to occupy some prefix of that tape. The \defn{space usage} of the data structure is defined as the size of the prefix it uses at any given moment.\footnote{Later, in Section \ref{sec:ram}, it will be helpful to give a formal name to this memory model (the \defn{Virtual Memory Model}). This name will make more sense in the context of Section \ref{sec:ram}, where we discuss techniques \cite{li2023dynamic,li2024dynamic} for implementing multiple logical instances of the Virtual Memory Model in a single physical memory.} 

Finally, there are several standard notations that we will use throughout the paper. We will use $[j, k)$ to denote the interval $\{j, \; j + 1, \ldots, k - 1\}$, and $[k]$ to denote interval $[0, k)$. Given a discrete probability distribution $D$, we will use $\supp(D)$ to denote the support of $D$, and $D(\circ)$ to denote the probability mass function for $D$. And, finally, given a parameter $\nmax$, we will say that an event $E$ occurs with \defn{high probability in $\nmax$} if the event occurs with probability $1 - O(1/\nmax^c)$ for a sufficiently large positive constant $c$ (typically, $c$ will be determined by other parameters in the algorithm).  

\paragraph{The main technical building block.} With these preliminaries in place, we are now ready to state the main technical result of the paper, which is the construction of a new data structure that we call the compressed tabulation-weighted treap. The properties of this data structure are stated in the following proposition, which is the formal version of \cref{prop:mainsummary} from Section \ref{sec:intro}. 

\begin{restatable}[Compressed Tabulation-Weighted Treaps]{proposition}{TreapLemma}
\label{lem:treap}
Let $\nmax$ and $\Umax$ be fixed upper bounds on $n$ and $U$, and let $U \le \Umax$ be an arbitrary universe size. Finally, let $w = \Theta(\log \Umax)$ denote the machine word size used in the word RAM model.
Then the compressed tabulation-weighted treap is a dynamic FID that can store up to $\nmax$ keys at a time from universe $U$, and that satisfies the following properties, where $n$ denotes the current number of keys:
\begin{itemize}
\item The data structure uses $\log \binom{U}{n} + O(w)$ bits of space. When $n$ changes due to insertions or deletions, the size of the data structure changes accordingly.
\item Each insertion or deletion takes $O(\log^3 n)$ expected time, and each rank or select query takes $O(\log^2 n)$ expected time.
\item The data structure requires access to a static lookup table of $\poly(\nmax, \log \Umax)$ words, depending only on $\nmax$, $\Umax$, and global randomness. This table can be shared among multiple data structure instances and can be computed in $\poly(\nmax, \log \Umax)$ time.
\item When an operation is being performed, the data structure is allowed to temporarily use up to $O(nw)$ bits of space.
\end{itemize}
\end{restatable}

The main feature of the compressed tabulation-weighted treap is that it is \emph{very} space efficient, using only $O(1)$ machine words of space beyond the information-theoretic optimum of $\log \binom{U}{n}$. In fact, these extra $O(1)$ machine words are only needed to store the parameters $n$, $\nmax$, $U$, and $\Umax$. If these parameters are given to us for free, then space bound becomes $\log \binom{U}{n} + O(1)$ bits. It should be emphasized that, in addition to being space efficient, the data structure is also surprisingly flexible: Rather than achieving a space bound based on $\nmax$, it achieves a bound dictated by the \emph{current} number $n$ of keys at any given moment. 

At the same time, as we discussed in Section \ref{sec:intro}, the data structure also comes with some nontrivial drawbacks: It requires seemingly quite large lookup tables of size $\poly(\nmax)$; it achieves time bounds of the form $\polylog n$, rather than the bound of $O(\log n)$ typically achieved by a treap, or the even better bound of $O(\log n / \log \log n)$ achieved by optimal FIDs (for the case of $U = \poly(n)$); and it requires a potentially large amount of intermediate space while operations are being performed.

Fortunately, all of these drawbacks can be removed with the same simple trick: Rather than using \Cref{lem:treap} to store an entire FID of some size $N$, we can use it just to store small subtrees of size roughly $2^{(\log n / \log w)^{1/3}}$.
With this idea, we can obtain the main result of the paper, Theorem \ref{thm:fid}, as a simple corollary of \Cref{lem:treap} (see Section \ref{sec:mainthm}). Thus, the vast majority of the paper is spent proving \cref{lem:treap} (Sections \ref{sec:treap}, \ref{sec:information}, and \ref{sec:ram}).

\section{The (Uncompressed) Tabulation-Weighted Treap}\label{sec:treap}

In this section, we present and analyze the (uncompressed) tabulation-weighted treap, a variation on the classical treap data structure that has been modified in order to enable the compression techniques that we develop later in the paper. We will also discuss concretely the properties of the tabulation-weighted treap that end up being important in later sections. 

\paragraph{Reviewing the classical treap.}
We begin by reviewing the classical treap data structure \cite{seidel1996randomized, motwani1995randomized}. The treap assumes access to a random \defn{weight function} $h(x)$ that maps every key $x \in [U]$ to an i.i.d.\ uniformly random weight $h(x) \in [W] \defeq [\poly(\nmax)]$, where the range size $W$ is chosen to be a sufficiently large power-of-two polynomial of $\nmax$. 

To store a set $S \subseteq [U]$, the structure of the treap is then defined as follows. Whichever key $p \in S$ has the largest weight in $S$ is called the \defn{pivot} of $S$,\footnote{For now, let us assume that the key with the largest weight is unique. In particular, in all of the constructions in this paper, if there are multiple keys that have the same largest weight, we will declare that an unlikely failure event has occurred, and we will switch to a more time-consuming way to encode the key set that does not make use of treaps (see, in particular, \cref{sec:failure_mode}).} and is used as the root of the tree. The remaining keys are divided into two subsets, $\Sleft$ and $\Sright$, denoting the sets of keys that are smaller than $p$ and larger than $p$, respectively. Finally, treaps for $\Sleft$ and $\Sright$ are constructed recursively, and used as the left and right subtrees of the root $p$.

Using this basic design, the treap can support insertions, deletions, and rank/select queries as follows. When inserting or deleting a key in the treap, the pivot of some subtree may change, in which case we rebuild that entire subtree in linear time.%
\footnote{Most classical implementations of treaps take a more careful approach than this, maintaining the tree structure with rotation operations, rather than rebuilding entire subtrees \cite{seidel1996randomized, motwani1995randomized}. However, for this paper, the rebuild-based variant will serve as a better starting point for us, as it will allow for time-efficient operations even in the compressed version of the tree that we build in Section \ref{sec:ram}.}
To support queries, we also record on each node $u$ the number of nodes in $u$'s subtree, so that both rank and select queries can be answered by traversing from the root of the treap down to some leaf node. All of these operations can be shown to take $O(\log n)$ expected time.

\paragraph{The tabulation-weighted treap.} One of the key insights in this paper is that, by just slightly changing the design of the treap, we can arrive at a data structure that is much more amenable to compression.

The difference between the classical treap and the tabulation-weighted treap is in how we define the weight function $h$. Whereas the classical treap assumes a fully random $h:[U] \rightarrow [W]$, the tabulation-weighted treap intentionally imposes additional combinatorial structure on $h$. This structure endows the treap with certain properties (which we will discuss concretely in a moment) that enable many of the arguments later in the paper. 

For every key $x \in [U]$, we define $\xlow$ as the $\log W$ least significant bits of $x$, define $\xmid$ as the $3 \log W$ bits immediately more significant than $\xlow$, and define $\xhigh$ as the remaining bits beyond $\xlow$ and $\xmid$. Formally, we write
\[
x = \xhigh \cdot W^4 + \xmid \cdot W + \xlow.
\]
For notational convenience, we also define
\[
\xhighmid \defeq \xhigh \cdot W^3 + \xmid, \qquad \xmidlow \defeq \xmid \cdot W + \xlow.
\]
If we fix a value of $\xhigh$ and iterate over all values of $\xmid$ and $\xlow$, then we get an interval of size $W^4$ in the universe $[U]$, and we call it a \defn{superchunk}. Similarly, fixing the values of $\xhigh$ and $\xmid$ defines an interval of size $W$, which we call a \defn{subchunk}.

To define the weight function, we create $\poly(W)$ auxiliary arrays of size $W^4$, denoted by $A_1, \ldots, A_{\poly(W)}$, where each array $A_i[0 \ldots W^4\!-\!1]$ is a concatenation of $W^3$ independent random permutations over $[W]$. We also take a pairwise-independent hash function $f : [U / W^4] \to [\poly(W)]$. Then, for each superchunk indexed by $\xhigh$, the $f(\xhigh)$-th auxiliary array $A_{f(\xhigh)}$ gives the sequence of weights for the $W^4$ elements in the superchunk. Formally, we define
\[
h(x) \defeq A_{f(\xhigh)}[\xmidlow]. \numberthis \label{eq:weight-function}
\]

One can think of $h(x)$ as being a variation on tabulation hashing  \cite{patrascu2012power}. Classical tabulation hashing would break the key into chunks $C_1, C_2, \ldots$ of $\log \nmax$ bits, and define $h(x)$ to be $\bigoplus_i A_i[C_i]$. Our weight function borrows the idea of using part of the key to index into an array, but uses the rest of the key (along with a pairwise-independent hash function $f$) to determine which array. As we will see, this simple structure is critical to the algorithmic approaches that we take in this paper.

\paragraph{Core properties. }Before continuing, it is worth emphasizing the key properties that the tabulation-weighted treap has. The three properties that end up being most important later in the paper are: 
\begin{enumerate}
    \item \textbf{Ability to store low-order bits implicitly.} For a given value of $\xhighmid$, each option for $\xlow$ results in a different weight $h(x)$. This means that, given $\xhighmid$ and $h(x)$, we can fully recover $\xlow$. Consequently, in our constructions, we will be able to design encodings that focus on storing $\xhighmid$ and $h(x)$ for each pivot $x$, and that then recover $\xlow$ indirectly.
    \item \textbf{A small number of options for how small intervals behave. } Define the \emph{weight profile} of an interval $[a, b) \subseteq [U]$ to be the tuple $(h(a), h(a + 1), \ldots, h(b - 1))$. Consider some interval size $s \le W^4$. If we used a fully random weight function, then every interval of the form $[a, a + s)$ might have a different weight profile. But, with our weight function, there are at most $\poly(W)$ possible weight profiles for intervals of size $s$. This will allow us, later in the paper, to use lookup-table techniques in order to construct very space-efficient encodings when we are compressing a subtree whose keys are all known to be in some very small range.
    \item \textbf{\boldmath{}Near-uniformity of $\xhighmid \mid h(x)$.} Given an interval $[aW, bW)$, where $a, b$ are integers, a uniformly random key $x \in [aW, bW)$, and a value $r \in [W]$ for $h(x)$, the conditional random variable $\xhighmid \mid h(x) = r$ is uniformly random in $[a, b)$, regardless of the value of $r$. This will be important in our construction because, in order to encode random variables space- and time-efficiently, we will need them to either (1) have a very small support or (2) have a large support but be uniformly random, like $\xhighmid \mid h(x) = r$.
\end{enumerate}

In addition to the above properties, we will also need the following lemma, which shows that (with high probability) the tabulation-weighted treap behaves in essentially the same way as the classical treap would.

\begin{lemma}
\label{lem:unique-weight}
For any set $S = \BK{x_1, x_2, \ldots, x_{n}}$ of $n \le \nmax$ keys, with probability at least $1 - 1/\poly(\nmax)$, $h(x_1), \ldots, h(x_n)$ are pairwise distinct.
\end{lemma}

\begin{proof}
We first analyze the probability of $h(x_i) = h(x_j)$ for a given pair $x_i \ne x_j$ of keys. This can happen in two cases:
\begin{enumerate}
\item $h(x_i)$ and $h(x_j)$ access the same entry of the same auxiliary array, which happens when $f(\xhigh[i]) = f(\xhigh[j])$ and $\xmidlow[i] = \xmidlow[j]$. This case requires a hash collision $f(\xhigh[i]) = f(\xhigh[j])$ of the function $f$ on two different inputs, which occurs with probability at most $1 / \poly(W) = 1 / \poly(\nmax)$ since $f$ is pairwise independent.
\item $h(x_i)$ and $h(x_j)$ access either different auxiliary arrays or different entries in the same array. In this case, $h(x_i)$ and $h(x_j)$ collide with probability only $1 / W = 1 / \poly(\nmax)$ by design of the auxiliary arrays.
\end{enumerate}
Adding these two cases together, we know that $\Pr[h(x_i) = h(x_j)] \le 1 / \poly(\nmax)$. Taking a union bound over all pairs of keys concludes the proof.
\end{proof}

\cref{lem:unique-weight} indicates that, when we construct a tabulation-weighted treap on a key set $S$, all of the keys in $S$ will have unique weights with high probability in $\nmax$. If this property holds, then, by symmetry, the relative order of the weights of the keys follows a uniform distribution over all $n!$ possible orders. This means that the tabulation-weighted treap behaves exactly the same as if it used a fully random weight function---we therefore know by the classic treap analysis that the depth of the treap is $O(\log n)$ with high probability in $n$ \cite{motwani1995randomized} and that the expected time to perform an insertion or deletion is also $O(\log n)$ (although this time bound will grow in later sections as we make the tree more space efficient). 

We remark that, if the property in \cref{lem:unique-weight} fails (i.e., if two keys share the same weight), the treap still performs correctly, though with a trivial time bound of $O(n)$; we handle this case in \cref{sec:failure_mode}. Since the latter bad case occurs only with probability $1 / \poly(n)$, the overall time complexity of the tabulation-weighted treap is $O(\log n)$ in expectation.

\section{A Space-Efficient Encoding in the Information Model}
\label{sec:information} \label{sec:information_model}

In this section, we show how to encode the tabulation-weighted treap in a simplistic model, which we call the \defn{information model}, where we allow the data structure to be encoded in terms of random variables instead of directly in bits. 

\paragraph{The information model. }In the information model, the data structure is given the parameters $n$, $\nmax$, and $U$, as well as the weight function $h$ used by the treap. (The data structure is not responsible for storing any of these.) The data structure must then represent the treap as a sequence of variables $X_1, \ldots, X_k$. For each variable $X_i$, the data structure gets to specify a probability distribution $D_i$ that is used to encode $X_i$. The distribution $D_i$ must be fully determined by the earlier variables $X_1, X_2, \ldots, X_{i - 1}$ and the known parameters $n, \nmax, U, h$. 

The information model defines the data structure's \defn{size} to be 
$$\sum_i \log \bigl(1/D_i(X_i)\bigr)$$
bits.\footnote{Throughout this paper, $D(\cdot)$ denotes the probability mass function of a discrete distribution $D$.} This quantity is the optimal number of bits needed to encode each $X_i$ under distribution $D_i$, allowing for fractional bits. Note that the model does not require the algorithm to actually encode each $X_i$ into memory bits---instead, the space usage for each $X_i$ is directly defined to be $\log \bigl(1/D_i(X_i)\bigr)$ bits. 

The data structure is permitted to specify a \defn{failure event} $E$ that depends on the set $S$ of keys and the weight function $h$. For a given set $S$, this failure event must occur with probability at most $1 / \poly(\nmax)$ over the randomness of $h$. Our goal is to construct a data structure (and a failure event) such that, if the failure event does not occur, then the data structure's size $\sum \log \bigl(1/D_i(X_i)\bigr)$ is at most 
\begin{equation}\log \binom{U}{n} + 1
\label{eq:spaceguar}
\end{equation}
bits. 

\paragraph{Requiring each $D_i$ to be efficiently encodable/decodable.}
A priori, the information model described above is too underconstrained to be useful. One could, for example, define the entire data structure to be a single variable $X_1$ encoded using the distribution $D_1$ that is uniformly random over all treaps of size $n$ with elements from universe $[U]$ and using weight function $h$. Such an encoding would trivially achieve an optimal space of $\log \binom{U}{n}$ bits, but would not be helpful for our larger goal of constructing a space-efficient data structure in the word RAM model.

When translating the data structure from the information model into the RAM model in Section \ref{sec:ram}, we will not be able to handle arbitrarily complicated distributions $D_i$. In order for us to be able to encode/decode a distribution efficiently, we will need it to either (1) be uniform over some range; or (2) have support size at most $\poly(\nmax)$. Additionally, encoding/decoding a small-support distribution will require storing a lookup table of size $\poly(\nmax)$, meaning that we will only be able to support $\poly(\nmax)$ such distributions. This means that, formally, we will need the distributions $D_1, D_2, \ldots, D_k$ to satisfy the following \defn{Simple-Distributions Property}: There must exist a family $\mathcal{F}$ of distributions such that
\begin{itemize}
    \item $|\mathcal{F}| \le \poly(\nmax)$;
    \item each $D \in \mathcal{F}$ satisfies $\supp(D) \le \poly(\nmax)$;
    \item and each $D_i$ is guaranteed to either be in $\mathcal{F}$ or to be a uniform distribution over some range. 
\end{itemize}
It should be emphasized that the restriction $\supp(D) \le \poly(\nmax)$ is a relatively strong condition since $\nmax$ may be much smaller than $U$ (and, indeed, for all of the main results of this paper, we will need $U = \nmax^{\omega(1)}$). Thus one should think of each $D_i$ as either being uniform or having a very small support. 

The main result that we will prove in this section is that, in the information model, there exists an encoding of the tabulation-weighted treap that satisfies the Simple-Distributions Property while also achieving the space bound \eqref{eq:spaceguar}. 

\paragraph{A remark on the role of randomness.}
Before continuing, it is worth remarking on the role of randomness in the constructions that we are about to present. 

From the perspective of the failure event $E$, the weight function $h$ is a random variable. 
But, from the perspective of the encoding algorithm (i.e., the algorithm that determines the $X_i$s and $D_i$s), the weight function $h$ is known and fixed. In fact, the encoding is completely deterministic as a function of $n, \nmax, U, h, S$. Likewise, for any set $S$ and any weight function $h$ that does not trigger the failure event $E$, the final space guarantee given by \eqref{eq:spaceguar} is also deterministic. The space bound does not make any distributional assumptions on the set $S$, or make use of the randomness in the weight function $h$ (besides using it to guarantee that the failure event $E$ is rare). The bound holds for any set $S$ and for any tabulation-weight function $h$, so long as $E$ does not occur. 

The fact that the space bound \eqref{eq:spaceguar} is deterministic may seem odd (at first glance) given that encoding is based on a sequence $D_1, D_2, \ldots$ of \emph{probability distributions}. It should be emphasized, however, that even though the encoding of $X_i$ is determined by a distribution $D_i$, the variable $X_i$ itself is not actually random---the entire sequence $X_1, \ldots, X_k$ is fully determined by the set $S$ being stored and the known variables $n, \nmax, U, h$. 

Although the encoding algorithm does not make any distributional assumptions on $S$, it will sometimes be helpful, in order to \emph{design} the distributions $D_1, D_2, \ldots$, to imagine what the distribution of $X_i$ \emph{would be} if the input set $S$ were to be drawn from some probability distribution. Thus, we will sometimes regard $S$ as being drawn from a specific distribution in order to motivate our choices for each $D_i$, even though the final data structure does not need any distributional assumption.

\paragraph{Algorithm overview.}
Our encoding scheme for the tabulation-weighted treap is recursive, leveraging the fact that any subtree of a treap is itself a treap. To encode the treap for a key set $S$, we first encode information at the root node, including the pivot $p \in S$ and its rank $r \defeq \rank_S(p)$, where the rank is the number of keys in $S$ that are smaller than $p$. We then recursively encode the left and right subtrees of the root, which are treaps on smaller key sets.

In this recursion, each subproblem requires encoding a key set $S$ of $n$ keys drawn from a key range $[a, b)$, where $0 \le a < b \le \Umax$, such that every key $x \in S$ satisfies $h(x) \le \hmax$. We define
\[
V \defeq V(a, b, \hmax) \defeq \#\BK{x \in [a, b) : h(x) \le \hmax}
\]
as the number of valid keys that may appear in $S$. Hence, the number of possible key sets $S$ is
\[
\mathcal N(n, a, b, \hmax) \,=\, \binom{V(a, b, \hmax)}{n}.
\]
We will show, by induction on $n$, that there is an encoding scheme that satisfies the Simple-Distributions Property and that has space usage
\[
H(n, a, b, \hmax) \,\le\, \log \mathcal N(n, a, b, \hmax) + n/\nmax
\numberthis \label{eq:induction_information}
\]
bits.

The base case of the induction is $n = 0$, for which \eqref{eq:induction_information} holds because both sides are zero. The induction concludes when $n$ matches the number of keys in the actual key set to be encoded; by choosing $[a, b) = [0, U)$ and $\hmax = W$ in \eqref{eq:induction_information}, we obtain an encoding that uses $\log \binom{U}{n} + 1$ bits in the information model, as desired.

We remark that, in addition to satisfying the Simple-Distributions Property, our construction also has another property that will be essential in later sections: In the recursive structure, the distributions used to encode the variables at a given node $u$ depend only on the variables at the $O(\log n)$ nodes on the root-to-$u$ path.
The result of this is that (1) queries only need to decode $O(\log n)$ variables; and that (2) modifications to the treap have an isolated effect, changing only the encodings for variables in the affected subtree and in the path from the root to that subtree. This, along with the Simple-Distributions Property, is what will make it possible in Section \ref{sec:ram} to construct a time-efficient encoding in the word RAM model. 

\paragraph{Selecting a failure event. } Throughout the construction, we assume the key set $S$ satisfies the following two properties:
\begin{itemize}
\item In each subproblem in the recursion, there is a unique key with the largest weight.
\item The weight of every key in $S$ is at least $\nmax^4$. As a result, we have $\hmax \ge \nmax^4$ for every subproblem.
\end{itemize}
Formally, we define the failure event $E$ for the construction to be the case where the two properties do not hold. 

Note that, for any given set $S$, and for a random tabulation-weight function $h$, we have that $\Pr[E] \le 1 / \poly(\nmax)$. In particular, the first property holds with probability $1 - 1/ \poly(\nmax)$ by \cref{lem:unique-weight}; and the second property follows from the fact that $W = \poly(\nmax)$, meaning that each of the $\le \nmax$ keys has probability at most $\nmax^4 / W = 1 / \poly(\nmax)$ of being less than $\nmax^4$.

\paragraph{Section outline.}
For the remainder of this section, we complete the inductive argument to prove \eqref{eq:induction_information} by considering two cases, the \emph{large-universe case} where $b- a \ge W^4$ (\cref{sec:info_large_universe}), and the \emph{small-universe case} where $b - a < W^4$ (\cref{sec:info_small_universe}). 

The main notations used in the inductive proof are summarized in \cref{table:notation_info}---these notations are also defined formally as they are introduced throughout the section.

\renewcommand{\arraystretch}{1.2}
\newcommand{\rowskip}{0.7ex}

\newcommand{\tablewrap}[1]{%
\begin{tabular}[t]{@{}l@{}}%
#1%
\end{tabular}}

\begin{table}[t]
\centering
\caption{Notations in \cref{sec:information}}\label{table:notation_info}
\begin{tabular}{l p{0.65\textwidth}}
\toprule
\textbf{Notation} & \textbf{Definition} \\
\midrule
$n$ & Number of keys in the current subproblem \\
$\nmax$ & Upper bound on the maximum $n$ for the inductive proof \\
$\Umax$ & Upper bound on the key universe size $U$ \\
$W$ & Range of weights, $W = \poly(\nmax)$ \\
$[a,b)$ & Key range for the current subproblem \\
$[\tilde{a},\tilde{b})$ & Key range with endpoints rounded to the nearest multiples of $W$ \\
$\ell \defeq (\tilde{b} - \tilde{a})/W$ & Number of subchunks in $[\tilde{a},\tilde{b})$ \\
$\hmax$ & Maximum allowed weight in the current subproblem \\
$h(\cdot)$ & Weight function (globally fixed) \\
$S$ & Key set to be encoded in the current subproblem \\
$V \defeq V(a,b,\hmax)$ & $\#\{x \in [a,b) : h(x) \le \hmax\}$, the number of valid keys \\
$\mathcal{N}(n,a,b,\hmax)$ & Number of possible key sets in the current subproblem, $\binom{V(a,b,\hmax)}{n}$ \\
$H(n,a,b,\hmax)$ & Size of the encoding (in bits) for the current subproblem \\
$p$ & Pivot of the key set $S$ \\
$\phighmid \defeq \floor{p/W}$ & High-middle part of the pivot $p$ \\
$r = \rank_S(p)$ & Rank of the pivot $p$ \\
\midrule
$\DistSet$ 
  & Distribution of $S$ taking all possible key sets (with unique max-weight keys) with equal probability \\[\rowskip]
$\DistWeight\bk[\big]{h(p)}$
  & Distribution of $h(p)$ for a random key set $S \sim D^S_{\tilde{a},\tilde{b},n,\hmax}$ \\[1.5ex]
$\DistWeightTd\bk[\big]{h(p)}$
  & Approximation to $\DistWeight$ that does not depend on $\tilde{a}, \, \tilde{b}$ \\[\rowskip]
$V_L, V_R$ 
  & \tablewrap{
    Number of valid keys for the left and right subtrees \\
    $V_L \defeq V(\tilde{a},\,p,\,h(p)-1)$, \quad
    $V_R \defeq V(p + 1,\,\tilde{b},\,h(p)-1)$ \\[\rowskip]
  } \\
$\DistRank(r)$ 
  & Distribution of $r = \rank_S(p)$ \\[\rowskip]
$\DistSmall(\Delta p,r)$
  & Distribution of $(\Delta p,r)$ in the small-universe case, where $\Delta p = p - a$ and the weight profile
  in the key range is $(A_i \circ A_j)[a' \ldots b'\!-\!1]$ \\
\bottomrule
\end{tabular}
\end{table}

\subsection{The Large-Universe Case}
\label{sec:info_large_universe}

In this subsection, we show how to achieve \eqref{eq:induction_information} when $b - a \ge W^4$. We present our algorithm step by step below.

\paragraph{Step 1: Rounding the key range.}

The first step of the encoding algorithm is to round $a$ (resp.\ $b$) down (resp.\ up) to the nearest multiple of $W$, denoted by $\tilde{a}$ (resp.\ $\tilde{b}$). Since the key range $b - a$ is sufficiently large, this adjustment does not significantly affect the subproblem's optimal encoding size. In particular, we have the following claim.

\begin{claim}
$\displaystyle \log \mathcal{N}\bigl(n,\tilde{a},\tilde{b},\hmax\bigr)
- \log \mathcal{N}\bigl(n,a,b,\hmax\bigr)
\le O\bk*{\frac{1}{n_{\max}^2}}.$
    \label{clm:rounding_loss}
\end{claim}
\begin{proof}
Expanding the left side gives    
\begin{align*}
&\log \mathcal{N}\bigl(n,\tilde{a},\tilde{b},\hmax\bigr)
- \log \mathcal{N}\bigl(n,a,b,\hmax\bigr)
= \log \binom{V(\tilde{a},\tilde{b},\hmax)}{n}
- \log \binom{V(a,b,\hmax)}{n} \\
& \le \log \bk*{\frac{\binom{V(a,b,\hmax)+2W}{n}}{\binom{V(a,b,\hmax)}{n}}}
\le n \log \bk*{\frac{V(a,b,\hmax) + 2W - n + 1}{V(a,b,\hmax) - n + 1}}. \numberthis \label{eq:intermediateloss}
\end{align*}
Since $V(a, b, \hmax) \ge V(a, b, 0) \ge (b - a - 2W)/W \ge \tfrac12 W^3$, we can further upper bound \eqref{eq:intermediateloss} by
\begin{align*}
& n \log \bk*{\frac{\tfrac12 W^3 + 2W - n + 1}{\tfrac12 W^3 - n + 1}},
\end{align*}
which, by the inequality $W > \nmax \ge n$, is at most
\begin{align*}
n \log \bk*{\frac{\tfrac12 W^3 + 2W - W}{\tfrac12 W^3 - W}} \le 
n \cdot \log \bk*{1 + O(1/W^2)}
\le O\bk*{\frac{n}{W^2}}
\le O\bk*{\frac{1}{n_{\max}^2}},
\end{align*}
as desired.
\end{proof}

\paragraph{Step 2: Encoding $h(p)$.}

We define $\DistSet$ to be the uniform distribution over all possible key sets $\hat{S}$ of $n$ keys from $[\tilde{a}, \tilde{b})$ with weight at most $\hmax$, under the condition that there is a unique key in $\hat{S}$ with the largest weight. Since we already assumed that $\hmax \ge \nmax^4$, a random key set satisfies the condition of having a unique largest-weight key with probability $\ge 1 - 1/\nmax^2$, thus
\[
\binom{V(\tilde a, \tilde b, \hmax)}{n} \cdot (1 - 1 / \nmax^2) \le \abs[\big]{\supp\bk[\big]{\DistSet}} \le \binom{V(\tilde a, \tilde b, \hmax)}{n}.
\numberthis \label{eq:distset_supp_size}
\]

As a way to motivate the distributions that we will use to encode $(p, r)$, in the next three steps of the algorithm, we regard $S$ (the key set to be encoded) as drawn from $\DistSet$. It is worth emphasizing (as discussed earlier) that the role of $\DistSet$ here is simply to help us design the distributions used in the encoding---once those distributions are selected, the final encoding works deterministically, without any distributional assumptions. 

We first encode the weight of the pivot $h(p) \in [0, \hmax]$. We denote by $\DistWeight$ the distribution of $h(p)$ when $S$ follows the distribution $\DistSet$. By definition,
\[
\DistWeight\bigl(h(p)\bigr) \,=\, \frac{\#\BK[\big]{p' \in [\tilde a, \, \tilde b) : h(p') = h(p)} \cdot \binom{V(\tilde a, \tilde b, h(p) - 1)}{n - 1}}{\abs[\big]{\supp\bk[\big]{\DistSet}}},
\numberthis \label{eq:distweight}
\]
where the first factor in the numerator is the number of ways to choose the pivot, and the second factor is the number of ways to choose the remaining $n - 1$ keys with weights strictly less than $h(p)$.

Ideally, we would encode $h(p)$ according to its true distribution $\DistWeight$. However, this distribution depends on the size of the key range $[\tilde{a}, \tilde{b})$, so there are $\poly(\Umax)$ different distributions of this type. If we used them directly, we would violate the first requirement of the Simple-Distributions Property. To resolve this issue, we derive an approximation to $\DistWeight$ that is independent of $[\tilde{a}, \tilde{b})$.

Recall that in our construction of the weight function $h$, the weights of the keys in a \emph{subchunk}\footnote{Recall that a subchunk is an aligned interval of length $W$.} of length $W$ form a permutation over $[W]$. This means that the weights $h(\tilde{a}), h(\tilde{a} + 1), \ldots, h(\tilde{b} - 1)$ are evenly distributed across $[W]$. If we were to select $n$ i.i.d.\ samples $v_1, \ldots, v_n$ uniformly at random from the keys in $[\tilde{a}, \tilde{b})$ with weights at most $\hmax$, then the weight $h(v_i)$ for each sample $v_i$ would follow a uniform distribution over $[0, \hmax]$, so the probability of exactly one of the samples having weight $h(p)$ and the rest having weights smaller than $h(p)$ would be
$$\sum_{i = 1}^n \Pr[h(v_i) = h(p)] \cdot \prod_{j \in [1, n] \setminus \{i\}} \Pr[h(v_j) < h(p)] = \sum_{i = 1}^n \frac{1}{\overline{h} + 1} \cdot \prod_{j \in [1, n] \setminus \{i\}} \frac{h(p)}{\overline{h} + 1} = \frac{n}{\overline{h} + 1} \cdot \left( \frac{h(p)}{\overline{h} + 1}\right)^{n - 1}.$$
This suggests that we should be able to use $\frac{n}{\overline{h} + 1} \cdot \left( \frac{h(p)}{\overline{h} + 1}\right)^{n - 1}$, which depends only on $\overline{h}$, $h(p)$, and $n$, as a good approximation for $\DistWeight(h(p))$. We analyze this approximation in the following claim.

\begin{claim}
$$\DistWeight\bigl(h(p)\bigr) = \frac{n}{\overline{h} + 1} \cdot \left( \frac{h(p)}{\overline{h} + 1}\right)^{n - 1} \cdot (1 \pm O(1/ \nmax^2)).$$
\label{clm:iidapprox}
\end{claim}
\begin{proof}
Letting $\l \defeq (\tilde b - \tilde a)/W \ge W^3$ be the number of subchunks in $[\tilde a, \, \tilde b)$, 
we have closed forms
\[
\#\BK[\big]{p' \in [\tilde a, \, \tilde b) : h(p') = h(p)} \,=\, \l, \qquad
V\bk[\big]{\tilde a, \tilde b, h(p) - 1} = \l \cdot h(p).
\]
Substituting into \eqref{eq:distweight} shows\footnote{We use $x^{\underline{k}} = x(x-1)\cdots(x-k+1)$ to denote the falling factorial of $x$ with $k$ factors.}
\begin{align*}
\DistWeight\bigl(h(p)\bigr) 
&= \frac{\l \cdot \binom{\l \cdot h(p)}{n - 1}}{\abs[\big]{\supp\bk[\big]{\DistSet}}} \numberthis \label{eq:dist_weight_defn} \\
&= \frac{\l \cdot \binom{\l \cdot h(p)}{n - 1}}{\binom{V(\tilde a, \tilde b, \hmax)}{n} \cdot (1 \pm O(1 / \nmax^2))}
= \frac{\l \cdot \binom{\l \cdot h(p)}{n - 1}}{\binom{\l \cdot (\hmax + 1)}{n}} \cdot (1 \pm O(1 / \nmax^2)) \\
&= \frac{\l \cdot n \cdot (\l \cdot h(p))^{\underline{n-1}}_{\phantom{M}}}{(\l \cdot (\hmax + 1))^{\underline{n}}_{\phantom{M}}} \cdot (1 \pm O(1 / \nmax^2)) \\
&= \frac{\l \cdot n \cdot (\l \cdot h(p))^{n-1}}{(\l \cdot (\hmax + 1))^{n}} \cdot (1 \pm O(1 / \nmax^2)) \\
&= \frac{n}{\hmax + 1} \cdot \bk*{\frac{h(p)}{\hmax + 1}}^{n-1} \cdot (1 \pm O(1 / \nmax^2)),
\numberthis \label{eq:distweight-simplify}
\end{align*}
where the second equality holds according to \eqref{eq:distset_supp_size};
and the second to last equality holds because for any integer $A \ge n$, $A^{\underline{n}}_{\phantom{M}} = A^n \cdot (1 \pm O(n^2 / A))$, and here we have $\l \ge W^3 \gg \nmax^4$. 
\end{proof}

Define the distribution $\DistWeightTd$, with support $[\overline{h}] \cup \{\emptyset\}$ so that 
$$\DistWeightTd(h(p)) \defeq \frac{n}{\hmax + 1} \cdot \bk*{\frac{h(p)}{\hmax + 1}}^{n-1},$$
for $h(p) \in [\overline{h}]$, and $$\DistWeightTd(\emptyset) = 1 - \sum_{i  = 0}^{\overline{h} - 1} \DistWeightTd(h(p)).$$
The final probability $\DistWeightTd(\emptyset)$ exists so that $\DistWeightTd$ is a valid probability distribution. 

Our algorithm encodes $h(p)$ using distribution $\DistWeightTd$. Since $\DistWeightTd$ depends only on $n$ and $\hmax$---both of which have $\poly(\nmax)$ possible values---and has a support of size $\overline{h} + 1 \le W  + 1\le \poly(\nmax)$, our use of the distribution $\DistWeightTd$ satisfies the Simple-Distributions Property.

\paragraph{Step 3: Encoding $\phighmid$.}

Next, we encode the pivot $p$ conditioned on its weight $h(p)$. Critically, this is equivalent to encoding $\phighmid \defeq \floor{p / W}$, because given $\phighmid$, there is a one-to-one correspondence between the lowest $\log W$ bits of $p$ (denoted $\plow$) and $h(p)$, according to our construction of the weight function (see \cref{sec:treap}). 

The value $\phighmid$ is an integer in $\bigl[\tilde{a}/W, \,\tilde{b}/W\bigr)$ and follows a uniform distribution. This is because, conditioned on $h(p)$, every key $p' \in [\tilde{a}, \tilde{b})$ with $h(p') = h(p)$ has the same probability
\[
\frac{\binom{V(\tilde{a}, \tilde{b}, h(p) - 1)}{n - 1}}
{\abs[\big]{\supp\bk[\big]{\DistSet}}}
\]
of being the pivot, and because there is exactly one such $p'$ for each possible value of $\phighmid \in \bigl[\tilde{a}/W, \,\tilde{b}/W\bigr)$. Consequently, in our algorithm, we encode $\phighmid$ under the uniform distribution, which takes $\log \frac{\tilde{b} - \tilde{a}}{W} = \log \l$ bits.

\paragraph{Step 4: Encoding the rank of $p$.}

Thus far, we have encoded the pivot $p$ for the key set. The next step is to encode its rank $r \defeq \rank_S(p) \defeq \#\BK{x \in S : x < p}$, which is an integer in $[n]$. Given $\tilde{a}$, $\tilde{b}$, and $p$, let $V_L$ and $V_R$ denote the number of keys with weight $< h(p)$ in $\bigl[\tilde{a}, p\bigr)$ and $\bigl[p + 1, \tilde{b}\bigr)$, respectively. Specifically,
\[
V_L \defeq V\bk[\big]{\tilde a, \, p, \, h(p) - 1}, \qquad V_R \defeq V\bk[\big]{p + 1, \, \tilde b, \, h(p) - 1}.
\]
We then define $\DistRank$ as the distribution of $r$ when $S$ is drawn from $\DistSet$, conditioned on $p$ being the pivot. By definition,
\[
\DistRank(r) \,=\, \frac{\binom{V_L}{r} \binom{V_R}{n - r - 1}}{\binom{V_L + V_R}{n - 1}}.
\numberthis \label{eq:def-distrank}
\]

Similar to Step 2, the true distribution $\DistRank$ of $r$ depends on $V_L$ and $V_R$, each of which may take $\Theta(\Umax)$ values. Directly using these distributions would require too many lookup tables for encoding $r$ into memory bits. Hence, we approximate $\DistRank$ by replacing $V_L$ and $V_R$ with their approximations.

Specifically, we define $\eps \defeq 1/\nmax^3$ and round up both $V_L$, $V_R$ to the (nearest) integer parts of powers of $(1 + \eps)$, written as
\[
\tilde{V}_L = \floor[\big]{(1 + \eps)^{t_L}}, \qquad \tilde{V}_R = \floor[\big]{(1 + \eps)^{t_R}} \qquad (t_L, t_R \in \N).
\numberthis \label{eq:rounding-rule-rank}
\]
Since there are only $\log_{(1 + \eps)} \Umax = O(\nmax^3 \log \Umax) = \poly(\nmax)$ possible values for $\tilde{V}_L$ and $\tilde{V}_R$, we encode $r$ under the approximate distribution $\DistRankTd$ in our algorithm, which requires only $\poly(\nmax)$ lookup tables in \cref{sec:ram} for further encoding. The following claim shows that encoding using $\DistRankTd$ incurs only negligible redundancy compared to encoding using $\DistRank$.

\begin{claim}
\EquationOnSameLine{%
\log \frac{1}{\DistRankTd(r)} - \log \frac{1}{\DistRank(r)} \le O\bk[\big]{1/\nmax^2}.
\phantom{\textbf{Claim X.X.}}}
\label{clm:Vapprox}
\end{claim}

\begin{proof}
By definition,
\[
\log \frac{1}{\DistRankTd(r)} - \log \frac{1}{\DistRank(r)} = \log \bk*{\frac{\binom{V_L}{r}}{\binom{\tilde{V}_L}{r}} \cdot \frac{\binom{V_R}{n - r - 1}}{\binom{\tilde{V}_R}{n - r - 1}} \cdot \frac{\binom{\tilde{V}_L + \tilde{V}_R}{n - 1}}{\binom{V_L + V_R}{n - 1}}}.
\numberthis \label{eq:distrank-distance}
\]
Since $\tilde{V}_L$ and $\tilde{V}_R$ are both rounded up, we have
$\binom{\tilde{V}_L}{r} \ge \binom{V_L}{r}$ and $\binom{\tilde{V}_R}{n - r - 1} \ge \binom{V_R}{n - r - 1}$.
For the last term in \eqref{eq:distrank-distance}, recalling that the rounding rule in \eqref{eq:rounding-rule-rank} ensures $\tilde{V}_L \le (1 + \eps) V_L$ and $\tilde{V}_R \le (1 + \eps) V_R$, we consider two cases:
\begin{itemize}
\item If $V_L + V_R \ge \nmax^3$, we have
\begin{align*}
\frac{\binom{\tilde{V}_L + \tilde{V}_R}{n - 1}}{\binom{V_L + V_R}{n - 1}} 
&\le \frac{\binom{(1+\eps)(V_L + V_R)}{n - 1}}{\binom{V_L + V_R}{n - 1}} = \prod_{i = 0}^{n - 2} \frac{(1 + \eps)(V_L + V_R) - i}{(V_L + V_R) - i} \\
&\le \left(\frac{(1 + \eps)(V_L + V_R) - (n-1)}{(V_L + V_R) - (n-1)}\right)^{n-1} \\
&\le \left(\frac{(1 + 2\eps)\bk[\big]{(V_L + V_R) - (n-1)}}{(V_L + V_R) - (n-1)}\right)^{n-1} \\
&= (1 + 2\eps)^{n-1} \le e^{2\eps(n-1)} \le e^{2\nmax/\nmax^3} = 2^{O(1/\nmax^2)},
\end{align*}
where the third inequality holds because $V_L + V_R \ge \nmax^3 \ge 2 (n - 1)$.
\item If $V_L + V_R < \nmax^3$, both $V_L$ and $V_R$ can be represented as integers of the form $\floor{(1 + \eps)^t}$ for $t \in \N$, so $\tilde{V}_L = V_L$ and $\tilde{V}_R = V_R$ are not changed by the rounding.
\end{itemize}
In both cases, we have
\[
\frac{\binom{\tilde{V}_L + \tilde{V}_R}{n - 1}}{\binom{V_L + V_R}{n - 1}} \le 2^{O(1/\nmax^2)}.
\]
Substituting these bounds into \eqref{eq:distrank-distance}, we get
\begin{align*}
\log \frac{1}{\DistRankTd(r)} - \log \frac{1}{\DistRank(r)} 
&= \log \bk*{\frac{\binom{V_L}{r}}{\binom{\tilde{V}_L}{r}} \cdot \frac{\binom{V_R}{n - r - 1}}{\binom{\tilde{V}_R}{n - r - 1}} \cdot \frac{\binom{\tilde{V}_L + \tilde{V}_R}{n - 1}}{\binom{V_L + V_R}{n - 1}}} \\
&\le \log(1 \cdot 1 \cdot 2^{O(1/\nmax^2)}) \\
&= O(1/\nmax^2).
\qedhere
\end{align*}
\end{proof}

\paragraph{Step 5: Encoding two subtrees recursively.}
Given the root information $(p, r)$ that has been encoded in the previous steps, it only remains to encode the two subtrees of the root node. The subproblem for the left (resp.\ right) subtree requires encoding a set of $r$ (resp.\ $n - r - 1$) keys in $\bigl[\tilde a, \, p\bigr)$ (resp.\ $\bigl[p + 1, \, \tilde b\bigr)$) that have weights at most $h(p) - 1$. Using our induction hypothesis \eqref{eq:induction_information}, the two subproblems can be encoded using
\begin{align*}
&{\phantom{{}\le{}}}H\bk[\big]{r,\, \tilde a,\, p,\, h(p) - 1} + H\bk[\big]{n - r - 1,\, p + 1,\, \tilde b,\, h(p) - 1} \\
&\le
\log \bk*{\mathcal{N}\bk[\big]{r,\, \tilde a,\, p,\, h(p) - 1} \cdot \mathcal{N}\bk[\big]{n - r - 1,\, p + 1,\, \tilde b,\, h(p) - 1}} + \frac{n - 1}{\nmax} \\
&= \log \bk*{\binom{V_L}{r} \binom{V_R}{n - r - 1}} + \frac{n - 1}{\nmax}
\numberthis \label{eq:space-usage-subtree}
\end{align*}
bits of space.

\paragraph{Space usage.}

Next, we calculate the space usage of the encoding algorithm. Recall that we have encoded $h(p)$, $\phighmid$, and $r$, which require $\log \bk[\big]{1 / \DistWeightTd(h(p))}$, $\log \ell$, and $\log\bk[\big]{1/\DistRankTd(r)}$ bits, respectively. We have also encoded the two subtrees with space usage given by \eqref{eq:space-usage-subtree}. Adding these terms together, the total space usage is
\begin{align*}
& \phantom{{}={}} \log \frac{1}{\DistWeightTd(h(p))} + \log \ell + \log\frac{1}{\DistRankTd(r)} + \log \bk*{\binom{V_L}{r} \binom{V_R}{n - r - 1}} + \frac{n - 1}{\nmax} \\
&= \log \frac{1}{\DistWeight(h(p))} + \log \ell + \log\frac{1}{\DistRank(r)} + \log \bk*{\binom{V_L}{r} \binom{V_R}{n - r - 1}} + \frac{n - 1}{\nmax} + O\bk[\big]{1/\nmax^2} \tag{by Claims \ref{clm:iidapprox} and \ref{clm:Vapprox}} \\
&= \log \bk*{
    \frac{\abs[\big]{\supp\bk[\big]{\DistSet}}}{\l \cdot \binom{\l \cdot h(p)}{n - 1}}
    \cdot
    \l
    \cdot
    \frac{\binom{V_L + V_R}{n - 1}}{\binom{V_L}{r} \binom{V_R}{n - r - 1}}
    \cdot
    \binom{V_L}{r} \cdot \binom{V_R}{n - r - 1}
} + \frac{n - 1}{\nmax} + O\bk[\big]{1/\nmax^2} \tag{by \eqref{eq:dist_weight_defn}, \eqref{eq:def-distrank}} \\
&= \log \, \abs[\big]{\supp\bk[\big]{\DistSet}} + \frac{n - 1}{\nmax} + O\bk[\big]{1/\nmax^2} \tag{by $V_L + V_R = V(\tilde a, \, \tilde b, \, h(p) - 1) = \l \cdot h(p)$} \\
&\le \log \binom{V(\tilde a, \, \tilde b, \, \hmax)}{n} + \frac{n - 1}{\nmax} + O\bk[\big]{1/\nmax^2} \tag{by \eqref{eq:distset_supp_size}} \\
&= \log \mathcal{N}\bk[\big]{n,\, \tilde a,\, \tilde b,\, \hmax} + \frac{n - 1}{\nmax} + O\bk[\big]{1/\nmax^2} \\
&\le \log \mathcal{N}\bk[\big]{n,\, a,\, b,\, \hmax} + \frac{n - 1}{\nmax} + O\bk[\big]{1/\nmax^2} \tag{by Claim \ref{clm:rounding_loss}} \\
&\le \log \mathcal{N}\bk[\big]{n,\, a,\, b,\, \hmax} + \frac{n}{\nmax} \numberthis \label{eq:space_usage_information_large_universe}
\end{align*}
bits, which implies the desired induction hypothesis \eqref{eq:induction_information}.

\subsection{The Small-Universe Case}
\label{sec:info_small_universe}

In this subsection, we prove \eqref{eq:induction_information} when the key range is small: $b - a < W^4$.
We use a similar strategy as in the large-universe case: We first encode the information $(p, r)$ of the root node, and then recursively encode the two subtrees. However, instead of rounding the key range $[a, b)$ to multiples of $W$ as in the large-universe case, we directly encode the information on the root $(p, r)$ given the \defn{weight profile} of the key range: $(h(a), \ldots, h(b-1))$.

Recall that for every aligned interval of size $W^4$ in the universe, we use an array $A_i[0 \ldots W^4-1]$ to assign weights to these keys. When $b - a < W^4$, only at most two arrays are accessed for keys in $[a, b)$. Thus, we can use a tuple $(i, j, a', b')$ to describe the weight profile, which represents the $a'$-th to $(b'-1)$-th elements in $A_i \circ A_j$ ($A_i$ concatenated with $A_j$), where $a'$ and $b'$ are integers in $[0, 2W^4]$. Additionally, so that we can parameterize our encoding by $i, j, a', b'$, rather than by $a$ and $b$, we will describe how to encode $\Delta p \defeq p - a$ (which lies in the range $[b' - a']$) rather than directly encoding $p$ (which lies in the range $[a, b)$). 

\paragraph{Step 1: Encoding $(\Delta p, r)$.}
Given the tuple $(i, j, a', b')$, determined by $a, b$ as above, let $\DistSmall(\Delta p, r)$ denote the distribution of $(\Delta p, r)$ when the key set $S$ is sampled as $S \sim \DistSetSmall$. By definition,
\[
\DistSmall(\Delta p, r) = \frac{\binom{V_L}{r} \binom{V_R}{n - r - 1}}{\abs[\big]{\supp\bk[\big]{\DistSetSmall}}},
\]
where $V_L = V(a, p, h(p) - 1)$ and $V_R = V(p + 1, b, h(p) - 1)$ are the number of valid keys for the left and right subtrees, respectively.
In the algorithm, we directly encode $(\Delta p, r)$ using this distribution $\DistSmall$, which takes $\log \bk[\big]{1 / \DistSmall(\Delta p, r)}$ bits.

Note that the distribution $\DistSmall(\Delta p, r)$ has a support of size $\poly(W)$, and that there are a $\poly(W)$ number of such distributions (one for each tuple $(i, j, a', b', n, \hmax)$). This is important so that we do not violate the Simple-Distributions Property.

\paragraph{Step 2: Encoding two subtrees recursively.}

The two subtrees are then encoded recursively using our induction hypothesis \eqref{eq:induction_information}, requiring
\[
\log \binom{V_L}{r} + \frac{r}{\nmax} + \log \binom{V_R}{n - r - 1} + \frac{n - r - 1}{\nmax} = \log \bk*{\binom{V_L}{r} \binom{V_R}{n - r - 1}} + \frac{n - 1}{\nmax}
\]
bits of space.

\paragraph{Space usage.}

Adding the space usage for $(p, r)$ and the subtrees, the total space usage of the encoding algorithm is
\begin{align*}
&\log \frac{1}{\DistSmall(\Delta p, r)} + \log \bk*{\binom{V_L}{r} \binom{V_R}{n - r - 1}} + \frac{n - 1}{\nmax}\\
&= \log \bk*{\frac{\abs[\big]{\supp\bk[\big]{\DistSetSmall}}}{\binom{V_L}{r} \binom{V_R}{n - r - 1}} \cdot \binom{V_L}{r} \cdot \binom{V_R}{n - r - 1}} + \frac{n - 1}{\nmax}\\
&= \log \, \abs[\big]{\supp\bk[\big]{\DistSetSmall}} + \frac{n - 1}{\nmax} \\
&\le \log \mathcal{N}\bk[\big]{n, a, b, \hmax} + \frac{n - 1}{\nmax}
\numberthis \label{eq:space_usage_information_small_universe}
\end{align*}
bits, which satisfies our induction hypothesis \eqref{eq:induction_information}.

This completes the proof of \eqref{eq:induction_information} for both the large-universe and small-universe cases. The entire treap can thus be encoded using $\log \binom{U}{n} + 1$ bits of space in the information model.

\section{A Space-Efficient Encoding in the Word RAM Model}
\label{sec:ram} \label{sec:ram_model}

Finally, in this section, we show how to implement the compressed tabulation-weighted treap in the word RAM model.
We start with several preliminaries that we will use in our encodings. 

\subsection{Preliminaries: VMs, Adapters, and Entropy Encoders}
\label{sec:prelim}
\label{sec:adapter}

\paragraph{Virtual memory model.}

Let $w$ denote the word size. The \defn{virtual memory (VM) model} extends the standard \emph{word RAM} model by allowing dynamic resizing of memory. In this model, the memory $m$ (referred to as \defn{virtual memory}) consists of a sequence of $M$ words, each containing exactly $w$ bits. The parameter $M$, also denoted as $|m|$, is called the \defn{size} of the VM. Words are indexed by integers $1, 2, \ldots, M$, which serve as their \defn{addresses}. As in the word RAM model, the user can \defn{access} any word in the VM by specifying its address, either reading or writing it, and each such operation takes constant time.

In addition to standard word accesses, the user can \defn{resize} the VM by either allocating a new word at the end (increasing $M$ by 1) or releasing the last word (decreasing $M$ by 1). When a word is allocated, its initial content is arbitrary, and the user typically performs a subsequent write to assign it meaningful data. Before releasing a word, the user usually reads its content to transfer the information elsewhere. In both allocation and release operations, the addresses and contents of all other words remain unchanged.

The VM can be viewed as an infinite tape of $w$-bit words with addresses $1, 2, \ldots$, although only the words with addresses up to $M$ are accessible. In this perspective, both allocation and release operations simply correspond to incrementing or decrementing the size $M$ by 1.

The VM is often used to accommodate a component of the data structure whose size changes over time. As the name suggests, the VM is a \emph{logical} memory segment with consecutive addresses, in contrast to the \defn{physical memory}, which refers to the word RAM where the final construction of the entire data structure is stored. Depending on the implementation, the words in the VM may not occupy a contiguous block of physical memory; there may be some address translation between virtual and physical memory. In our construction, we design a sequence of \emph{address translation subroutines}, where each subroutine stores one or more VMs inside another VM at a lower abstraction level, ultimately forming the words that are directly stored in physical memory. There are two types of subroutines that we will use extensively in our construction.

\paragraph{Subroutine 1: Adapters.}
The first subroutine is called a \defn{(two-way) adapter}, introduced by Li et al.~\cite{li2023dynamic, li2024dynamic}. It provides an efficient way to maintain multiple variable-sized components within contiguous memory space while supporting fast updates and queries.
Specifically, an adapter takes two input VMs and merges them into a single, larger output VM without wasting any space. The key property of an adapter is that every operation performed on either input VM can be simulated efficiently by operations on the output VM, as stated in \cref{thm:adapter}.

\begin{theorem}[\cite{li2023dynamic}]
\label{thm:adapter}
Let $\mIn^{(1)}$ and $\mIn^{(2)}$ be two input VMs consisting of $L_1$ and $L_2$ words, respectively. There exists an adapter that maintains both input VMs within a single output VM $\mOut$ of $L = L_1 + L_2$ words, with the following properties:
\begin{enumerate}
    \item\label{item:adapter-sim-1} Any read/write of a word in either $\mIn^{(1)}$ or $\mIn^{(2)}$ can be simulated by a read/write of a word in $\mOut$. 
    \item\label{item:adapter-sim-2} Any allocation or release in either $\mIn^{(1)}$ or $\mIn^{(2)}$ can be simulated by a corresponding allocation or release in $\mOut$, followed by $O(\log L)$ additional word reads and writes in $\mOut$. 
    \item\label{item:adapter-local-computation} In each of the above simulations, define the \textbf{local computation time} to be the time spent on determining which words to read/write in $\mOut$ (as opposed to the time spent actually reading/writing those words, which may depend on the implementation of the VM $\mOut$). Then, the local computation times are $O(1)$ and $O(\log L)$ for the two simulations, respectively.
    \item The adapter requires $L_1$ and $L_2$ to be stored externally and provided to the adapter before each operation.
    \item The adapter requires an additional lookup table of $O(L_{\max}^3)$ words, where $L_{\max}$ is an upper bound on $L_1$ and $L_2$. The lookup table can be computed in linear time (in its size) and only depends on $L_{\max}$, so it can be shared among multiple adapters.

\end{enumerate}
\end{theorem}

Informally speaking, in our algorithm (which we will present later), each subtree is encoded within its own VM. For every node $u$, we use an adapter to merge the VMs of $u$'s two children into a single VM at $u$. This merged VM then serves as an input to the adapter at $u$'s parent. By recursively applying this process up the tree, all information is ultimately combined into a single VM at the root node, which is then stored directly in physical memory. The efficient maintenance of this recursive structure relies on \cref{thm:adapter}.

\paragraph{Virtual memory with spillover representation.}

The VM is designed to store a variable-length component of our data structure. However, the size of such a component is not necessarily a multiple of $w$ bits; it may not even be an integer number of bits. To address this, we use the \defn{spillover representation}, which was first introduced by \Patrascu~\cite{patrascu2008succincter} and then extensively used in many succinct data structures. In the spillover representation, each component is represented by a memory part $m \in \{0,1\}^{M} = \{0,1\}^{w \cdot L}$ and an integer $k \in [K]$, referred to as the \defn{spill}. $K$ is called the \defn{spill universe}. The space usage of this component is defined as $M + \lg K$ bits, which is typically fractional. In this paper, we allow $K$ to be any number of the form $K = 2^{O(w)}$, meaning that we only consider spills that can be stored in $O(1)$ machine words. The memory part $m$ is stored in a VM, while the spill $k$ is stored separately. When a component is stored in this form, we say that it is stored in a VM $m$ with a spill $k$.

In a VM with a spill, it is helpful to think of the spill as a ``special word'' located outside the main VM, whose size can be a fractional number (up to $O(w)$) of bits, rather than exactly $w$ bits. Like the other words in the VM, accessing the spill can be performed in constant time. The presence of the spill also provides more flexibility in managing the component's size: in addition to allocating or releasing words in the VM, the user can change the spill universe $K$ to any value in $2^{O(w)}$ with a single resizing operation, typically followed by a write to the spill to assign it a meaningful value.

In our data structure, we will use the spillover representation to encode the random variables described in \cref{sec:information}, each following specific, complicated distributions. The technique for encoding such variables into a spillover representation was initially developed by \cite{patrascu2008succincter} and later extended to the dynamic settings in \cite{li2023dynamic}. This encoding process incurs almost no space overhead while supporting efficient access and modification operations. For modularity, we encapsulate this encoding method into a subroutine called an \emph{entropy encoder}.

\paragraph{Subroutine 2: Entropy encoders.}
Let $D$ be a distribution and let $\phi \in \supp(D)$ be a given symbol, which can be viewed as a random variable drawn from $D$. Intuitively, encoding $\phi$ optimally requires $\log (1 / D(\phi))$ bits. Consider a variable-sized component $(\mIn, \kIn) \in \{0, 1\}^{w \cdot \MIn^D(\phi)} \times [\KIn^D(\phi)]$ stored in a VM with a spill, where the size depends on $\phi$. An \defn{entropy encoder} takes as input both the value $\phi$ and the component $(\mIn, \kIn)$, and encodes them into a new VM with a spill, in which $\phi$ is encoded optimally with respect to $D$.

The full formalization of an entropy encoder is provided in the following theorem.

\begin{restatable}{theorem}{thmEntropyEncoder}
\label{thm:entropy-encoder}
Let $\mathcal{F}$ be a family of distributions where each distribution $D' \in \mathcal{F}$ is associated with input size functions $\MIn^{D'}(\cdot)$ and $\KIn^{D'}(\cdot)$.\footnote{There can be distributions with identical probability mass functions but different input size functions. They are considered distinct members of $\mathcal{F}$.} Let $D \in \mathcal{F}$ be a distribution specified by the user, and let $\phi \in \supp(D)$ be a given symbol. Suppose there exist an input VM $\mIn$ of $\MIn^D(\phi)$ words and an input spill $\kIn \in [\KIn^D(\phi)]$, and suppose the parameters satisfy
\[
|\mathcal{F}| \leq 2^{O(w)}, \qquad \max_{D' \in \mathcal{F}} \abs[\big]{\supp(D')} \leq 2^{O(w)}, \qquad \max_{D' \in \mathcal{F}, \, \psi \in \supp(D')} \KIn^{D'}(\psi) \leq 2^{O(w)}.
\]
Then, we can construct an entropy encoder that encodes both the symbol $\phi$ and $(\mIn, \kIn)$ together into an output VM $\mOut$ of $\MOut^D$ words and an output spill $\kOut \in [\KOut^D]$ with the following properties:
\begin{enumerate}
    \item\label{item:entropy-encoder-size} The size of the output VM and spill only depends on $D$ but not on $\phi$, and 
    \[
    w \cdot \MOut^D + \log \KOut^D = \max_{\psi \in \supp(D)} \BK*{\log \frac{1}{D(\psi)} + w \cdot \MIn^D(\psi) + \log \KIn^D(\psi)} + O\bk*{\frac{1}{q}},
    \]
    where $q \leq 2^{O(w)}$ is a redundancy parameter that we can choose. Moreover, we have $\KOut^D \leq 2^{O(w)}$.
    \item\label{item:entropy-encoder-simulate-mIn} Each word read or write to $\mIn$ can be simulated either by a single read/write to some word in $\mOut$ or by reading/writing to $\kOut$.
    \item\label{item:entropy-encoder-simulate-kIn} Reading/writing $\kIn$ can be simulated by reading/writing $\kOut$.
    \item\label{item:entropy-encoder-simulate-phi} Reading $\phi \in \supp(D)$ can be simulated by reading $\kOut$.
    \item\label{item:entropy-encoder-local-computation} In each of the above simulations, define the \textbf{local computation time} to be the time spent on determining which words to read/write in $\mOut$ and $\kOut$. Then, each of the above simulations incur $O(1)$ local computation time.
    \item\label{item:entropy-encoder-distribution-change} The user may change the distribution $D$ to another distribution $\widehat{D} \in \mathcal{F}$ and the symbol $\phi$ to a new symbol $\hat{\phi} \in \supp(\widehat{D})$. This changes the sizes of the input VM and spill universe from $\MIn^D(\phi), \KIn^D(\phi)$ to $\MIn^{\widehat{D}}(\hat{\phi}), \KIn^{\widehat{D}}(\hat{\phi})$,\footnote{As described in the virtual memory model, when the input VM resizes to a larger size, the new words are allowed to contain arbitrary initial content. When it resizes to a smaller size, the last words are removed. Similarly, when the input spill universe size changes, we assume the new input spill contains arbitrary content, and the user typically performs subsequent writes to assign meaningful values to the new spill and any new words in the input VM.} and the sizes of the output VM and spill universe from $\MOut^D, \KOut^D$ to $\MOut^{\widehat{D}}, \KOut^{\widehat{D}}$. This operation can be simulated by $|\MOut^{\widehat{D}} - \MOut^D|$ allocations or releases on $\mOut$, followed by $O(|\MOut^{\widehat{D}} - \MOut^D|)$ reads and writes to words in $\mOut$ and to $\kOut$. The local computation time for this operation is $O(1 + |\MOut^{\widehat{D}} - \MOut^D|)$.
    \item\label{item:entropy-encoder-size-query} The entropy encoder can calculate any of the following quantities in constant time: $\MOut^D$, $\KOut^D, \MIn^D(\phi)$, $\KIn^D(\phi)$, and $D(\phi)$.
    \item\label{item:entropy-encoder-distribution-index} Each distribution $D' \in \mathcal{F}$ is described by an index of at most $O(w)$ bits, which needs to be stored externally and provided to the entropy encoder before each operation.
    \item\label{item:entropy-encoder-lookup-table} The entropy encoder requires access to an additional lookup table of
    \[
    \poly\bk[\Big]{|\mathcal{F}|, \, q, \, \max_{D' \in \mathcal{F}} \abs[\big]{\supp(D')}}
    \]
    words that can be shared among multiple instances of entropy encoders. Given the input size functions $\MIn^{D'}(\psi)$ and $\KIn^{D'}(\psi)$ and the probability mass function $D'(\psi)$ for all $D' \in \mathcal{F}$ and $\psi \in \supp(D')$, the lookup table can be computed in linear time in its size.
\end{enumerate}
\end{restatable}

It is worth making some comments on how \cref{thm:entropy-encoder} will be used in our applications. We will ensure that the family $\mathcal{F}$ contains at most $\poly(\nmax)$ distributions, each with support size at most $\poly(\nmax)$. We set the redundancy parameter $q$ to $\nmax^2$. With these parameters, the lookup table will occupy only $\poly(\nmax)$ words due to Property~\ref{item:entropy-encoder-lookup-table}, which matches the need in \cref{lem:treap}. During initialization, the user provides a table of $\poly(\nmax)$ words containing the input size functions $\MIn^{D'}(\psi)$ and $\KIn^{D'}(\psi)$, as well as the probability mass function $D'(\psi)$ for all $D' \in \mathcal{F}$ and $\psi \in \supp(D')$. The lookup table can then be constructed in time linear in its size, i.e., in $O(\poly(\nmax))$ time. In addition, each distribution in $\mathcal{F}$ can naturally be represented by an index of $O(\log \nmax) \le O(w)$ bits, as required by Property~\ref{item:entropy-encoder-distribution-index}. 

We also take a moment to comment on a subtle point concerning dynamic resizing. Note that although VMs are designed to support dynamic resizing, the entropy encoder does not allow arbitrary allocations or releases on the input VM. This is because the size of the input VM (and spill) is entirely determined by the current distribution $D$ and the encoded symbol $\phi$. As a result, when using an entropy encoder, the way to resize the input VM is to change $D$ and $\phi$ to new values, rather than by directly invoking allocation or release operations.

\smallskip

Finally, we also present a variant of \cref{thm:entropy-encoder} for uniform distributions. Critically, this version of the theorem does not need to make use of any lookup tables.

\begin{restatable}{theorem}{thmEntropyEncoderUniform}
\label{thm:entropy-encoder-uniform}
Let $\mathcal{F}$ be a family of distributions where each distribution $D' \in \mathcal{F}$ is uniform over an interval $[a, b)$ of integers, and is associated with input size functions $\MIn^{D'}(\cdot)$ and $\KIn^{D'}(\cdot)$. Let $D \in \mathcal{F}$ be a distribution specified by the user, and let $\phi \in \supp(D)$ be a given symbol. Suppose there exist an input VM $\mIn$ of $\MIn^D(\phi)$ words and an input spill $\kIn \in [\KIn^D(\phi)]$, and suppose the parameters satisfy
\[
|\mathcal{F}| \leq 2^{O(w)}, \qquad \max_{D' \in \mathcal{F}} \abs[\big]{\supp(D')} \leq 2^{O(w)}, \qquad \max_{D' \in \mathcal{F}, \, \psi \in \supp(D')} \KIn^{D'}(\psi) \leq 2^{O(w)}.
\]
Then, we can construct an entropy encoder that encodes both the symbol $\phi$ and $(\mIn, \kIn)$ together into an output VM $\mOut$ of $\MOut^D$ words and an output spill $\kOut \in [\KOut^D]$. The entropy encoder satisfies Properties~\ref{item:entropy-encoder-simulate-mIn}--\ref{item:entropy-encoder-distribution-index} in \cref{thm:entropy-encoder} with the following additional properties:
\begin{enumerate}
\item Before each operation, the user must provide the following information to the entropy encoder:
\begin{itemize}
    \item The index of the distribution $D$, which has at most $O(w)$ bits.
    \item The support interval $[a, b)$ for $D$.
    \item A real number $\HMax^D \ge 0$ depending only on $D$ such that
    \[
    \HMax^D \ge \max_{\psi \in \supp(D)} \BK*{w \cdot \MIn^D(\psi) + \log \KIn^D(\psi)}.
    \]
    \item The current input VM size $\MIn^D(\phi)$ and input spill universe size $\KIn^D(\phi)$. The only exception is when reading $\phi$, in which case the user is not required to provide $\MIn^D(\phi)$ and $\KIn^D(\phi)$.
\end{itemize}
\item The size of the output VM and spill only depends on $D$ but not on $\phi$, and
\[
w \cdot \MOut^D + \log \KOut^D \le \HMax^D + \log {\abs{\supp(D)}} + 2^{-\Theta(w)},
\]
where the hidden constant in the $\Theta(w)$ term is a large positive constant of our choice. 
\item The entropy encoder does not require any lookup table.
\end{enumerate}
\end{restatable}

It is worth taking a moment to expand on the role of $\HMax^D$ in \cref{thm:entropy-encoder-uniform}, since it does not appear in \cref{thm:entropy-encoder}.
In the uniform-distribution case, the family $\mathcal{F}$ may contain many distributions, each with a potentially large support. As a result, there is no way for the entropy encoder to use lookup tables in order to store the values for $\MOut^D$ and $\KOut^D$; and, in general, the entropy encoder also cannot hope to calculate these quantities efficiently on the fly. To resolve this issue, \cref{thm:entropy-encoder-uniform} asks that the user provides an \emph{upper bound} $\HMax^D$ on $\max_{\psi \in \supp(D)} \{w \cdot \MIn^D(\psi) + \log \KIn^D(\psi)\}$, which the encoder can then use to generate $\MOut^D$ and $\KOut^D$.  This $\HMax^D$ quantity must be provided by the user on every operation (and must be the same every time, for a given $D$).

Finally, there is one other point worth emphasizing. Before each operation, the user is asked to provide $\MIn^D(\phi)$ and $\KIn^D(\phi)$. To do this, the user may wish to first retrieve $\phi$ from the entropy encoder---this is why it is important that the encoder not require $\MIn^D(\phi)$ and $\KIn^D(\phi)$ for the specific operation of retrieving $\phi$.

The proofs of \cref{thm:entropy-encoder,thm:entropy-encoder-uniform} are similar to part of the analysis in \cite{li2023dynamic}. We provide the proofs in \cref{sec:proof-entropy-encoder} for completeness.

\subsection{The Proof Framework}
\label{sec:ram_framework}

We now begin the proof of \cref{lem:treap}. We will use consistent notations with those in \cref{sec:information} without repeating their definitions. See \cref{table:notation_ram} for the main notations used in this section.
From now on, we first present an algorithm that encodes a key set $S$ with little space overhead, and then in \cref{sec:ram_operations}, we demonstrate how to efficiently support queries and updates.

\begin{table}[tp]
\centering
\caption{Notations in \cref{sec:ram}}\label{table:notation_ram}
\begin{tabular}{l p{0.65\textwidth}}
\toprule
\textbf{Notation} & \textbf{Definition} \\
\midrule
$n$ & Number of keys in the current subproblem \\
$\nmax$ & Upper bound on the maximum $n$ for the inductive proof \\
$\Umax$ & Upper bound on the key universe size $U$ \\
$W$ & Range of weights, $W = \poly(\nmax)$ \\
$[a, b)$ & Key range for the current subproblem \\
$[\tilde{a}, \tilde{b})$ & Key range with endpoints rounded to the nearest multiples of $W$ \\
$\hmax$ & Maximum allowed weight in the current subproblem \\
$h(\cdot)$ & Weight function (globally fixed) \\
$S$ & Key set to be encoded in the current subproblem \\
$V \defeq V(a, b, \hmax)$ & $\#\{x \in [a,b) : h(x) \le \hmax\}$, the number of valid keys \\
$\mathcal{N}(n, a, b, \hmax)$ & Number of possible key sets in the current subproblem \\
$M(n, a, b, \hmax)$ & Size of the VM (in words) in our encoding for the current subproblem \\
$K(n, a, b, \hmax)$ & Size of the spill universe in our encoding for the current subproblem \\
$p$ & Pivot of the key set $S$ \\
$\phighmid \defeq \floor{p/W}$ & High-middle part of the pivot $p$ \\
$r = \rank_S(p)$ & Rank of the pivot $p$ \\
$d$ & Depth of the treap \\
\midrule
$\DistWeightTd\bk[\big]{h(p)}$
  & Approximate distribution of $h(p)$ for a random key set $S \sim \DistSet$ \\[\rowskip]
$\tilde{V}_L, \tilde{V}_R, \tilde{V}$
  & Number of valid keys for the left and right subtrees, and in the current subproblem, all rounded up to the nearest integer of the form $\floor[\big]{(1 + 1 / \nmax^3)^t}$ for $t \in \N$ \\[\rowskip]
$\DistRankTd(r)$ 
  & Approximate distribution of $r = \rank_S(p)$ \\[\rowskip]
$\mLeft, \kLeft, \mRight, \kRight$
  & VMs and spills returned by recursive encodings of the left and right subtrees \\
$\mCat, \, \kCat$
  & VM and spill obtained by concatenating the left and right subtree representations \\
$\MCat(p,r), \, \KCat(p,r)$
  & Size of $\mCat$ and the spill universe size of $\kCat$. Similar notations apply to other VMs and spills. \\
$\mRank, \, \kRank$
  & VM and spill after encoding the rank $r$ \\
$\mPivot, \, \kPivot$
  & VM and spill after encoding $\phighmid$ \\
$\mWeight, \, \kWeight$
  & VM and spill after encoding the weight $h(p)$ \\
$\mSpill$
  & A short sequence of $O(1)$ words extracted from $\kWeight$ (in the large-universe case) or $\kAll$ (in the small-universe case) \\
$\mOut, \, \kOut$
  & Final output of the encoding algorithm \\
\midrule
$\DistSmall(\Delta p,r)$
  & Distribution of $(\Delta p,r)$ in the small-universe case, where $\Delta p = p - a$ and the weight profile
  in the key range is $(A_i \circ A_j)[a' \ldots b'\!-\!1]$ \\
$\mAll, \, \kAll$
  & VM and spill after encoding $p$ and $r$ in the small-universe case \\
\bottomrule
\end{tabular}
\end{table}

Our encoding algorithm follows the same recursive structure described in \cref{sec:information}: To encode a treap for a key set $S$, we first recursively encode the left and right subtrees of the root, then combine them with the root information to construct the complete encoding of $S$, which includes both the pivot $p \in S$ and its rank $r \defeq \rank_S(p)$. Each subproblem in the recursion requires encoding a key set $S$ containing $n$ keys within a range $[a, b)$, where $0 \le a < b \le \Umax$, with the constraint that every key $x \in S$ has weight $h(x) \le \hmax$.

We will prove by induction that for any such subproblem, the key set $S$ can be encoded within a VM of $M(n, a, b, \hmax)$ words and a spill with universe size $K(n, a, b, \hmax) \le 2^w \cdot \nmax^2$, using
\[
w \cdot M(n, a, b, \hmax) + \log K(n, a, b, \hmax) \le \log \mathcal{N}(n, a, b, \hmax) + \frac{n}{\nmax}
\numberthis \label{eq:space_usage_ram}
\]
bits of space,
where we recall that $\mathcal{N}(n, a, b, \hmax)$ represents the number of possible key sets in the current subproblem, and $\nmax$ is a fixed upper bound on $n$ in the entire induction.

The base case of the induction is when $n = 0$, in which case we use an empty encoding $M(0, a, b, \hmax) = 0$ and $K(0, a, b, \hmax) = 1$, and \eqref{eq:space_usage_ram} holds trivially.

In the following subsections, we will prove the induction hypothesis \eqref{eq:space_usage_ram} for both the large-universe case ($b - a \ge W^4$) and the small-universe case ($b - a < W^4$), by properly encoding the random variables described in \cref{sec:information}.

Similar to \cref{sec:information}, throughout the proof, we assume that the key set $S$ satisfies the following two properties, which hold with high probability in $\nmax$:
\begin{itemize}
\item In each subproblem in the recursion, there is a unique key with the largest weight.
\item The weight of every key in $S$ is at least $\nmax^4$. As a result, we have $\hmax \ge \nmax^4$ for every subproblem.
\end{itemize}
If these two properties do not hold, we say that a \defn{failure} has occurred, and we will handle this case in \cref{sec:failure_mode}.

\subsection{The Large-Universe Case}
\label{sec:ram_large_universe}

\begin{figure}[t]
    \centering
    \includegraphics[width=0.6\textwidth]{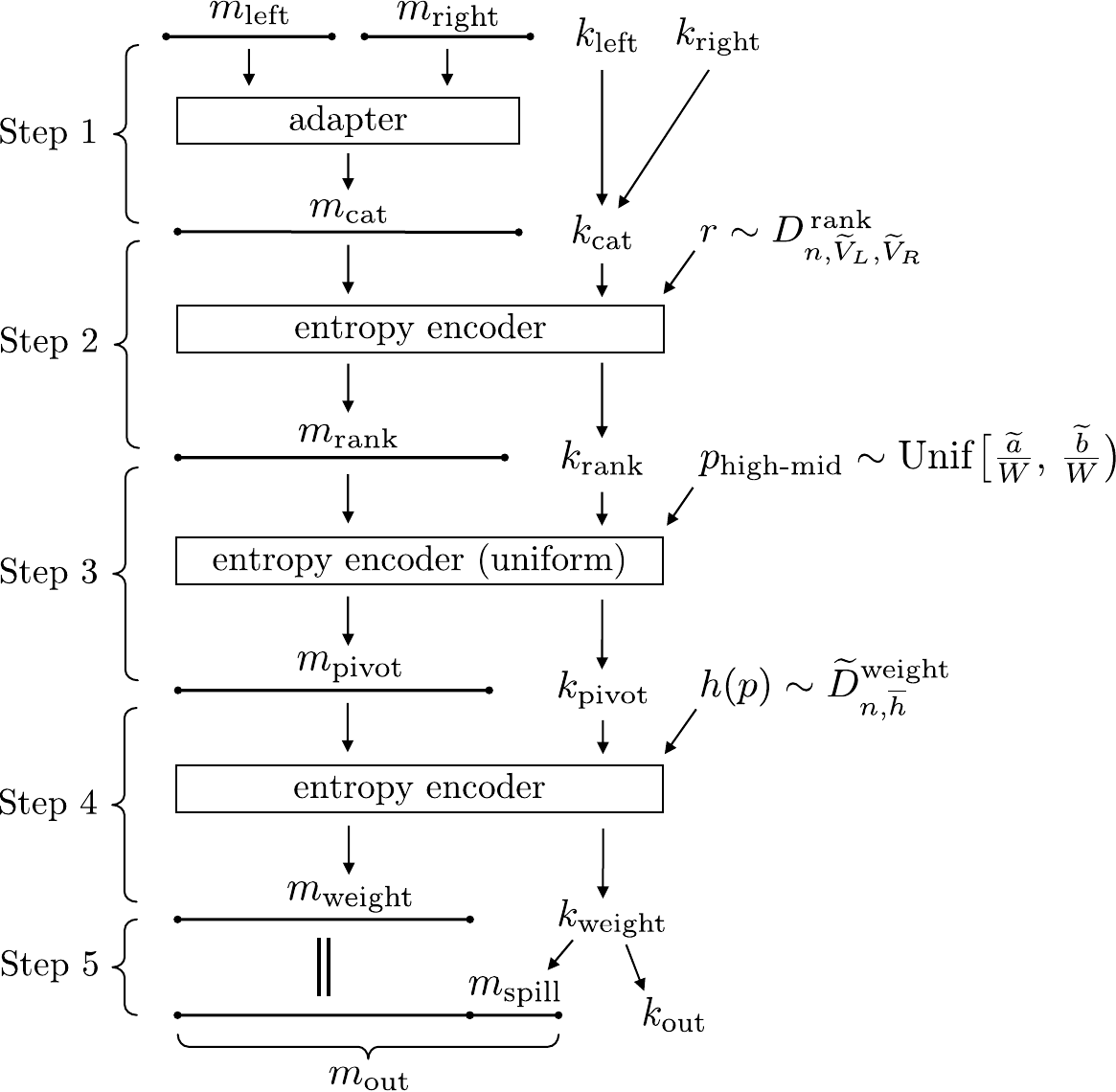}
    \caption{Encoding process in the large-universe case. In Step 1, we start with the spillover representations of the two subtrees, and combine them using an adapter. In Steps 2--4, we use three entropy encoders to add $r$, $\phighmid$, and $h(p)$ to the representation, respectively. In Step 5, we reduce the size of the spill and form the final output.}
    \label{fig:ram_encoding}
    \label{fig:ram_encoding_large_universe}
\end{figure}

In this subsection, we prove \eqref{eq:space_usage_ram} for subproblems where $b - a \ge W^4$. Similar to \cref{sec:information}, the first step is to round $a$ down and round $b$ up to the nearest multiples of $W$, denoted by $\tilde{a}$ and $\tilde{b}$, respectively. \cref{clm:rounding_loss} states that the rounding does not significantly affect the subproblem's optimal encoding size. Then, we perform the encoding steps in \cref{sec:information} in the reversed order, starting with the spillover representation of the two subtrees, and then combine them with $r$, $\phighmid$, and $h(p)$, respectively, to obtain the encoding of the entire key set $S$. See \cref{fig:ram_encoding} for an overview of the encoding process.

\paragraph{Step 1: Connecting two subtrees using an adapter.}

The induction hypothesis provides spillover representations $(\mLeft, \kLeft)$ and $(\mRight, \kRight)$ for the two subtrees, where $\mLeft$ and $\mRight$ are VMs, and $\kLeft$ and $\kRight$ are spills. The sizes of the VMs and spills are given by
\begin{align*}
(\mLeft, \kLeft) &\in \{0, 1\}^{w \cdot M(r, \, \tilde{a}, \, p, \, h(p)-1)} \times [K(r, \, \tilde{a}, \, p, \, h(p)-1)], \\
(\mRight, \kRight) &\in \{0, 1\}^{w \cdot M(n-r-1, \, p+1, \, \tilde{b}, \, h(p)-1)} \times [K(n-r-1, \, p+1, \, \tilde{b}, \, h(p)-1)].
\end{align*}

We concatenate the VMs $\mLeft$ and $\mRight$ using an adapter (see \cref{thm:adapter}) to form a combined VM, denoted by $\mCat$. This combined VM has
\[
\MCat(p,r) \defeq M(r, \, \tilde{a}, \, p, \, h(p)-1) + M(n-r-1, \, p+1, \, \tilde{b}, \, h(p)-1)
\]
words. Note that $\MCat(p, r)$ also depends on the parameters $n, \tilde{a}, \tilde{b}$ of the current subproblems, but we omit them for brevity.

We also directly merge the two spills $\kLeft$ and $\kRight$ into a larger spill $\kCat$. Its universe size is
\[
\KCat(p,r) \defeq K(r, \, \tilde{a}, \, p, \, h(p)-1) \cdot K(n-r-1, \, p+1, \, \tilde{b}, \, h(p)-1) \le \bk*{2^{w + 1} \cdot \nmax^3}^2 = 2^{O(w)}.
\]

The space usage for this combined representation $(\mCat, \kCat)$ is the sum of the space used by the two subproblems. By the induction hypothesis \eqref{eq:space_usage_ram}, it is at most
\begin{align*}
&{\phantom{{}\le{}}} w \cdot M(r, \, \tilde{a}, \, p, \, h(p)-1) + \log K(r, \, \tilde{a}, \, p, \, h(p)-1) \\
&\quad + w \cdot M(n-r-1, \, p+1, \, \tilde{b}, \, h(p)-1) + \log K(n-r-1, \, p+1, \, \tilde{b}, \, h(p)-1) \\
&\le \bk*{\log \mathcal N(r, \, \tilde{a}, \, p, \, h(p)-1) + \frac{r}{\nmax}}
+ \bk*{\log \mathcal N(n-r-1, \, p+1, \, \tilde{b}, \, h(p)-1) + \frac{n-r-1}{\nmax}} \\
&= \log \mathcal N(r, \, \tilde{a}, \, p, \, h(p)-1) + \log \mathcal N(n-r-1, \, p+1, \, \tilde{b}, \, h(p)-1) + \frac{n-1}{\nmax}
\numberthis \label{eq:space_usage_ram_step1}
\end{align*}
bits.

\paragraph{Step 2: Encoding the rank $r$.}

So far, the combined representation $(\mCat, \kCat)$ includes the information of the key set $S$ conditioned on the pivot $p$ and the pivot's rank $r$. The next step is to incorporate the rank $r$ into our representation.

We fix the value of $p$, and view $r$ as a symbol following the distribution $\DistRankTd$, which is the approximate distribution of the rank $r$ for a random key set $S$, as defined in \eqref{eq:def-distrank} and \eqref{eq:rounding-rule-rank}. Then, the size of $(\mCat, \kCat)$ is a function of $r$. We use an entropy encoder (\cref{thm:entropy-encoder}) to encode $r$ together with the combined representation $(\mCat, \kCat)$ with the following parameters:
\begin{itemize}
    \item The redundancy parameter is chosen as $q = \nmax^2$.
    \item The distribution $D$ is $\DistRankTd$, which is indexed by three numbers $n, \tilde{V}_L, \tilde{V}_R \le 2^{O(w)}$. The support size $\abs[\big]{\supp\bk[\big]{\DistRankTd}}$ is at most $\nmax$.
    \item The family $\mathcal{F}$ of distributions is the collection of all distributions of the form $\DistRankTd$. Recalling that $\tilde{V}_L$ and $\tilde{V}_R$ are integers of the form $\floor{(1 + 1 / \nmax^3)^t}$ for $t \in \N$ (see \eqref{eq:rounding-rule-rank}), and there are only $O(\nmax^3 \log \Umax)$ possible values for $\tilde{V}_L$ and $\tilde{V}_R$, the family $\mathcal{F}$ has at most $\poly(\nmax, \, \log \Umax) = \poly(\nmax)$ distributions.
\end{itemize}

This entropy encoder outputs a new VM $\mRank$ of $\MRank(p)$ words and a spill $\kRank$ with universe size $\KRank(p) = 2^{O(w)}$, which uses at most
\begin{align*}
& \phantom{{}\le{}} w \cdot \MRank(p) + \log \KRank(p) \\
&\le \max_{r' \in [0, n-1]} \Biggl\{ \log \frac{1}{\DistRankTd[n, \tilde{V}_L, \tilde{V}_R](r')} + w \cdot \MCat(p, r') + \log \KCat(p, r') \Biggr\} + O(1/q) \\
&\le \max_{r' \in [0, n-1]} \Biggl\{ \log \mathcal N(r', \, \tilde{a}, \, p, \, h(p)-1) + \log \mathcal N(n-r'-1, \, p+1, \, \tilde{b}, \, h(p)-1) \\
&\hspace{6em} + \frac{n-1}{\nmax} + \log \frac{1}{\DistRankTd[n, \tilde{V}_L, \tilde{V}_R](r')} \Biggr\} + O(1/\nmax^2)
\numberthis \label{eq:space_usage_ram_step2}
\end{align*}
bits of space.

\paragraph{Step 3: Encoding $\phighmid$.}

Next, we encode the high-middle part of the pivot, denoted $\phighmid \defeq \floor{p / W}$. Given $h(p)$, $\phighmid$ uniquely determines the pivot $p$. The value $\phighmid$ is an integer in $\bigl[\tilde{a}/W, \, \tilde{b}/W\bigr)$ and follows a uniform distribution.

Using an entropy encoder for uniform distributions (\cref{thm:entropy-encoder-uniform}), we encode $\phighmid$ together with the VM $\mRank$ and spill $\kRank$ from Step 2. This entropy encoder requires us to provide a real number $\HMax\bk[\big]{n,\, \tilde{a},\, \tilde{b},\, \hmax,\, h(p)}$ greater than or equal to
\begin{align*}
  & \phantom{{}\le{}} \max_{p' \in [\tilde{a},\, \tilde{b}) \,:\, h(p') = h(p)} \BK*{w \cdot \MRank(p') + \log \KRank(p')}. \numberthis \label{eq:Hreq}
\end{align*}
In order for $\HMax\bk[\big]{n,\, \tilde{a},\, \tilde{b},\, \hmax,\, h(p)}$ to be at least \eqref{eq:Hreq}, it suffices for it to be at least
\begin{align*}
  & \max_{\substack{r' \in [0, n-1] \\ p' \in [\tilde{a}, \tilde{b}) \,:\, h(p')=h(p)}} \Biggl\{
    \log \mathcal N(r', \, \tilde{a}, \, p', \, h(p)-1)
    + \log \mathcal N(n-r'-1, \, p'+1, \, \tilde{b}, \, h(p)-1). \\
    & \hspace{9em} + \frac{n-1}{\nmax}
    + \log \frac{1}{\DistRankTd(r')}
    \Biggr\}
  + O(1/\nmax^2), \numberthis \label{eq:hmax_bound_1}
\end{align*}
since, by \eqref{eq:space_usage_ram_step2}, we have that \eqref{eq:Hreq} is at most \eqref{eq:hmax_bound_1}.

By \eqref{eq:space_usage_information_large_universe}, for any $p'$ with $h(p') = h(p)$ and rank $r'$, we have
\begin{align*}
& \log \mathcal{N}(r', \tilde{a}, p', h(p)-1) + \log \mathcal{N}(n-r'-1, p'+1, \tilde{b}, h(p)-1) + \log\frac{1}{\DistRankTd(r')} \\
&\le \log \mathcal{N}(n, a, b, \hmax) - \log \frac{1}{\DistWeightTd(h(p))} - \log \frac{\tilde{b}-\tilde{a}}{W} + O(1/\nmax^2).
\end{align*}
Substituting this inequality into \eqref{eq:hmax_bound_1}, we know that a valid choice of $\HMax$ is
\[
\HMax(n, \tilde{a}, \tilde{b}, \hmax, h(p)) \defeq \log \mathcal{N}(n, a, b, \hmax) + \frac{n-1}{\nmax} - \log \frac{1}{\DistWeightTd(h(p))} - \log \frac{\tilde{b}-\tilde{a}}{W} + \Theta(1/\nmax^2) \numberthis \label{eq:hmax_defn}
\]
with a sufficiently large $\Theta(1/\nmax^2)$ term. The following claim shows that we can compute $\HMax$ efficiently.

\newcommand{\HMaxDefinition}{Given $n,\, \tilde{a},\, \tilde{b},\, \hmax,\, h(p)$, we can compute $\HMax\bk[\big]{n,\, \tilde{a},\, \tilde{b},\, \hmax,\, h(p)}$ defined in \eqref{eq:hmax_defn} in constant time using a lookup table of $\poly(\nmax, \log \Umax)$ words.}

\begin{restatable}{claim}{HMaxClaim}
  \label{claim:hmax}
  \HMaxDefinition
\end{restatable}

\renewcommand{\HMaxDefinition}{Given $n,\, \tilde{a},\, \tilde{b},\, \hmax,\, h(p)$, we can compute
\[\HMax(n, \tilde{a}, \tilde{b}, \hmax, h(p)) \defeq \log \mathcal{N}(n, a, b, \hmax) - \log \frac{1}{\DistWeightTd(h(p))} - \log \frac{\tilde{b}-\tilde{a}}{W} + \Theta(1/\nmax^2) \tag{\ref{eq:hmax_defn}}\]
in constant time using a lookup table of $\poly(\nmax, \log \Umax)$ words.}

We defer the proof of \cref{claim:hmax} to \cref{sec:proof-output-size}.

\smallskip

The output of the entropy encoder is a VM $\mPivot$ of $\MPivot\bk[\big]{h(p)}$ words and a spill $\kPivot$ with universe size $\KPivot\bk[\big]{h(p)}$, which uses at most
\begin{align*}
& \phantom{{}\le{}} w \cdot \MPivot\bk[\big]{h(p)} + \log \KPivot\bk[\big]{h(p)} \\
& \le \HMax\bk[\big]{n,\, \tilde{a},\, \tilde{b},\, \hmax,\, h(p)} + \log \frac{\tilde{b}-\tilde{a}}{W} + O(1/\nmax^2) \\
& = \log \mathcal{N}(n, a, b, \hmax) + \frac{n-1}{\nmax} - \log \frac{1}{\DistWeightTd(h(p))} + O(1/\nmax^2)
\numberthis \label{eq:space_usage_ram_step3}
\end{align*}
bits of space, where the last equality holds by \eqref{eq:hmax_defn}.

\paragraph{Step 4: Encoding $h(p)$.}

We use the entropy encoder again (with $q = \nmax^2$) to encode the weight $h(p)$ of the pivot, which approximately follows the distribution $\DistWeightTd$, together with $(\mPivot, \kPivot)$. The encoder outputs a VM $\mWeight$ of $\MWeight$ words and a spill $\kWeight \in [\KWeight]$. Note that $\MWeight$ and $\KWeight$ do not depend on $p$ or $r$. The space usage of $(\mWeight, \kWeight)$ is at most
\begin{align*}
& \phantom{{}\le{}} w \cdot \MWeight + \log \KWeight \\
& \le \max_{h(p) \in [0, \hmax]} \BK*{w \cdot \MPivot\bk[\big]{h(p)} + \log \KPivot\bk[\big]{h(p)} + \log \frac{1}{\DistWeightTd\bk[\big]{h(p)}} + O(1/\nmax^2)} \\
& \le \log \mathcal{N}\bk[\big]{n,\, a,\, b,\, \hmax} + \frac{n - 1}{\nmax} + O\bk[\big]{1/\nmax^2}
\numberthis \label{eq:space_usage_ram_step4}
\end{align*}
bits, where the last inequality follows from \eqref{eq:space_usage_ram_step3}.

\paragraph{Step 5: Adjusting VM and spill size.}

So far, we have encoded all information about the key set $S$ into $(\mWeight, \allowbreak \kWeight)$ by applying the entropy encoder three times. Each use of the entropy encoder may increase the spill universe size to $2^{O(w)}$, with a potentially larger leading constant in the $O(w)$ in the exponent. Therefore, before outputting, we must reduce the size of the spill. In this step, we also ensure that the sizes of the output VM and spill, $M(n, a, b, \hmax)$ and $K(n, a, b, \hmax)$, can be computed efficiently. We begin by assuming an appropriate choice of $M(n, a, b, \hmax)$ and $K(n, a, b, \hmax)$ as follows.

\begin{restatable}{claim}{OutputSizeClaim}
\label{claim:output_size}
There exist functions $M(n, a, b, \hmax)$ and $K(n, a, b, \hmax)$ for subproblems in the large-universe case, such that $K(n, a, b, \hmax) \le \nmax^2 \cdot 2^w$,
\begin{equation}
w \cdot M(n, a, b, \hmax) + \log K(n, a, b, \hmax) = \log \mathcal{N}(n, a, b, \hmax) + \frac{n - 1}{\nmax} + \frac{1}{\nmax^{3/2}} + O(1/\nmax^2),
\label{eq:output_size_prop}
\end{equation}
and for any given $n, a, b, \hmax$, the values $M(n, a, b, \hmax)$ and $K(n, a, b, \hmax)$ can be computed in constant time using a lookup table of size $\poly(\nmax, \log \Umax)$.
\end{restatable}

The proof of this claim is deferred to \cref{sec:proof-output-size}.

\smallskip

The final step in the encoding process is to adjust the size of $(\mWeight, \kWeight)$ to obtain a VM of $M(n, a, b, \hmax)$ words and a spill with universe size $K(n, a, b, \hmax)$, forming the output encoding $(\mOut, \kOut)$. Specifically, we compare $\MWeight$ to $M(n, a, b, \hmax)$:
\begin{itemize}
\item If $M(n, a, b, \hmax) = \MWeight + L$ for some integer $L \ge 0$, we \emph{extract} $wL$ bits from the spill $\kWeight$, forming a sequence $\mSpill \in [2^{wL}]$ of $O(1)$ words and a new spill $\kOut \in \Bk[\big]{\ceil{\KWeight / 2^{wL}}}$. We then attach $\mSpill$ to the end of $\mWeight$ to form the output VM $\mOut$.
\item If $M(n, a, b, \hmax) = \MWeight - L$ for some integer $L > 0$, we take the last $L$ words of $\mWeight$ and merge them with $\kWeight$ to form a larger spill $\kOut \in [\KWeight \cdot 2^{wL}]$. The remaining words of $\mWeight$ form the output VM $\mOut$. In this case, we define $\mSpill$ to be empty.
\end{itemize}
In both cases, the new spill $\kOut$ has universe size at most $K(n, a, b, \hmax)$, because $w \cdot M(n, a, b, \hmax) + \log K(n, a, b, \hmax)$ is an upper bound on the space usage of $(\mWeight, \kWeight)$, according to \eqref{eq:space_usage_ram_step4} and \eqref{eq:output_size_prop}.
Hence, we can directly output $\kOut$ as a spill in the universe $[K(n, a, b, \hmax)]$.

The total space usage of our encoding $(\mOut, \kOut)$, by \eqref{eq:output_size_prop}, is
\[
w \cdot M(n, a, b, \hmax) + \log K(n, a, b, \hmax) \le \log \mathcal{N}(n, a, b, \hmax) + \frac{n - 1}{\nmax} + O(1/\nmax^{3/2}) \le \log \mathcal{N}(n, a, b, \hmax) + \frac{n}{\nmax}
\]
bits, which concludes the proof of \eqref{eq:space_usage_ram} in the large-universe case.

\subsection{The Small-Universe Case}
\label{sec:ram_small_universe}

\begin{figure}[t]
    \centering
    \includegraphics[width=0.6\textwidth]{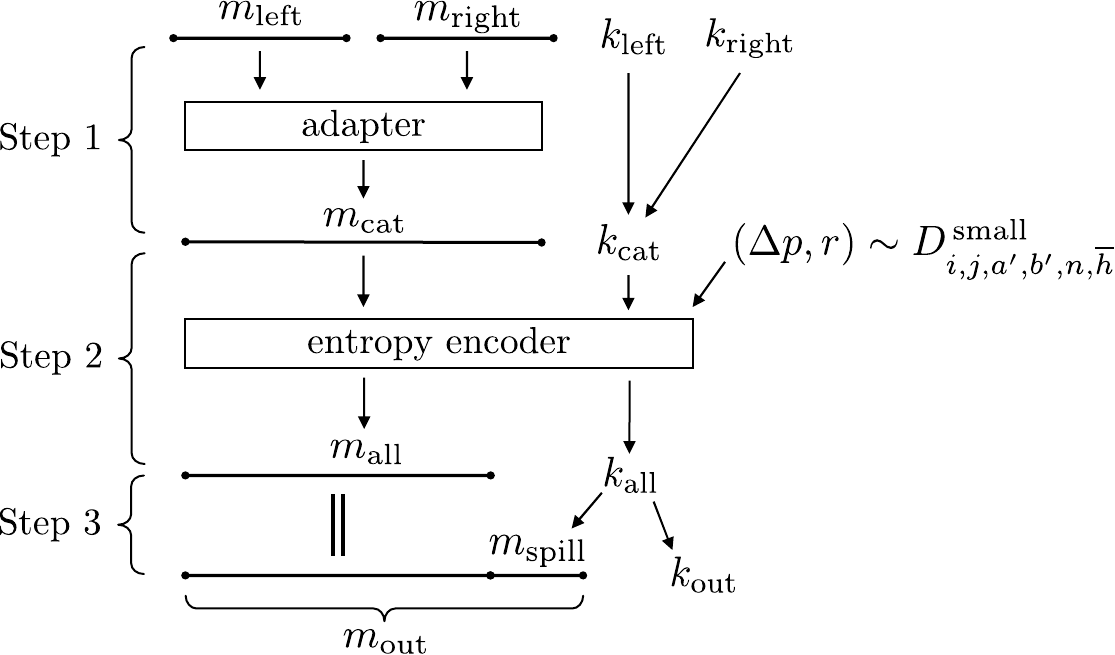}
    \caption{Encoding process in the small-universe case. We start with the spillover representations of the two subtrees, and combine them using an adapter. Then, we use an entropy encoder to add $(\Delta p,r)$ to the representation. Finally, we reduce the size of the spill and form the final output.}
    \label{fig:ram_small_universe}
\end{figure}

For subproblems with $b - a < W^4$, we use a different approach to prove \eqref{eq:space_usage_ram} guided by \cref{sec:info_small_universe}. Recall that the sequence of weights for the keys in $[a,b)$, written $\bk[\big]{h(a), \dots, h(b-1)}$, can be described using $A_i \circ A_j[a' \ldots b'-1]$, where $i, j \in [\poly(W)]$ and $a', b' \in [0, 2W^4]$. Hence, the weight sequence can be indexed by a tuple $(i,j,a',b')$ that has only $\poly(W)$ possible values. This enables us to encode the pivot $p$ and its rank $r$ directly using an entropy encoder without introducing lookup tables of unaffordable sizes. Similarly to \cref{sec:information}, we will describe how to encode $\Delta p \defeq p - a$ (which lies in the range $[b' - a']$) rather than directly encoding $p$ (which lies in the range $[a, b)$). See \cref{fig:ram_small_universe}.

\paragraph{Encoding algorithm.}

Similar to the large-universe case, we first recursively encode the left and right subtrees, obtaining $(\mLeft, \kLeft)$ and $(\mRight, \kRight)$. We use an adapter to connect the VMs $\mLeft$ and $\mRight$ into $\mCat$, and concatenate the spills $\kLeft$ and $\kRight$ into $\kCat$. By the induction hypothesis, the space usage for $(\mCat, \kCat)$ is
\begin{align*}
&{\phantom{{}\le{}}} w \cdot \MCat(\Delta p,r) + \log \KCat(\Delta p,r) \\
&\le \log \mathcal{N}\bk[\big]{r,\, a,\, p,\, h(p)-1} + \frac{r}{\nmax} + \log \mathcal{N}\bk[\big]{n-r-1,\, p+1,\, b,\, h(p)-1} + \frac{n-r-1}{\nmax} \\
&= \log \mathcal{N}\bk[\big]{r,\, a,\, p,\, h(p)-1} + \log \mathcal{N}\bk[\big]{n-r-1,\, p+1,\, b,\, h(p)-1} + \frac{n-1}{\nmax}
\end{align*}
bits.

Next, we apply an entropy encoder (with $q = \nmax^2$) to encode the pair $(\Delta p, r)$ at the root, together with $(\mCat, \kCat)$. The pair $(\Delta p, r)$ follows the distribution $\DistSmall[i,j,a',b',n,\hmax]$, where $(i, j, a', b')$ describes the weight profile for the range $[a, b)$. This step produces a VM $\mAll$ and a spill $\kAll$, whose sizes are denoted $\MAll$ and $\KAll$, respectively.

Finally, we reduce the spill universe size of $\kAll$ if necessary. If $\KAll > \nmax^2 \cdot 2^w$, we split $\kAll$ into $\mSpill$, consisting of $O(1)$ words, and a shorter spill $\kOut$ with universe size in $[\nmax^2, \, \nmax^2 \cdot 2^w]$. If $\KAll$ is already at most $\nmax^2 \cdot 2^w$, we do nothing in this step and set $\mSpill$ to be empty. The output VM is $\mOut = \mAll \circ \mSpill$, and the output spill is $\kOut$. The sizes of the output VM and spill, $M(n, a, b, \hmax)$ and $K(n, a, b, \hmax)$, depend only on $i, j, a', b', n, \hmax$, which are integers bounded by $\poly(\nmax)$. Therefore, these sizes can be computed in constant time using a lookup table of size $\poly(\nmax)$.

\paragraph{Space usage.}
Now, we calculate the space usage for our encoding $(\mOut, \kOut)$. The space usage for $(\mAll, \kAll)$ is at most
\begin{align*}
& {\phantom{{}\le{}}} w \cdot \MAll + \log \KAll \\
&\le \max_{\Delta p',r'} \left\{ \log \frac{1}{\DistSmall[i,j,a',b',n,\hmax](\Delta p',r')} + w \cdot \MCat(\Delta p',r') + \log \KCat(\Delta p',r') \right\} + O\bk[\bigg]{\frac{1}{\nmax^2}} \\
&\le \max_{\Delta p',r'} \Bigg\{
    \log \frac{1}{\DistSmall[i,j,a',b',n,\hmax](\Delta p',r')}
    + \log \mathcal{N}(r', \, a, \, p', \, h(p')-1) \\
&\hspace{7em}
    + \log \mathcal{N}(n-r'-1, \, p'+1, \, b, \, h(p')-1)
    + \frac{n-1}{\nmax}
\Bigg\} + O\bk[\bigg]{\frac{1}{\nmax^2}} \\
&\le \log \mathcal{N}(n, a, b, \hmax) + \frac{n-1}{\nmax} + O\bk[\bigg]{\frac{1}{\nmax^2}} \numberthis \label{eq:space_usage_ram_small_temp}
\end{align*}
bits, where the third inequality holds according to \eqref{eq:space_usage_information_small_universe}.

When $\KAll > \nmax^2 \cdot 2^w$, the final step of reducing the spill size adds at most
\[
\log K(n, a, b, \hmax) - \log \bk[\big]{K(n, a, b, \hmax) - 1} \le \log (\nmax^2 + 1) - \log \nmax^2 \le O(1/\nmax^2)
\]
bits of space usage; when $\KAll \le \nmax^2 \cdot 2^w$, the final step does not change the space usage.
Thus, the total space for $(\mOut, \kOut)$ is at most
\[ w \cdot M(n,a,b,\hmax) + \log K(n,a,b,\hmax) \le \log \mathcal{N}(n, a, b, \hmax) + \frac{n - 1}{\nmax} + O\bk*{\frac{1}{\nmax^2}} \le \log \mathcal{N}(n, a, b, \hmax) + \frac{n}{\nmax} \]
bits.
This concludes the proof of the induction hypothesis \eqref{eq:space_usage_ram} in the small-universe case.

\subsection{Queries, Insertions, and Deletions}
\label{sec:ram_operations}

On a standard, uncompressed treap where each node maintains its subtree size, insertions, deletions, and rank/select queries can be completed in $O(\log n)$ expected time. In this subsection, we show how to simulate these operations on our succinct encoding of the treap, and we analyze the expected time complexity of each operation.

\paragraph{Some conventions for the section.} To simplify our discussion, we will adopt the following conventions in this section. For a given vertex $u$, we will use $m_u$ to denote the output VM $\mOut$ at node $u$, and we will use $k_u$ to denote the output spill $\kOut$ at node $u$. For intermediate VMs and spills (e.g., $\mCat, \kCat, \mRank, \kRank, \ldots$), we will continue to leave the node $u$ implicit, as it will be clear from context which node $u$ they pertain to. 

In several places in the constructions, the user must calculate the sizes of the VMs and spills that they are interacting with. In some cases, this can be done with a simple query to an encoder (i.e., using \cref{item:entropy-encoder-size-query} of \cref{thm:entropy-encoder}), but in other cases (e.g., calculating the sizes of $\mOut$ and $\kOut$, or calculating the parameter $\HMax$ needed for Theorem \ref{thm:entropy-encoder-uniform}), a more substantial calculation is necessary (using  \cref{claim:output_size,claim:hmax}). To streamline the exposition, we will for most of the subsection ignore the issue of calculating VM and spill sizes, and then we will address the issue in detail at the end of the subsection. 

Finally, when computing expected running times, we will often make use of the fact that the depth of the treap is at most $O(\log n)$ with high probability in $n$. In all of our calculations, the $1 / \poly(n)$-probability event in which the depth is $\omega(\log n)$ has negligible effect on the expected values that we are computing. Thus, as a slight abuse of notation, we will sometimes treat $O(\log n)$ as a worst-case upper bound on the depth when computing expected values. 

\paragraph{Nested address translation.} Before describing how to implement queries, let us begin with a simpler operation: reading the VM $m_v$ at a given node $v$. 

Recall that only the VM of the root is explicitly stored in physical memory; all other VMs must be accessed via address translation subroutines, including both adapters and entropy encoders. We now describe how to perform these translations in time linear in $v$'s depth $d$.

\begin{claim}
Consider a node $v$ at depth $d$, and suppose that we have already calculated the output spills of all the entropy encoders on the root to leaf path to $v$. Then, we can implement an access to the VM $m_v$ in time $O(d)$.
\label{claim:translation}
\end{claim}
\begin{proof}
Suppose $u$ is $v$'s parent. To streamline the discussion, we will assume that $u$ is in the large-universe case---the small-universe case follows from the same arguments but with different variables. 

We can access the $i$-th word in $v$'s VM, denoted $m_v[i]$, as follows:
\begin{enumerate}
\item Recall that $v$'s VM is fed into an adapter as an input VM, with the output VM being $\mCat$ at node $u$. By \cref{item:adapter-sim-1,item:adapter-local-computation} of \cref{thm:adapter}, we can compute an address $j_{\text{cat}}$ in constant time, such that accessing $m_v[i]$ can be simulated by accessing $\mCat[j_{\text{cat}}]$. 
\item $\mCat$ is fed into an entropy encoder as the input VM. By \cref{item:entropy-encoder-simulate-mIn} of \cref{thm:entropy-encoder}, there is an address $j_{\text{rank}}$ such that accessing $\mCat[j_{\text{cat}}]$ can be simulated by accessing either $\mRank[j_{\text{rank}}]$ or the output spill $\kRank$. Since $\kRank$ is already known before visiting node $v$, we only need to access $\mRank[j_{\text{rank}}]$. By \cref{item:entropy-encoder-local-computation} of \cref{thm:entropy-encoder}, $j_{\text{rank}}$ can be calculated in constant time.
\item Similarly, accessing $\mRank[j_{\text{rank}}]$ can be simulated by accessing $\mPivot[j_{\text{pivot}}]$, which can in turn be simulated by accessing $\mWeight[j_{\text{weight}}]$. Each of the translations, going from $j_{\text{rank}}$ to $j_{\text{pivot}}$, and from $j_{\text{pivot}}$ to $j_{\text{weight}}$, can be computed in constant time.
\item Finally, since $\mWeight$ is a prefix of $u$'s VM $m_u$, we can conclude that $m_u[j_{\text{weight}}] = \mWeight[j_{\text{weight}}]$, meaning that we have reduced the task of accessing $m_v[i]$ to that of accessing $m_u[j]$ where $j = j_{\text{weight}}$.
\end{enumerate}
In summary, accessing $m_v[i]$ reduces in constant time to accessing some $m_u[j]$, where $u$ is $v$'s parent. This address translation also exists when $u$ is in the small-universe case. By repeatedly applying this translation, we can implement any access to $m_u$ in time $O(d)$ (see \cref{alg:access_vm}), as desired.
\end{proof}

\begin{algorithm}[t]
    \caption{Accessing Virtual Memory Words}
    \label{alg:access_vm}
    \DontPrintSemicolon

    \SetKwFunction{Access}{Access}
    \SetKwProg{Fn}{Function}{:}{}

    \Fn(\tcp*[f]{Access the $i$-th word in $v$'s VM}){\Access{$m_v[i]$}} {
        \uIf{$v$ is the root} {
            Directly access $m_v[i]$ on physical memory\;
        }
        \Else{
            $u \gets v$'s parent\;
            $j \gets$ the translated address of $m_v[i]$ on $m_u$\;
            \Access{$m_u[j]$}\;
        }
    }
\end{algorithm}

\paragraph{Performing a query.} Next we describe how to efficiently implement a query to some node $v$.

\begin{claim}
We can perform a rank/select query in expected time $O(\log^2 n)$.     
\end{claim}
\begin{proof}
The query algorithm starts at the root and traverses down some root-to-leaf path. Before visiting any node $u$, we already know the parameters describing the subproblem for $u$'s subtree: the key range $[a_u, b_u)$, the number of keys $n_u$, and the maximum allowed weight $\hmax_u$. We also know the spill $k_u$ of the encoding for the treap. The remaining information for $u$'s subtree is encoded in $u$'s VM $m_u$.

The main task when we visit node $u$ is to retrieve all intermediate spills and variables used in the encoding process at $u$. Assume for now that $u$ is in the large-universe case (see \cref{fig:ram_encoding_large_universe}). We begin by retrieving $\mSpill$ via $O(1)$ word reads to $m_u$ (each of which takes $O(\log n)$ expected time by \cref{claim:translation}), and then obtain $\kWeight$ by combining $\mSpill$ and $k_u$. Recall that node $u$ uses three entropy encoders to encode $r$, $\phighmid$, and $h(p)$. A key property of entropy encoders is that, once the output spill is known, both the input spill and the encoded symbol can be inferred in constant time without further VM accesses (\cref{item:entropy-encoder-simulate-kIn,item:entropy-encoder-simulate-phi} of \cref{thm:entropy-encoder}). By repeatedly applying this property, we can retrieve $\kPivot$, $\kRank$, $\kCat$, as well as $h(p)$, $\phighmid$, and $r$, all from $\kWeight$ in constant time.

Recall that $h(p)$ and $\phighmid$ determine $\plow$, as $\plow$ is the unique value for which $A_{f(\phighmid)}[\plow] = h(p)$. We can therefore recover $\plow$ using a lookup table of size $\poly(\nmax)$ that maps $(f(\phighmid), h(p))$ to $\plow$. Thus we can recover $\plow$, and therefore $p$, in $O(1)$ time. 

At this point, we have enough information $(p, r)$ at node $u$ to determine whether to recurse to the left or right child $v$, and to infer the key range $[a_v, b_v)$, the number of keys $n_v$, and the maximum allowed weight $\hmax_v = h(p) - 1$ for $v$. Additionally, we can decode $(\kLeft, \kRight)$ from $\kCat$, in order to get $k_v$. We then recurse into child $v$ and repeat the process until the query completes. 

The process is similar when $u$ is in the small-universe case: We combine $\mSpill$ and $k_u$ to obtain $\kAll$, and then retrieve both $(\Delta p, r)$ and $\kCat$ from $\kAll$ using the entropy encoder. We then obtain the pivot $p$ by computing $p = a + \Delta p$. The remaining steps are the same as in the large-universe case.

Finally, having described the query algorithm, we can now analyze its time complexity. For each of the $O(\log n)$ nodes $u$ that the query algorithm visits, the time spent by the algorithm is dominated by two sources: the $O(\log n)$ time to perform $O(1)$ word reads to the VM $m_u$ (in order to retrieve $\mSpill$); and the time needed to compute the sizes of various VMs and spills, which we will show at the end of the section can be done in constant time per node. Since the query visits a total of $O(\log n)$ nodes (with high probability in $n$), the total expected query time is $O(\log^2 n)$.
\end{proof}

\paragraph{Insertion.}
We now describe how to perform insertions. We begin by recalling the insertion strategy for the non-compressed tabulation-weighted treap, which is based on subtree rebuilds.

To insert a new key $x$ into the treap, we first search for $x$, traversing a path $v_1, \ldots, v_{d_x}$ in the treap, where $v_1$ is the root and where $d_x = O(\log n)$ with high probability in $n$. We then find the first node $\rho$ on the path whose weight is less than $h(x)$, meaning that $\rho$'s position will be replaced by $x$ after the insertion. To perform the insertion, we rebuild the entire subtree rooted at $\rho$, spending time linear in the size of the subtree (assuming we are operating a non-compressed treap). If no such $\rho$ exists, we insert $x$ as a leaf node.

Before continuing, we prove the following standard claim about the expected size of the subtree that is rebuilt, when a given element $x$ is inserted:
\begin{claim}
    The expected number of nodes rebuilt during an insertion is $O(\log n)$. 
    \label{claim:rebuild}
\end{claim}
\begin{proof}
To analyze the expected number of nodes rebuilt, recall a property of our weight function from \cref{sec:treap}: With high probability in $n$, the weights of a fixed set of keys are pairwise distinct, and their relative order is uniformly random over all total orders. From the perspective of the current claim, this is equivalent to the keys having i.i.d.\ uniformly random real-valued weights in $[0, 1)$. To simplify the exposition, we will adopt this $[0, 1)$-weight perspective for the rest of the proof.

Let $v_1, v_2, \ldots, v_{O(\log n)}$ be the root-to-leaf path that we follow during the insertion. For each $v_i$, let $S_i$ denote the size of the subtree rooted at $v_i$. Define $R_i$ to be $S_i$ if the subtree rooted at $v_i$ is rebuilt, and to be $0$ otherwise. We will argue that, conditioned on the weight $w := h(v_{i - 1}) \in [0, 1)$ of $v_i$'s parent, conditioned on $h(x) < w$, and conditioned on any fixed value $s_i$ for $S_i$, we have $\E[R_i] = O(1)$. 

To see this, observe that conditioning on the weight $w$ of $v_{i - 1}$ and conditioning on $h(x) < w$ results in all of the $S_i + 1$ elements in $v_i$'s subtree (including $x$) having i.i.d.\ uniformly random weights in $[0, w)$. It follows that the probability of $x$ having the largest weight in the subtree (and therefore causing a rebuild) is $O(1 / s_i)$. Since a rebuild causes $R_i = s_i$, the conditional expected value $\E[R_i \mid w, s_i, (h(x) < w)]$ is $O(1)$.

Since  $\E[R_i \mid w, s_i, (h(x) < w)]$ is $O(1)$ regardless of $w$ and $s_i$, it follows that 
$$\E[R_i \mid h(x) < h(v_{i - 1})] = O(1).$$
Finally, since $h(x) < h(v_{i - 1})$ is a necessary condition for the subtree rooted at $v_i$ to be rebuilt (otherwise, a higher subtree would be rebuilt), we can conclude that the expected cost (i.e., number of nodes rebuilt) from rebuilding the tree rooted at $v_i$ is $O(1)$. Summing over the $v_i$'s gives the desired bound of $O(\log n)$.

\end{proof}

We now show how to perform the rebuild time-efficiently, in order to obtain insertions in expected time $O(\log^3 n)$.
\begin{claim}
We can perform an insertion in expected time $O(\log^3 n)$. 
\end{claim}
\begin{proof}
To implement the insertion on our encoding, we first follow the query algorithm to find the path $v_1, \ldots, v_{d_x}$, retrieving all spills and encoded symbols along the path, and locate the node $\rho$ whose subtree needs to be rebuilt. Assuming direct access to all intermediate VMs and spills, we proceed as follows:

\begin{enumerate}
\item First, we decode the subtree currently rooted at $\rho$. This requires $O(n_\rho)$ reads to $m_\rho$, which by \cref{claim:translation} takes $O(n_\rho \log n)$ time.
\item Next, we construct the new (uncompressed) treap that will be rooted at $\rho$. This takes time $O(n_\rho \log n_\rho)$ (or time $O(n_\rho)$ if implemented carefully).
\item Next, we compute the compressed treap rooted at $\rho$. This takes expected time $O(n_\rho \log n)$, because given a parent $u$ and two children $v_1, v_2$, we can build the compressed treap rooted at $u$ using the compressed treaps for the two children in time linear in the size of the compressed treap rooted at $u$. This step produces a new VM $\hat m_\rho$ and a new spill $\hat k_\rho$ to be used in the larger data structure. 
\end{enumerate}

At this point, there are two steps that must be completed in order to actually update the data structure:
\begin{enumerate}
\item The first is to write the new VM $\hat m_\rho$ and spill $\hat k_\rho$ in place of the VM and spill that formerly represented the subtree rooted at $\rho$. This will require $\Theta(n_\rho)$ word writes to the VM $m_\rho$. 
\item The second is to update the metadata encoded along the root-to-$\rho$ path. Specifically, for every node $u$ on the path from the root to $\rho$, we will need to update the stored subproblem size $n_u$ to $n_u + 1$, and we may also need to update the pivot rank $r_u$ depending on whether $x$ is inserted to the left or right of $u$. To implement these changes, the entropy encoders must update their distributions and encoded symbols. According to \cref{item:entropy-encoder-distribution-change} of \cref{thm:entropy-encoder}, this can be completed by performing $O(\text{VM size difference})$ allocations/releases and word accesses on the output VM of the entropy encoder, and updating the output spill. Note that the size of the output VM of any entropy encoder at node $u$ is within $O(w)$ bits of the encoding size $\log \mathcal{N}(n_u, a_u, b_u, \hmax_u)$ of $u$'s subproblem, since the spill, redundancy, and symbols to be encoded are all within $O(w)$ bits. Therefore, the number of allocations/releases and word accesses per node in this step is bounded by
\begin{align*}
O(\text{VM size difference}) &= \abs*{\frac{\log \mathcal{N}(n_u + 1, a_u, b_u, \hmax_u) - O(w)}{w} - \frac{\log \mathcal{N}(n_u, a_u, b_u, \hmax_u) - O(w)}{w}} \\
&= \frac{1}{w} \cdot \abs*{\log \binom{V(a_u, b_u, \hmax_u)}{n_u + 1} - \log \binom{V(a_u, b_u, \hmax_u)}{n_u}} \pm O(1) \\
&= \frac{1}{w} \cdot \abs*{\log \bk[\big]{V(a_u, b_u, \hmax_u) - n_u} - \log (n_u + 1)} \pm O(1) \\
&= \frac{1}{w} \cdot \abs{O(w) - O(w)} \pm O(1) \\
&= O(1).
\end{align*}
\end{enumerate}

At this point, we have generated a sequence of word allocations/releases, word updates, and spill updates required to complete the insertion. However, as described so far, these updates take place in VMs and spills that are nested deep within the tree. To actually implement the updates, we perform a sequence of translations, where we start with the updates at whichever node $v_\ell$ satisfies $v_\ell = \rho$; we translate the updates at $v_{\ell}$ into updates at $v_{\ell - 1}$; we translate the updates at $v_{\ell - 1}$, including the updates to $v_{\ell - 1}$'s metadata, into updates at $v_{\ell - 2}$; and so on, until we have translated all of our updates into word writes that can be directly implemented at the root of the tree. Specifically, these translations work as follows: 
\begin{itemize}
\item \textbf{Translating Word Writes:} Suppose $v$ is a child of $u$. By the same analysis as in Claim \ref{claim:translation}, a write to $m_v$ (or to $k_v$) can, in constant time, be translated into either a single write to $m_u$, or into an update to both $k_u$ and $\mSpill$ (at $u$). It follows that, if we start with $t$ word writes to $m_v$, this translates to at most $t + O(1)$ word writes to $m_u$ (where the $O(1)$ accounts for updating $\mSpill$, which is in $m_u$), along with an update to $k_u$.
\item \textbf{Translating Allocation/Releases: }Suppose $v$ is a child of $u$. Recall that $v$'s VM $m_v$ is fed into the adapter at node $u$ as an input VM. By \cref{thm:adapter}, each allocation/release on $m_v$ can be simulated by $O(\log |\mCat|) = O(\log n)$ word reads/writes on the VM $\mCat$ at node $u$, which can in turn be simulated by $O(\log n)$ word read/writes on $m_u$.
\item \textbf{Translating Metadata Updates at the Current Node:} Finally, consider the intermediate VMs used in the encoding process of $u$. As in the query algorithm, each word update on $\mCat$, $\mRank$, $\mPivot$, $\mWeight$, and $\mAll$ at node $u$ can be simulated by either a single write to $m_u$, or by an update to both $k_u$ and $\mSpill$ (at $u$). Additionally, all of the intermediate spills at node $u$ can be accessed and modified through $\mSpill$ and $k_u$. Thus, the metadata updates that take place within the encoding of each node $u$ can be translated into $O(1)$ word writes to $m_u$, along with an update to $k_u$.
\end{itemize}

Finally, we analyze the time complexity of this bottom-up update. The starting point is the rebuilt subtree rooted at $\rho$, which initiates $|m_\rho| = O(n_\rho)$ word updates to rewrite the entire VM. Let $v$ be an ancestor of $\rho$ and $u$ be $v$'s parent. As discussed above, if $v$ receives $t$ word updates on $m_v$, then the translation step from $v$ to $u$ takes $O(t + \log n)$ time and produces $t + O(\log n)$ word updates at $m_u$. Since the initial number of word updates at $\rho$ is $O(n_\rho)$, and since the depth of $\rho$ is at most $O(\log n)$, it follows that each of the $O(\log n)$ translation steps (including the updates performed at the root at the end of the translation) takes time at most
\[O(n_\rho + \log^2 n).\]
Summing over the $O(\log n)$ translation steps, the total time for the bottom-up update is 
\[O(n_\rho \log n + \log^3 n).\]

Combining this with the $O(n_\rho \cdot \log n)$ time spent on preparing the update, the total time to perform the insertion is 
\[O(n_\rho \log n + \log^3 n).\]
Since $\E[n_\rho] = O(\log n)$, this gives an expected time bound of $O(\log^3 n)$.
\end{proof}

\paragraph{Deletion.}

The deletion algorithm is similar to the insertion algorithm. We first use the query algorithm to locate the key $x$ to be deleted, then rebuild the subtree rooted at $x$. Finally, we propagate all VM and spill updates bottom-up, repeatedly translating these updates into word accesses to the next level. The expected time complexity for a deletion is $O(\log^3 n)$ by the same argument as for insertions.

\paragraph{Calculating intermediate VM and spill sizes.}
Finally, as noted at the beginning of the subsection, we have assumed for ease of discussion that the intermediate VM and spill sizes are already known. We conclude the subsection by describing how to compute them efficiently.

Suppose we need to access a node $u$ in the large-universe case. Before accessing $u$, we already know the parameters $n_u, a_u, b_u, \hmax_u$ of $u$'s subproblem. We compute the size of the intermediate VMs and spills in $u$'s encoding process as follows:
\begin{enumerate}
\item We apply \cref{claim:output_size} to compute the size of $u$'s output VM $\mOut$ and spill $\kOut$ in constant time.
\item $n_u$ and $\hmax_u$ determine the distribution $\DistWeightTd[n_u, \hmax_u]$ used in the entropy encoder for Step 4. We use the index of this distribution to query the output sizes $\MWeight$ and $\KWeight$ of the entropy encoder for Step 4.
\item We apply \cref{claim:hmax} to compute the parameter $\HMax(n_u, \tilde{a}_u, \tilde{b}_u, \hmax_u, h(p))$ that we provide to the entropy encoder in Step 3, which takes constant time.
\item The distribution $\mathrm{Unif}\,[\tilde{a} / W, \, \tilde{b} / W)$ and input size functions used by the entropy encoder in Step 3 depend on $\tilde{a}$, $\tilde{b}$, and $h(p)$.\footnote{Although $\tilde{a}$ and $\tilde{b}$ already determine the range $[\tilde{a} / W, \, \tilde{b} / W)$ of the uniform distribution, the associated input sizes depend on $h(p)$.} By providing the index of the distribution and $\HMax$ to the entropy encoder in Step 3, we can retrieve the encoded symbol $\phighmid$ in $O(1)$ time. We further obtain $p$ from $\phighmid$ and $h(p)$.
\item $p$ determines $\tilde{V}_L$ and $\tilde{V}_R$, which further determine the distribution $\DistRankTd[n_u, \tilde{V}_L, \tilde{V}_R]$ used in the entropy encoder for Step 2. We use the index of this distribution to query the output sizes $\MRank(p)$ and $\KRank(p)$ of the entropy encoder for Step 2. They are also the input sizes for the entropy encoder in Step 3. We also query the entropy encoder in Step 2 to obtain its input sizes $\MCat(p, r)$ and $\KCat(p, r)$.
\item Finally, since we already know $p$ and $r$ on the current node $u$, we can obtain the subproblem parameters $n_v, a_v, b_v, \hmax_v$ for both of $u$'s children. By \cref{claim:output_size}, we can further compute the sizes of both children's output VMs and spills in constant time, which equal the sizes of $\mLeft, \mRight, \kLeft, \kRight$ in $u$'s encoding process.
\end{enumerate}

It is also easy to compute the intermediate VM and spill sizes for the small-universe case. The distribution $\DistSmall[i, j, a', b', n, \hmax]$ used in the only entropy encoder depends only on $i, j, a', b', n, \hmax$, which can be directly inferred from the subproblem parameters. Then we query the entropy encoder to obtain its output sizes $\MAll, \KAll$ and input sizes $\MCat(\Delta p, r), \KCat(\Delta p, r)$.

It is worth remarking that the calculations above contribute at most $O(1)$ time at each node that we visit during a given query or update. In expectation, this contributes an additional $O(\log n)$ time for queries, and an additional $O(\log n + \E[n_\rho]) = O(\log n)$ time for insertions and deletions. Thus the time spent on calculating VM and spill sizes does not affect the expected-time bound claimed earlier in the section.

\subsection{The Failure Mode}
\label{sec:failure_mode}

So far, we have assumed that the key set $S$ satisfies the two properties described in \cref{sec:ram_framework}. If either of these properties is violated, we say that the data structure enters \defn{failure mode}. In this case, we use an encoding scheme for the key set $S$ with at most $O(w)$ bits of redundancy, while allowing $\poly(n)$ time per operation. Since failure occurs with probability $1 / \poly(n)$ for a sufficiently large polynomial factor, the expected time overhead incurred by the failure-mode encoding is $o(1)$.

The encoding scheme in failure mode depends on whether $n \ge \sqrt{w} = \Theta(\sqrt{\log \Umax})$.
When $n \ge \sqrt{w}$, we represent each key set $S$ as a sequence of $n$ increasing keys $x_1 < \cdots < x_n$. Let $z_S$ denote the rank of $(x_1, \ldots, x_n)$ in lexicographic order among all $\binom{U}{n}$ possible key sequences. To encode $S$, we simply write the binary representation of $z_S$ in memory using $\ceil[\big]{\log \binom{U}{n}}$ bits. This encoding wastes at most one bit due to rounding.

Given a key set $S$, we can compute $z_S$ in polynomial time as follows. First, we compute the number of key sets $S'$ whose smallest key $x'_1$ is less than $x_1$, which is $\binom{U}{n} - \binom{U - x_1}{n}$. All such key sets are lexicographically smaller than $(x_1, \ldots, x_n)$. The remaining task is to compute the rank of $(x_2, \ldots, x_n)$ among all key sets with $n - 1$ keys from the range $(x_1, U)$, which can be done recursively using the same algorithm.
Given $z_S$, we can also decode $S$ in polynomial time using binary search and the above procedure, which takes $\poly(n) \cdot \log \Umax = \poly(n)$ time.
With these two algorithms, we can implement insertions, deletions, and rank/select queries on the key set $S$, each in $\poly(n)$ time.

When $n < \sqrt{w}$, we directly write down the binary representations of all keys, which takes 
\[
n \ceil[\big]{\log U} = \log \binom{U}{n} + O(n \log n) \le \log \binom{U}{n} + O(\sqrt{w} \log w)
\]
bits of space. This means that the redundancy of this encoding scheme is $O(\sqrt{w} \log w) \le O(w)$ bits. Under this encoding scheme, we can implement insertions, deletions, and rank/select queries in $O(n)$ time, as desired.

To combine the failure-mode encoding with the normal-mode treap encoding, we use a special bit to indicate whether the data structure is in failure mode. When performing an insertion on a normal-mode treap, we check whether the insertion would violate either of the two assumed properties, and switch to failure mode if so. When performing a deletion on a failure-mode data structure, we spend polynomial time to check if the two properties are restored, and switch back to normal mode if so. Since the probability of failure is small, this additional check incurs only negligible time overhead.

\subsection{Putting the Pieces Together}
\label{sec:ram_summary}

In this subsection, we combine all pieces together to form the complete proof of \cref{lem:treap}.

\TreapLemma*

\begin{proof}[Proof of \cref{lem:treap}]
We begin by constructing a weight function as introduced in \cref{sec:treap}. This weight function generates $\poly(W) = \poly(\nmax)$ auxiliary arrays, each of size $W^4 = \poly(\nmax)$, which we store in the lookup table.

Next, we use $O(1)$ words to store $n$, $U$, and a single bit indicating whether the data structure is in failure mode. If a failure occurs, we use the failure-mode encoding for the key set $S$ described in \cref{sec:failure_mode}, which requires at most $\log \binom{U}{n} + O(w)$ bits of space.

If there is no failure, we apply the recursive construction from \cref{sec:ram_large_universe,sec:ram_small_universe} to encode the key set as a treap. This construction encodes $S$ in a VM with $M(n, \, 0, \, U, \, W - 1)$ words and a spill with universe size $K(n, \, 0, \, U, \, W - 1) = 2^{O(w)}$. By \eqref{eq:space_usage_ram}, the space usage for the treap is at most
\[
w \cdot M(n, \, 0, \, U, \, W - 1) + \log K(n, \, 0, \, U, \, W - 1) \le \log \mathcal{N}(n, \, 0, \, U, \, W - 1) + \frac{n}{\nmax} \le \log \binom{U}{n} + 1
\]
bits.
Finally, we store the binary representation of the spill directly in $O(1)$ memory words, incurring at most $O(w)$ bits of redundancy.

Combining all of the above, the total space usage of the data structure is at most $\log \binom{U}{n} + O(w)$ bits, as desired.

As discussed in \cref{sec:ram_operations}, when there is no failure, the data structure supports queries in $O(\log^2 n)$ expected time, and insertions and deletions in $O(\log^3 n)$ expected time. A failure occurs with probability $1 / \poly(n)$ for a sufficiently large polynomial factor, in which case the data structure spends $\poly(n)$ time per operation. Taking both cases into account yields the desired time complexity for all operations.

\paragraph{Lookup tables.}

It remains to analyze the size of the lookup tables used in our data structure. We focus first on the lookup tables for the entropy encoders.

Recall that each entropy encoder is constructed with a family $\mathcal{F}$ of distributions and encodes a symbol $\phi$ following some distribution $D \in \mathcal{F}$. By \cref{thm:entropy-encoder}, the lookup table size for each entropy encoder is
$
\poly\bk[\big]{|\mathcal{F}|, \, q, \, \max_{D \in \mathcal{F}} \abs[\big]{\supp(D)}}
$
words, where $q = \nmax^2$ is our chosen parameter. The only exception is the entropy encoder for $\phighmid$, which requires no lookup table since $\phighmid$ follows a uniform distribution. For all other entropy encoders, we verify that $|\mathcal{F}| \le \poly(\nmax, \log \Umax)$ and $\abs[\big]{\supp(D)} \le \poly(\nmax)$ for every $D \in \mathcal{F}$, ensuring that each lookup table uses at most $\poly(\nmax, \log \Umax)$ words, as desired.
\begin{itemize}
\item For the entropy encoder that encodes $r \sim \DistRankTd$, each distribution $\DistRankTd$ in the family is parameterized by three numbers $n, \tilde{V}_L, \tilde{V}_R \le 2^{O(w)}$. Since $\tilde{V}_L$ and $\tilde{V}_R$ are integers of the form $\floor{(1 + 1 / \nmax^3)^t}$ for $t \in \N$ (see \eqref{eq:rounding-rule-rank}), there are only $O(\nmax^3 \log \Umax)$ possible values for $\tilde{V}_L$ and $\tilde{V}_R$. Thus, the family contains at most $\poly(\nmax, \, \log \Umax)$ distributions. Additionally, $\abs[\big]{\supp\bk[\big]{\DistRankTd}} = O(\nmax)$ since $r \in [0, n - 1]$.
\item For the entropy encoder that encodes $h(p) \sim \DistWeightTd$, each distribution $\DistWeightTd$ is parameterized by two integers $n, \hmax \le \poly(\nmax)$, so the family has size $\poly(\nmax)$. We also have $\abs[\big]{\supp\bk[\big]{\DistWeightTd}} = \poly(\nmax)$ since $h(p) \in [W] = [\poly(\nmax)]$.
\item In the small-universe case, we use an entropy encoder for $(\Delta p, r) \sim \DistSmall$, where each distribution is parameterized by a tuple $(i, j, a', b', n, \hmax)$ of integers bounded by $\poly(\nmax)$, so the family size is $\poly(\nmax)$. We also have $\abs[\big]{\supp\bk[\big]{\DistSmall}} = \poly(\nmax)$ since both $\Delta p$ and $r$ are bounded by $\poly(\nmax)$ in the small-universe case.
\end{itemize}
Besides entropy encoders, our data structure also uses lookup tables at the following places:
\begin{itemize}
\item The adapters in the recursive construction, which concatenates VMs of at most $O(\nmax)$ words. By \cref{thm:adapter}, they use a lookup table of $O(\nmax^3)$ words.
\item Auxiliary arrays to describe the weight function, which uses $\poly(\nmax)$ words.
\item A lookup table for \cref{claim:output_size,claim:hmax} to compute $M(n, a, b, \hmax)$, $K(n, a, b, \hmax)$, and $\HMax(n, a, b, \hmax, h(p))$ in the large-universe case, and a lookup table in \cref{sec:ram_small_universe} to compute $M(n, a, b, \hmax)$ and $K(n, a, b, \hmax)$ in the small-universe case. These tables occupy $\poly(\nmax, \log \Umax)$ words.
\item A lookup table that maps $(f(\phighmid), h(p))$ to $\plow$, which is used in the query algorithm. This table occupies $\poly(\nmax)$ words.
\end{itemize}
All these tables occupy $\poly(\nmax, \log \Umax)$ words of space, as desired.
\end{proof}

\section{Proof of Theorem \ref{thm:fid}}\label{sec:mainthm}
\label{sec:proof-main-theorem}

So far, we have proved \cref{lem:treap}, which shows how to use treaps to construct dynamic FIDs that support $O(\log^3 \nmax)$-time operations and incur $O(w)$ bits of redundancy. In this section, we use \cref{lem:treap} as a subroutine to prove the main theorem in this paper. The proof uses the standard idea of partitioning the key universe into many blocks, and maintaining a separate FID for each block.

\MainTheoremFID*

\begin{proof}
Let $B$ be a parameter set to satisfy $$2^{2 \cdot (\log n / \log w)^{1/3}} \le B \le 2^{O(\log n / \log w)^{1/3}},$$ and assume for now that $n \ge \polylog U$. Additionally, for this first part of the proof, we will assume that $n$ stays within some fixed constant-factor range, that we can use a fixed $B$ even as $n$ changes. At the end of the proof, we will extend the data structure to allow $n$ to change arbitrarily.

We partition the key universe $[0, U)$ into disjoint intervals $I_1, I_2, \dots$, each containing at least $B$ keys and at most $4B$ keys. We call each interval a \defn{block}. Let $U^{(i)}$ denote the size of the part of the key universe in the $i$-th block, and let $n^{(i)}$ be the number of keys in the $i$-th block.

All keys in the $i$-th block are stored in a small FID provided by \cref{lem:treap}, with parameters $\nmax = 4B$ and $\Umax = U$. By \cref{lem:treap}, the small FID on block $I_i$ uses
\[
\log \binom{U^{(i)}}{n^{(i)}} + O(w)
\]
bits of space. Summing over all blocks, the total space usage of the small FIDs is at most
\[
\sum_{i=1}^{\Theta(n / B)} \bk*{\log \binom{U^{(i)}}{n^{(i)}} + O(w)} \le \log \binom{\sum U^{(i)}}{\sum n^{(i)}} + O(wn / B) = \log \binom{U}{n} + O(wn / B)
\numberthis \label{eq:small-fid-size-reduction}
\]
bits.

In addition to the FIDs, we also use a \emph{dynamic fusion tree} \cite{patrascu2014dynamic} to maintain the collection of blocks, recording the interval boundaries and the number of keys in each block. Since there are $O(n / B)$ blocks, the dynamic fusion tree uses $O(w n / B)$ bits of space, and supports insertions, deletions, and rank/select queries in $O(\log n / \log w)$ time.

\paragraph{Rank/select queries.}
To answer a rank query, i.e., the number of keys smaller than $x$, we first find the block $I_k$ that contains $x$ using the dynamic fusion tree. Then, we perform a rank query on the dynamic fusion tree to calculate the number of keys in the first $k - 1$ blocks, and perform a rank query on the FID on block $I_k$ to calculate the number of keys in $I_k$ that are smaller than $x$. Adding these two results gives the final answer.

Similarly, for a select query, we first find the block $I_k$ that contains the $i$-th smallest key using the dynamic fusion tree, and then perform a select query on the FID on block $I_k$ to find the $i$-th smallest key.

Since both rank and select queries perform a constant number of queries on the dynamic fusion tree and the small FIDs, the expected time complexity of each query is $O(\log n / \log w)$.

\paragraph{Insertions and deletions.}
To perform insertions and deletions, we first use the dynamic fusion tree to find the block $I_k$ that contains the key to be inserted or deleted. Then, we perform an insertion or deletion on the FID on block $I_k$, which takes $O(\log^3 B) = O(\log n / \log w)$ time. We also need to update the number of keys in the block $I_k$ in the dynamic fusion tree, which also takes $O(\log n / \log w)$ time.

When an insertion causes the number of keys in block $I_k$ to exceed $4B$, we split the block into two new blocks, taking $O(B \cdot \log n / \log w)$ time. Similarly, when a deletion causes the number of keys in block $I_k$ to drop below $B$, we merge the block with its neighboring block, forming a new block with at most $5B$ keys, and further partition the new block into two halves if necessary, which also takes $O(B \cdot \log n / \log w)$ time. Since splits and merges of blocks only occur after $O(B)$ insertions and deletions to each block, the amortized expected time complexity of each insertion and deletion is $O(\log n / \log w)$.

Finally, it is worth recalling that updates to a given block $I_k$ may temporarily use up to $O(n^{(i)} w)$ bits of intermediate space. Because $n \ge \polylog U = \poly(w)$ and $n^{(i)} = n^{o(1)}$, this intermediate space is negligible in our overall space bound.

\paragraph{Concatenation of FIDs and space usage.}

By \cref{lem:treap}, the small FID on each block is stored on a VM that resizes as insertions and deletions occur. To concatenate these VMs (along with the VM storing the dynamic fusion tree), we use a standard ``chunking trick'', which is captured in the following lemma (see, e.g., Lemma 5.1 of \cite{li2023dynamic}, for an example of a similar lemma in prior work).
\begin{lemma}
\label{lem:concatenate-vm}
Assume there are $K$ VMs of $\ell_1, \ldots, \ell_K$ bits respectively, where $\sum_{i=1}^K \ell_i \leq L$ always holds. We can store all these $K$ VMs and their lengths within $\sum_i \ell_i + O(\sqrt{LK \log L} + K \log L)$ memory bits under the word RAM model with word size $w = \Omega(\log L)$. Each word access or allocation/release in any VM takes $O(1)$ amortized time to complete.
\end{lemma}
\begin{proof}
The algorithm breaks physical memory into chunks of size $Q := \Theta(\sqrt{(L / K) \cdot \log L})$ bits. It maintains the invariant that each VM $i$ is allocated $\left\lfloor \ell_i / Q \right\rfloor$ chunks that the VM uses 100\% of, up to one chunk that the VM uses some prefix of, and up to one chunk that the VM does not (currently) use any of. It also maintains the invariant that, for every time a chunk is allocated/removed from a VM, the VM has performed at least $\Omega(Q)$ allocations/releases to warrant that chunk allocation/removal. The algorithm can therefore afford to incur $O(Q)$ time performing a chunk allocation/removal.

Finally, the algorithm maintains the invariant that the allocated chunks form a prefix of physical memory. When a chunk is allocated, it is added to the prefix. And when a chunk $i$ is removed, whichever chunk $j$ is currently at the end of the prefix gets moved to position $i$ (in time $O(Q)$). 

This construction incurs $O(1)$ amortized time per VM allocation/release. Moreover, beyond the $\sum_i \ell_i$ bits that the VMs require, the construction has two sources of space:
\begin{itemize}
    \item \textbf{Metadata: } This includes both the sizes of the VMs, and the metadata needed to store the mapping between chunks and VMs. Since there are at most $O(L/Q)$ chunks and at most $K$ VMs, the metadata can be stored using a total of $O((K + L/Q) \cdot \log L)$ bits.
    \item \textbf{Wasted Chunks: } The empty space in the up to $O(K)$ chunks that are not fully utilized takes a total of at most $O(KQ)$ bits of space.
\end{itemize}
The total space redundancy is therefore at most
$$O((K + L/Q) \cdot \log L + QK) = O(\sqrt{LK \log L} + K \log L).$$
bits.
\end{proof}

To apply \cref{lem:concatenate-vm}, we set $K = \Theta(n / B)$ to be the number of VMs we are combining, and we use $L = \Theta(nw)$ to be the maximum size of their concatenation. By \eqref{eq:small-fid-size-reduction}, the total space overhead of using \cref{lem:concatenate-vm} is
\begin{align*}
  O\bk*{K \log L + \sqrt{LK \log L}} &\le O\bk*{(n/B) \log (nw) + \sqrt{nw \cdot (n/B) \log (nw)}} \\
  &\le O\bk[\big]{nw/B + nw / \sqrt{B}} = O\bk[\big]{nw / \sqrt{B}}
  \numberthis \label{eq:chunkingoverhead}
\end{align*}
bits.

Adding together the sizes of the small FIDs (given by \eqref{eq:small-fid-size-reduction}), the size of the dynamic fusion tree ($O(wn / B)$ bits), the extra space needed to use \cref{lem:concatenate-vm} ($O(wn/\sqrt{B}) $ bits by \eqref{eq:chunkingoverhead}), and the $\poly(B \log U)$ bits of space used by the lookup tables from \cref{lem:treap}, the total space usage of the dynamic FID is
\begin{align*}
& \log \binom{U}{n} + O\bk[\big]{wn / \sqrt{B}} + \poly(B \log U)  \\
& \le \log \binom{U}{n} + O\left(w n \Big/ 2^{(\log n / \log w)^{1/3}}\right) \tag{since $B = n^{o(1)}$ and $n \ge \polylog U$}
\end{align*}
bits, as desired. 

\paragraph{Allowing $n$ to change arbitrarily.}
So far, we have operated under the assumption that $n$ stays within some fixed constant-factor range $[n_{\text{min}}, O(n_{\text{min}})]$, so that $B$ can be treated as a fixed parameter. We now describe how to support arbitrary $n$. Note that, when $n \le \polylog U$, our space bound allows us to revert to a space-inefficient solution (e.g., directly use a dynamic fusion tree \cite{patrascu2014dynamic}). Thus, we will focus on the case of $n \ge \polylog U$.

Each time that $n$ changes by a constant factor, we will rebuild the entire data structure to use a new value of $B$. This rebuild takes time $O(n \log n / \log w)$, and since the rebuild only happens when $n$ changes by a constant factor, the rebuild cost amortizes to $O(\log n / \log w)$ per insertion/deletion. In addition to rebuilding the data structure, we must also rebuild the lookup tables for the FIDs that are constructed with \cref{lem:treap}. This takes time $\poly(B \log U)$, which by the assumption $n \ge \polylog U = \poly(w)$ and by the fact $B = n^{o(1)}$, is $o(n \log n / \log w)$, and is therefore a low-order term. 

One minor point that one must be careful about is to perform the rebuild of the data structure without using a large amount of temporary intermediate space. The part of the rebuild that might, a priori, require a large amount of intermediate space is the step in which we must reformat the data structure to use a different chunk size in the proof of \cref{lem:concatenate-vm}. Note that the chunk size is a power of two without loss of generality, so the challenge is either to split chunks into two (which is easy) or to merge logically contiguous pairs of chunks into physically contiguous new chunks (which is slightly harder). The latter type of merge can be done space efficiently as follows: Given two small chunks $i$ and $j$ that need to be combined into a larger chunk, and given a pair of physically adjacent small chunks $k$ and $\ell$, we can swap small chunks $i$ and $j$ with small chunks $k$ and $\ell$, and update the data structure so that the (now physically adjacent) small chunks $i$ and $j$ become a single large chunk. This allows us to reformat the data structure to use larger (or smaller) chunks without ever using a large amount of intermediate memory (e.g., $O(1)$ chunks worth of temporary extra memory suffice, which is negligible to our space bounds). 
\end{proof}

\section{Open Questions and Directions for Future Work}

We conclude the paper with several open questions and directions for future work that we view as especially important. 
\begin{enumerate}
    \item \textbf{\boldmath{}Achieving redundancy $n^{1 - \Omega(1)}$ bits.} It remains open whether the space bounds in the current paper are optimal. For example, is it possible to get $n^{1 - \Omega(1)}$ bits of redundancy, or even $O(1)$ bits of redundancy, in the regime where we are storing $n$ keys that are each $\Theta(\log n)$ bits? Is it possible to prove lower bounds prohibiting such a solution? Even proving very weak lower bounds appears to be currently out of reach for known techniques.

    \item \textbf{Succinct dynamic predecessor search.} Given a dynamically changing set of $\Theta(n)$ keys $S$, where each key $s \in S$ is $O(\log n)$ bits, a van Emde Boas tree (or, similarly, a $y$-fast trie) can support $O(\log \log n)$-time updates and predecessor queries, while incurring redundancy $O(S \log n)$ bits. Here, the state of the art is a redundancy of $O(n)$ bits \cite{pibiri2017dynamic, pibiri2020succinct}. Can one hope for a significantly smaller redundancy, for example, of the form $n / \polylog n$? Such a result would appear to require a significantly different approach than the one taken in this paper.
    \item \textbf{Succinct sequences with insertions and deletions.} In this paper, we showed how to use succinct binary search trees to maintain a set of integers in a space- and time-efficient way. Another important application of binary trees is to maintain a \emph{dynamic sequence} of numbers \cite{li2023dynamic}, modified via insertions and deletions, where the numbers are not necessarily distinct. 
    
    The fact that the numbers are not necessarily distinct makes it quite difficult to apply the techniques from the current paper to this setting. Indeed, the tabulation-weighted treap relies heavily on the fact that different keys have different hashes, and therefore can be hashed to obtain pivots. Whether or not it is possible to design a time-efficient solution to the dynamic-sequence problem with very low redundancy remains a quite interesting open question.

    \item \textbf{Succinct fusion nodes. } Many applications of FIDs (e.g., in \cite{patrascu2014dynamic}) use the parameter regime where $n = \polylog U$, allowing them to support $O(1 + \log n/\log w) = O(1)$-time operation using a data structure known as a \emph{dynamic fusion node} \cite{patrascu2014dynamic, fredman1993surpassing, ajtai1984hash, fredman1994transdichotomous}. 
    
    Dynamic fusion nodes rely on quite different machinery from standard binary search trees, and the construction in the current paper does not offer any interesting space guarantees for this parameter regime. Whether it is possible to build succinct $O(1)$-time dynamic fusion nodes remains open. 

    \item \textbf{History independence as an algorithmic tool for more space-efficient data structures. }At the heart of the current paper is the introduction of a new history-independent data structure---the tabulation-weighted treap---that is especially amenable to dynamic compression. Are there other natural problems where, by introducing new history-independent data structures, one can unlock otherwise unobtainable improvements in space efficiency?
\end{enumerate}

\bibliographystyle{alpha}
\bibliography{library.bib}

\newcommand{\etalchar}[1]{$^{#1}$}
\begin{thebibliography}{HHM{\etalchar{+}}05}

\bibitem[AFK84]{ajtai1984hash}
Mikl{\'o}s Ajtai, Michael~L. Fredman, and J{\'a}nos Koml{\'o}s.
\newblock Hash functions for priority queues.
\newblock {\em Information and Control}, 63(3):217--225, December 1984.

\bibitem[AK74]{amble1974ordered}
Ole Amble and Donald~Ervin Knuth.
\newblock Ordered hash tables.
\newblock {\em The Computer Journal}, 17(2):135--142, January 1974.

\bibitem[BBJ{\etalchar{+}}16]{bender2016antipersistence}
Michael~A. Bender, Jonathan~W. Berry, Rob Johnson, Thomas~M. Kroeger, Samuel McCauley, Cynthia~A. Phillips, Bertrand Simon, Shikha Singh, and David Zage.
\newblock Anti-persistence on persistent storage: History-independent sparse tables and dictionaries.
\newblock In {\em Proc. 35th ACM SIGMOD-SIGACT-SIGAI Symposium on Principles of Database Systems (PODS)}, pages 289--302, 2016.

\bibitem[BCF{\etalchar{+}}24]{bender2024online}
Michael~A. Bender, Alex Conway, Mart{\'i}n {Farach-Colton}, Hanna Koml{\'o}s, William Kuszmaul, and Nicole Wein.
\newblock Online list labeling: Breaking the {$\log^2 n$} barrier.
\newblock {\em SIAM Journal on Computing}, pages FOCS22--60, April 2024.

\bibitem[BCS16]{bajaj2016practical}
Sumeet Bajaj, Anrin Chakraborti, and Radu Sion.
\newblock Practical foundations of history independence.
\newblock {\em IEEE Transactions on Information Forensics and Security}, 11(2):303--312, February 2016.

\bibitem[BDH{\etalchar{+}}19]{behnezhad2019fully}
Soheil Behnezhad, Mahsa Derakhshan, MohammadTaghi Hajiaghayi, Cliff Stein, and Madhu Sudan.
\newblock Fully dynamic maximal independent set with polylogarithmic update time.
\newblock In {\em Proc. 60th IEEE Symposium on Foundations of Computer Science (FOCS)}, pages 382--405, 2019.

\bibitem[BFGK24]{bender2024historyindependent}
Michael~A. Bender, Mart{\'i}n {Farach-Colton}, Michael~T. Goodrich, and Hanna Koml{\'o}s.
\newblock History-independent dynamic partitioning: Operation-order privacy in ordered data structures.
\newblock {\em Proceedings of the ACM on Management of Data}, 2(2), May 2024.

\bibitem[BG07]{blelloch2007strongly}
Guy~E. Blelloch and Daniel Golovin.
\newblock Strongly history-independent hashing with applications.
\newblock In {\em Proc. 48th IEEE Symposium on Foundations of Computer Science (FOCS)}, pages 272--282, 2007.

\bibitem[BK22]{bansal2022balanced}
Nikhil Bansal and William Kuszmaul.
\newblock Balanced allocations: The heavily loaded case with deletions.
\newblock In {\em Proc. 63rd IEEE Symposium on Foundations of Computer Science (FOCS)}, 2022.

\bibitem[BKP{\etalchar{+}}22]{berger2022memoryless}
Aaron Berger, William Kuszmaul, Adam Polak, Jonathan Tidor, and Nicole Wein.
\newblock Memoryless worker-task assignment with polylogarithmic switching cost.
\newblock In {\em Proc. 49th International Colloquium on Automata, Languages and Programming (ICALP)}, pages 19:1--19:19, 2022.

\bibitem[BP06]{buchbinder2006lower}
Niv Buchbinder and Erez Petrank.
\newblock Lower and upper bounds on obtaining history independence.
\newblock {\em Information and Computation}, 204(2):291--337, February 2006.

\bibitem[BS13]{bajaj2013hifs}
Sumeet Bajaj and Radu Sion.
\newblock {HIFS}: History independence for file systems.
\newblock In {\em Proc. 20th ACM conference on Computer and Communications Security (CCS)}, pages 1285--1296, 2013.

\bibitem[CLM85]{celis1985robin}
Pedro Celis, Per-Ake Larson, and J.~Ian Munro.
\newblock {Robin Hood} hashing.
\newblock In {\em Proc. 26th IEEE Symposium on Foundations of Computer Science (FOCS)}, pages 281--288, 1985.

\bibitem[Eli74]{elias1974efficient}
Peter Elias.
\newblock Efficient storage and retrieval by content and address of static files.
\newblock {\em Journal of the ACM}, 21(2):246--260, April 1974.

\bibitem[Fan71]{fano1971number}
Robert~Mario Fano.
\newblock {\em On the number of bits required to implement an associative memory}.
\newblock Massachusetts Institute of Technology, Project MAC, 1971.

\bibitem[FS89]{fredman1989cell}
Michael~L. Fredman and Michael~E. Saks.
\newblock The cell probe complexity of dynamic data structures.
\newblock In {\em Proc. 21st ACM Symposium on Theory of Computing (STOC)}, pages 345--354, 1989.

\bibitem[FW93]{fredman1993surpassing}
Michael~L. Fredman and Dan~E. Willard.
\newblock Surpassing the information theoretic bound with fusion trees.
\newblock {\em Journal of Computer and System Sciences}, 47(3):424--436, December 1993.

\bibitem[FW94]{fredman1994transdichotomous}
Michael~L. Fredman and Dan~E. Willard.
\newblock Trans-dichotomous algorithms for minimum spanning trees and shortest paths.
\newblock {\em Journal of Computer and System Sciences}, 48(3):533--551, June 1994.

\bibitem[GHSV07]{gupta2007compressed}
Ankur Gupta, Wing-Kai Hon, Rahul Shah, and Jeffrey~Scott Vitter.
\newblock Compressed data structures: Dictionaries and data-aware measures.
\newblock {\em Theoretical Computer Science}, 387(3):313--331, 2007.

\bibitem[HHM{\etalchar{+}}05]{hartline2005characterizing}
Jason~D. Hartline, Edwin~S. Hong, Alexander~E. Mohr, William~R. Pentney, and Emily~C. Rocke.
\newblock Characterizing history independent data structures.
\newblock {\em Algorithmica}, 42(1):57--74, May 2005.

\bibitem[Kus23]{kuszmaul2023strongly}
William Kuszmaul.
\newblock Strongly history-independent storage allocation: New upper and lower bounds.
\newblock In {\em Proc. 64th IEEE Symposium on Foundations of Computer Science (FOCS)}, pages 1822--1841, 2023.

\bibitem[LLYZ23]{li2023dynamic}
Tianxiao Li, Jingxun Liang, Huacheng Yu, and Renfei Zhou.
\newblock Dynamic ``succincter''.
\newblock In {\em Proc. 64th IEEE Symposium on Foundations of Computer Science (FOCS)}, pages 1715--1733, 2023.

\bibitem[LLYZ24]{li2024dynamic}
Tianxiao Li, Jingxun Liang, Huacheng Yu, and Renfei Zhou.
\newblock Dynamic dictionary with subconstant wasted bits per key.
\newblock In {\em Proc. 35th ACM-SIAM Symposium on Discrete Algorithms (SODA)}, pages 171--207, 2024.

\bibitem[LZ25]{liang2025optimal}
Jingxun Liang and Renfei Zhou.
\newblock Optimal static fully indexable dictionaries.
\newblock In {\em Proc. 52nd International Colloquium on Automata, Languages and Programming (ICALP)}, pages 114:1--114:20, 2025.

\bibitem[MR95]{motwani1995randomized}
Rajeev Motwani and Prabhakar Raghavan.
\newblock {\em Randomized Algorithms}.
\newblock Cambridge University Press, Cambridge, 1995.

\bibitem[NSW08]{naor2008historyindependent}
Moni Naor, Gil Segev, and Udi Wieder.
\newblock History-independent cuckoo hashing.
\newblock In {\em Proc. 35th International Colloquium on Automata, Languages and Programming (ICALP)}, pages 631--642, 2008.

\bibitem[NT01]{naor2001antipersistence}
Moni Naor and Vanessa Teague.
\newblock Anti-persistence: History independent data structures.
\newblock In {\em Proc. 33rd ACM Symposium on Theory of Computing (STOC)}, pages 492--501, 2001.

\bibitem[P{\v a}t08]{patrascu2008succincter}
Mihai P{\v a}tra{\c s}cu.
\newblock Succincter.
\newblock In {\em Proc. 49th IEEE Symposium on Foundations of Computer Science (FOCS)}, pages 305--313, 2008.

\bibitem[PT06]{patrascu2006timespace}
Mihai P{\v a}tra{\c s}cu and Mikkel Thorup.
\newblock Time-space trade-offs for predecessor search.
\newblock In {\em Proc. 38th ACM Symposium on Theory of Computing (STOC)}, pages 232--240, 2006.

\bibitem[PT12]{patrascu2012power}
Mihai P{\v a}tra{\c s}cu and Mikkel Thorup.
\newblock The power of simple tabulation hashing.
\newblock {\em Journal of the ACM}, 59(3):14:1--14:50, June 2012.

\bibitem[PT14]{patrascu2014dynamic}
Mihai P{\v a}tra{\c s}cu and Mikkel Thorup.
\newblock Dynamic integer sets with optimal rank, select, and predecessor search.
\newblock In {\em Proc. 55th IEEE Symposium on Foundations of Computer Science (FOCS)}, pages 166--175, 2014.

\bibitem[PV10]{patrascu2010cellprobe}
Mihai P{\v a}tra{\c s}cu and Emanuele Viola.
\newblock Cell-probe lower bounds for succinct partial sums.
\newblock In {\em Proc. 21st ACM-SIAM Symposium on Discrete Algorithms (SODA)}, pages 117--122, 2010.

\bibitem[PV17]{pibiri2017dynamic}
Giulio~Ermanno Pibiri and Rossano Venturini.
\newblock Dynamic {Elias-Fano} representation.
\newblock In {\em Proc. 28th Annual Symposium on Combinatorial Pattern Matching (CPM)}, pages 30:1--30:14, 2017.

\bibitem[PV20]{pibiri2020succinct}
Giulio~Ermanno Pibiri and Rossano Venturini.
\newblock Succinct dynamic ordered sets with random access.
\newblock Preprint arXiv:2003.11835, March 2020.

\bibitem[SA96]{seidel1996randomized}
Raimund Seidel and Cecilia~R. Aragon.
\newblock Randomized search trees.
\newblock {\em Algorithmica}, 16(4):464--497, October 1996.
\newblock See also FOCS 1989.

\bibitem[Tzo12]{tzouramanis2012historyindependence}
Theodoros Tzouramanis.
\newblock History-independence: A fresh look at the case of {R}-trees.
\newblock In {\em Proc. 27th ACM Symposium on Applied Computing (SAC)}, pages 7--12, 2012.

\bibitem[Vio23]{viola2023new}
Emanuele Viola.
\newblock New sampling lower bounds via the separator.
\newblock In {\em Proc. 38th Computational Complexity Conference (CCC)}, pages 26:1--26:23, 2023.

\end{thebibliography}

\appendix

\section{Entropy Encoders}
\label{sec:proof-entropy-encoder}

In this section, we prove \cref{thm:entropy-encoder,thm:entropy-encoder-uniform}. See \cref{table:notation_entropy_encoder} for the main notations that we define and use throughout the proof.

\thmEntropyEncoder*

\begin{table}[tp]
\centering
\caption{Notations in \cref{sec:proof-entropy-encoder}}\label{table:notation_entropy_encoder}
\begin{tabular}{l p{0.65\textwidth}}
\toprule
\textbf{Notation} & \textbf{Definition} \\
\midrule
$\phi \sim D$ & Symbol and distribution needs to be encoded \\
$\mathcal{F}$ & Collection of possible distributions $D$ \\
$\mIn, \kIn$ & Input VM and spill \\
$\MIn^D(\phi)$ & Number of words in $\mIn$ \\
$\KIn^D(\phi)$ & Spill universe size for $\kIn$ \\
$\mOut, \kOut$ & Output VM and spill \\
$\MOut^D$ & Number of words in $\mOut$ \\
$\KOut^D$ & Spill universe size for $\kOut$ \\
$\HIn^D$ & \tablewrap{Information-theoretic optimal encoding length for the entropy encoder, \\
$\displaystyle \HIn^D \defeq \max_{\psi \in \supp(D)} \BK*{\log \frac{1}{D(\psi)} + w \MIn^D(\psi) + \log \KIn^D(\psi)}$} \\
$\hat{\phi} \sim \widehat{D}$ & New symbol and distribution after an update operation \\
$q$ & Parameter controlling redundancy \\
\midrule
$\beta$ & \tablewrap{Large constant integer with \\
$\displaystyle \beta w \ge 4 \left( \log \max_{\psi, D'} \KIn^{D'}(\psi) + \log q + \log \max_{D'} {\abs[\big]{\supp(D')}} + 10 \right)$
} \\
$\MMax^D$ & Maximum $|\mIn| = \MIn^D(\phi)$ in the current distribution $D$ \\
$\MFix^D$ & $\MFix^D := \max\{\MMax^D - \beta,\ 0\} = \MOut^D$ \\
$\mFix$ & First $\MFix^D$ words of $\mIn$ (padded with arbitrary words if $|\mIn| < \MFix^D$). This is directly output as $\mOut$ \\
$\mRem$ & Portion of $\mIn$ starting from the $(\MFix^D+1)$-th word. If $|\mIn| \leq \MFix^D$, then $\mRem$ is empty. \\
$\mSpill, \, \kShort$ & $\kIn$ written as a bitstring $\mSpill$ and a shorter spill $\kShort$ with universe size $\KShort^D(\phi) \leq 2q$ \\
$\NSpill^D(\phi)$ & Number of bits in $\mSpill^D(\phi)$ \\
$\mShort$ & Concatenation of $\mRem$ and $\mSpill$. The pair $(\mShort, \kShort)$ is also the input to \cref{lem:succincter-lem5} \\
\midrule
$\NShort^D(\phi)$ & Number of bits in $\mShort$ \\
$\KShort^D(\phi)$ & Spill universe size for $\kShort$ \\
$\tilde{D}$ & Distribution perturbed from $D$. We are encoding the symbol $\phi$ with respect to $\tilde{D}$ in \cref{lem:succincter-lem5} \\
$\HShort^{D}$ & $\displaystyle \HShort^{D} := \max_{\psi \in \supp(\tilde{D})} \left\{ \log \frac{1}{\tilde{D}(\psi)} + \NShort^{D}(\psi) + \log \KShort^{D}(\psi) \right\}$ \\
$\mEnc,\, \kEnc$ & Output VM and spill of \cref{lem:succincter-lem5} \\
$\MFix^{\widehat{D}}$ & $\MFix$ calculated based on the new distribution $\widehat{D}$ \\
\bottomrule
\end{tabular}
\end{table}

\begin{proof}
  In \cite{patrascu2008succincter}, \Patrascu{} shows the following lemma, which is a \emph{static} version of \cref{thm:entropy-encoder} for the case where the variables to be encoded fit within $O(1)$ words.

  \begin{lemma}[Lemma 5 in \cite{patrascu2008succincter}, restated]
    \label{lem:succincter-lem5}
    Assume we need to encode a variable $\phi \in \supp(\tilde{D})$ and a pair $(\mShort, \kShort) \in \{0,1\}^{\NShort^D(\phi)} \times [\KShort^D(\phi)]$. Let
    \[
      \HShort^D = \max_{\psi \in \supp(D)} \left\{ \log \frac{1}{\tilde{D}(\psi)} + \NShort^D(\psi) + \log \KShort^D(\psi) \right\}.
    \]
    We further assume the word size $w = \Omega(\log {\abs{\supp(D)}} + \log q + \log \max_{\psi \in \supp(D)} \KShort^D(\psi))$ for a chosen parameter $q$, and that $\HShort^D = O(w)$.
    Then we can encode $(\phi, \mShort, \kShort)$ using spillover representation $(\mEnc, \kEnc) \in \{0,1\}^{\NEnc^D} \times [\KEnc^D]$ such that
    \begin{enumerate}
      \item $\KEnc^D \le 2q$.
      \item $\NEnc^D + \log \KEnc^D \le \HShort^D + 4/q$.
      \item Given a precomputed lookup table of $O(\abs{\supp(D)} \cdot q \cdot \max_{\psi \in \supp(D)} \KShort^D(\psi))$ words that only depends on the input functions $\NEnc^D, \KEnc^D, D$, both encoding $(\psi, \mShort, \kShort)$ to $(\mEnc, \kEnc)$ and decoding from $(\mEnc, \kEnc)$ to $(\psi, \mShort, \kShort)$ takes $O(1)$ time in the word RAM model. The lookup table can be computed in linear time.
    \end{enumerate}
  \end{lemma}

  Based on this lemma, the entropy encoder works as follows. See \cref{fig:entropy-encoder} for an overview.

  \begin{figure}[ht]
    \centering
    \includegraphics[width=0.6\textwidth]{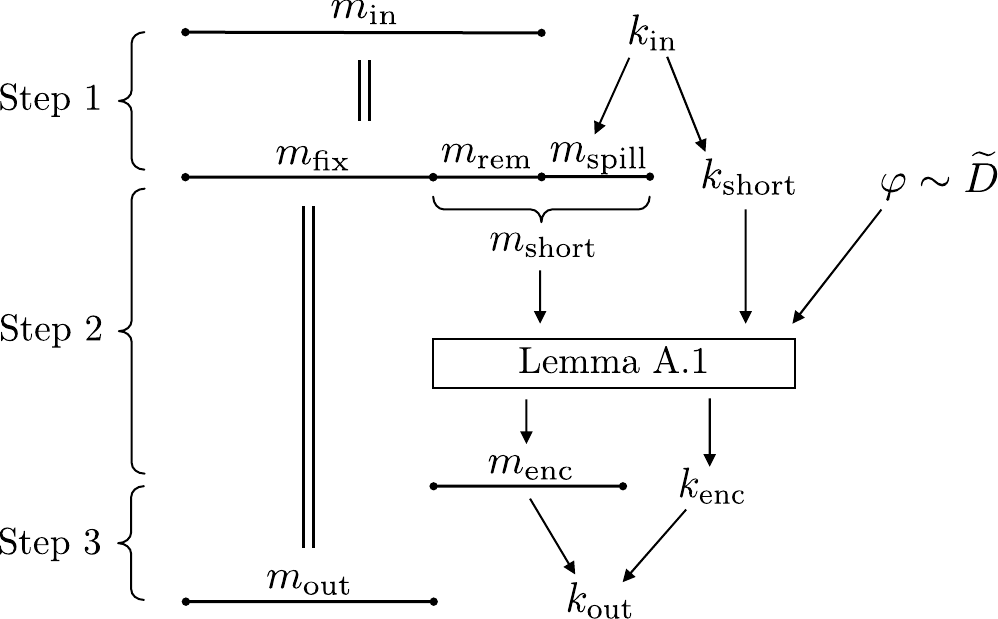}
    \caption{Encoding process of the entropy encoder. We first partition $\mIn$ into $\mFix$ and $\mRem$, and split $\kIn$ into $\mSpill$ and $\kShort$. We then encode $(\phi, \mRem, \mSpill, \kShort)$ to $(\mEnc, \kEnc)$ using \cref{lem:succincter-lem5} under the perturbed distribution $\tilde{D}$. Finally, we output $(\mOut, \kOut)$, where $\mOut = \mFix$ and $\kOut$ is composed of $\mEnc$ and $\kEnc$.}
    \label{fig:entropy-encoder}
  \end{figure}

  \paragraph{Step 1.}

  Define
  \[
    \MMax^D \defeq \max_{\phi \in \supp(D)} \MIn^D(\phi), \qquad \MFix^D = \max\{\MMax^D - \beta,\ 0\},
  \]
  where $\beta$ is a sufficiently large constant integer to be determined later.

  We partition $\mIn$ into two segments: the first $\MFix^D$ words, denoted $\mFix$, and the remaining $O(1)$ words, denoted $\mRem$. If $|\mIn| < \MFix^D$, we pad $\mIn$ with arbitrary content to reach length $\MFix^D$, leaving $\mRem$ empty. Importantly, the length of $\mFix$ is independent of $\phi$.

  Next, choose the unique integer $\NSpill^D \ge 0$ such that $\frac{\KIn^D(\phi)}{2^{\NSpill^D}} \in [q, 2q)$. We then split $\kIn$ into two parts: a string of $\NSpill^D$ bits, denoted $\mSpill$, and an integer $\kShort \in \Bk[\big]{\ceil[\big]{\KIn^D(\phi) / 2^{\NSpill^D}}}$. The concatenation of $\mRem$ and $\mSpill$ forms $\mShort$. This division introduces at most $2/q$ bits of redundancy, since
  \[
    \NSpill^D + \log \KShort^D(\phi) \le \NSpill^D + \log\left( \frac{\KIn^D(\phi)}{2^{\NSpill^D}} \right) + \log(1 + 1/q)
    \le \log \KIn^D(\phi) + 2/q. \numberthis \label{eq:entropy-encoder-step1-redundancy}
  \]

  \paragraph{Step 2.}

  We perturb $D$ to $\tilde{D}$ over the same support, defined by
  \[
    \tilde{D}(\phi) := \left(1 - \frac{1}{2q}\right) D(\phi) + \frac{1}{2q \abs{\supp(D)}}.
  \]
  This perturbed distribution has two key properties:
  \begin{enumerate}
    \item For any $\phi \in \supp(D)$,
      \[
        \log \frac{1}{\tilde{D}(\phi)} - \log \frac{1}{D(\phi)} \le -\log\left(1 - \frac{1}{2q}\right) < \frac{1}{q}.
      \]
      This means that encoding $\phi$ with respect to $\tilde{D}$ instead of $D$ incurs only $1/q$ bits of redundancy.
    \item For any $\phi \in \supp(D)$,
      \[
        \log \frac{1}{\tilde{D}(\phi)} \le \log \bk[\big]{2q \cdot \abs{\supp(D)}} = O(w).
      \]
      That is, encoding any symbol $\phi$ under the new distribution $\tilde{D}$ requires at most $O(1)$ words.
  \end{enumerate}

  We then apply \cref{lem:succincter-lem5} to encode $\phi \sim \tilde{D}$ and $(\mShort, \kShort)$ together, producing output $(\mEnc, \kEnc)$. To bound the size of this output, we show the following claim.
  \begin{claim}
    \EquationOnSameLine{
      \HShort^D \le \max_{\psi \in \supp(D)} \left\{ \log \frac{1}{D(\psi)} + w \MIn^D(\psi) + \log \KIn^D(\psi) \right\} - w\MFix^D + \frac{3}{q}. \phantom{Claim X.X.}
    }
    \label{claim:entropy-encoder-step2}
  \end{claim}

  \begin{proof}
    By the definition of $\HShort^D$ and \eqref{eq:entropy-encoder-step1-redundancy}, it suffices to verify that for every $\psi \in \supp(D)$,
    \begin{equation}
      \hspace{-0.4em} \log \frac{1}{\tilde{D}(\psi)} + \NShort^D(\psi) + \log \KShort^D(\psi) 
      \le \max_{\psi' \in \supp(D)} \left\{ \log \frac{1}{\tilde{D}(\psi')} + w \MIn^D(\psi') - w\MFix^D + \log \KIn^D(\psi') \right\} + \frac{2}{q} \hspace{0.4em}
      \label{eq:entropy-encoder-step2-claim-goal}
    \end{equation}
    When we perform Steps 1 and 2 with $\psi$, we have $w|\mRem| = \max\{0, \, w(\MIn^D(\psi) - \MFix^D)\}$.
    From \eqref{eq:entropy-encoder-step1-redundancy}, we have $\NSpill^D(\psi) + \log \KShort^D(\psi) = \log \KIn^D(\psi) + 2/q$.
    Adding them up, the left-hand side of \eqref{eq:entropy-encoder-step2-claim-goal} is bounded by
    \begin{align*}
      \NShort^D(\psi) + \log \KShort^D(\psi) + \log \frac{1}{\tilde{D}(\psi)} &= w|\mRem| + \NSpill^D(\psi) + \log \KShort^D(\psi) + \log \frac{1}{\tilde{D}(\psi)} \\
      &\le w|\mRem| + \log \KIn^D(\psi) + \log \frac{1}{\tilde{D}(\psi)} + \frac{2}{q}.
      \numberthis \label{eq:entropy-encoder-step2-claim-have}
    \end{align*}

    We consider two cases for $w|\mRem|$:

    \begin{itemize}
      \item \textbf{Case 1.} $\MIn^D(\psi) \ge \MFix^D$, i.e., no padding was needed to form $\mFix$.

        In this case, $w|\mRem| = w(\MIn^D(\psi) - \MFix^D)$, so \eqref{eq:entropy-encoder-step2-claim-have} becomes
        \[
          w \MIn^D(\psi) + \log \KIn^D(\psi) + \log \frac{1}{\tilde{D}(\psi)} + \frac{2}{q} - w\MFix^D,
        \]
        and therefore \eqref{eq:entropy-encoder-step2-claim-goal} holds.

      \item \textbf{Case 2.} $\MIn^D(\psi) < \MFix^D$, i.e., padding was used to form $\mFix$.

        In this case, $w|\mRem| = 0$. Let $\psi'' = \argmax_{\psi' \in \supp(D)} \MIn^D(\psi')$, so $\MIn^D(\psi'') = \MMax^D$. Then the right-hand side of \eqref{eq:entropy-encoder-step2-claim-goal} is at least
        \[
          w(\MIn^D(\psi'') - \MFix^D) = w(\MMax^D - \MFix^D) = \beta w.
        \]

        On the other hand, by \eqref{eq:entropy-encoder-step2-claim-have}, the left-hand side of \eqref{eq:entropy-encoder-step2-claim-goal} is at most
        \[
          w|\mRem| + \log \KIn^D(\psi) + \log \frac{1}{\tilde{D}(\psi)} + \frac{2}{q} \le 0 + O(w) + O(w) + \frac{2}{q} \le \beta w,
        \]
        where the last inequality holds by choosing $\beta$ to be a sufficiently large constant satisfying
        \[
          \beta w \ge 4 \left( \log \max_{\psi', D'} \KIn^{D'}(\psi') + \log q + \log \max_{D'} {\abs[\big]{\supp(D')}} + 10 \right).
        \]
        Thus \eqref{eq:entropy-encoder-step2-claim-goal} also holds in this case.
    \end{itemize}
    In summary, \eqref{eq:entropy-encoder-step2-claim-goal} holds for every $\psi \in \supp(D)$, which concludes the proof of the claim.
  \end{proof}

  The bound in \eqref{eq:entropy-encoder-step2-claim-goal} also ensures the required property $\HShort^D = O(w)$ in \cref{lem:succincter-lem5}, since
  \[
    \HShort^D \le \log \frac{1}{\tilde{D}(\psi)} + w \MIn^D(\psi) - w\MFix^D + \log \KIn^D(\psi) + \frac{3}{q} \le O(w) + \beta w + O(w) + \frac{3}{q} = O(w).
  \]

  \paragraph{Step 3.}

  We output $\mOut := \mFix$, and merge $\mEnc$, $\kEnc$ into a single number $\kOut \in [2^{\NEnc^D} \cdot \KEnc^D]$. The final output of the entropy encoder is $(\mOut, \kOut)$, which uses space
  \begin{align*}
    w \MOut^D + \log \KOut^D &= w\MFix^D + \NEnc^D + \log \KEnc^D \\
    &= w\MFix^D + \HShort^D + \frac{4}{q} \tag{by \cref{lem:succincter-lem5}} \\
    &\le \max_{\psi \in \supp(D)} \left\{ \log \frac{1}{D(\psi)} + w \MIn^D(\psi) + \log \KIn^D(\psi) \right\} + \frac{7}{q} \tag{by \cref{claim:entropy-encoder-step2}} \\
    &= \HIn^D + \frac{7}{q}
  \end{align*}
  bits, as desired. We can also verify that
  \begin{align*}
    \log \KOut^D &= \NEnc^D + \log \KEnc^D \le \HShort^D + \frac{4}{q} \le O(w) \tag{by \cref{lem:succincter-lem5}}
  \end{align*}
  bits as desired.

  \paragraph{Operations.}

  We now describe how to implement the required operations efficiently given the entropy encoder construction. Before any operation, the user provides the distribution $D$ (stored externally) and accesses the output spill $\kOut$. The entropy encoder extracts $(\mEnc, \kEnc)$ from $\kOut$ and uses \cref{lem:succincter-lem5} to decode $(\phi, \mRem, \mSpill)$ from $(\mEnc, \kEnc)$.

  When the user performs a word read/write operation on $\mIn$, we determine whether the accessed word is among the first $\MFix^D$ words of the input VM. If so, the word appears directly in $\mOut$ at the corresponding position, so the operation translates to a single read/write of that word in $\mOut$. Otherwise, the accessed word belongs to $\mRem$ and has been encoded into $\kOut$. In this case, we retrieve the word by accessing the corresponding position in the decoded $\mRem$. To update such a word, we modify $\mRem$ accordingly and re-execute Steps 2 and 3 to form a new $\kOut$. Operations on the input spill $\kIn$ are handled similarly by reading or updating the output spill $\kOut$ directly.

  Changing the distribution and symbol to $(\widehat{D}, \hat{\phi})$ requires careful handling of the boundary between $\mFix$ and $\mRem$. The new distribution determines a new value $\MFix^{\widehat{D}} = \MOut^{\widehat{D}}$, which may differ from $\MFix^D$. If $\MOut^{\widehat{D}} > \MOut^D$, we transfer $\MOut^{\widehat{D}} - \MOut^D$ words from $\mRem$ to $\mFix$; if $\MOut^{\widehat{D}} < \MOut^D$, we move $\MOut^D - \MOut^{\widehat{D}}$ words from $\mFix$ to $\mRem$. This requires $|\MOut^{\widehat{D}} - \MOut^D|$ word read/write operations. We then re-execute Steps 2 and 3 to encode the new symbol $\hat{\phi} \sim \widehat{D}$ and form the updated $\kOut$ in $O(1)$ time.

  Note that when we update the distribution and symbol, the input VM size may change substantially from $\MIn^D(\phi)$ to $\MIn^{\widehat{D}}(\hat{\phi})$, but this change incurs no additional cost according to the VM model: When $\MIn^{\widehat{D}}(\hat{\phi}) > \MIn^D(\phi)$, newly allocated words contain arbitrary initial content; when $\MIn^{\widehat{D}}(\hat{\phi}) < \MIn^D(\phi)$, words beyond position $\MIn^{\widehat{D}}(\hat{\phi})$ are released. So the output VM $\mOut$ only needs to agree with the input VM $\mIn$ on the first $\min\{\MIn^D(\phi), \MIn^{\widehat{D}}(\hat{\phi}), \MFix^{\widehat{D}}\}$ words. Since $\mOut$ and $\mIn$ already agree on the first $\min\{\MIn^D(\phi), \MFix^D\}$ words before the operation, maintaining this invariant requires updating only $|\MFix^{\widehat{D}} - \MFix^D| = |\MOut^{\widehat{D}} - \MOut^D|$ words.
  
  \paragraph{Lookup tables.}

  To answer queries to the quantities $\MOut^D, \KOut^D, \MIn^D(\phi), \KIn^D(\phi)$, and $D(\phi)$ in constant time, we use a lookup table to store them for all $D \in \mathcal{F}$ and $\phi \in \supp(D)$. This lookup table takes $O(\abs{\mathcal{F}} \cdot \max_{D \in \mathcal{F}} {\abs{\supp(D)}})$ words and can be computed in linear time in its size.

  We also need a lookup table for \cref{lem:succincter-lem5}. Since $D$ may change over time, we must create a lookup table for every $D \in \mathcal{F}$, each consisting of
  \[O\bk*{\abs{\supp(D)} \cdot q \cdot \max_{\psi \in \supp(D)} \KShort^D(\psi)}\]
  words. The total space usage for these lookup tables is
  \[O\bk*{\abs{\mathcal{F}} \cdot \max_{D \in \mathcal{F}} {\abs{\supp(D)}} \cdot q \cdot \max_{D \in \mathcal{F}, \psi \in \supp(D)} \KShort^D(\psi)}\]
  words. By \cref{lem:succincter-lem5}, this lookup table can also be computed in linear time in its size.
  This completes the proof of \cref{thm:entropy-encoder}.
\end{proof}

Finally, we present the entropy encoder for uniform distributions.

\thmEntropyEncoderUniform*

\begin{proof}
  We begin by describing the encoding process.

  \paragraph{Step 1.} We compute a pair of integers $\MMax^D \ge 0$ and $\KMax^D \ge 0$ that satisfy:
  \begin{enumerate}
    \item $\KMax^D = 2^{O(w)}$;
    \item $w \cdot \MMax^D + \log \KMax^D \le \HMax^D + 2^{-\Theta(w)}$;
    \item $w \cdot \MMax^D + \log \KMax^D \ge w \cdot \MIn^D(\psi) + \log \KIn^D(\psi)$ for all $\psi \in \supp(D)$;
    \item\label{item:entropy-encoder-uniform-condition-MMax-lb} $\MMax^D \le \MIn^D(\psi)$ for all $\psi \in \supp(D)$.
  \end{enumerate}
  This is completed by considering two cases:
  \begin{itemize}
    \item When $\HMax^D \le \Theta(w)$, we set $\MMax^D = 0$ and $\KMax^D = \floor[\big]{2^{\HMax^D}}$.
    \item Otherwise, we set $\MMax^D = \floor[\big]{\HMax^D / w - \beta}$, where $\beta$ is a sufficiently large constant satisfying $\beta w \ge 10 \log \max_{\psi \in \supp(D)} \KIn^D(\psi)$, and $\KMax^D = \ceil[\big]{2^{\HMax^D - w \cdot \MMax^D}}$.
  \end{itemize}
  It is easy to verify that these choices satisfy the above conditions.

  \paragraph{Step 2.} We transform the input VM $\mIn$ and spill $\kIn$ into a new VM $\mAdj$ with $\MMax^D$ words and a new spill $\kAdj$ with universe size $\KMax^D$. The size of $(\mAdj, \kAdj)$ depends only on $D$ and not on $\phi$. By condition~\ref{item:entropy-encoder-uniform-condition-MMax-lb}, we have $\MMax^D \le \MIn^D(\phi)$, so we can complete this step by transferring $\MIn^D(\phi) - \MMax^D$ words from $\mIn$ to the spill, forming the VM $\mAdj$ and a larger spill $\kAdj$.

  \paragraph{Step 3.} Finally, we encode $\phi \in \supp(D)$ as an integer in $[\abs{\supp(D)}]$, and combine this integer with $\kAdj$ to form a larger spill $\kOut \in [\KMax^D \cdot \abs{\supp(D)}] \eqdef [\KOut^D]$. The VM $\mAdj$ from the previous step becomes the output VM $\mOut$. The final output of the entropy encoder $(\mOut, \kOut)$ uses
  \[
    w \cdot \MOut^D + \log \KOut^D = w \cdot \MMax^D + \log \KMax^D + \log {\abs{\supp(D)}} \le \HMax^D + \log {\abs{\supp(D)}} + 2^{-\Theta(w)}
  \]
  bits of space, as desired.

  \paragraph{Operations.} All operations are implemented similarly to the non-uniform case (\cref{thm:entropy-encoder}) as long as the input sizes $\MIn^D(\phi)$ and $\KIn^D(\phi)$ are provided to the entropy encoder. The only operation that does not require input sizes is reading $\phi$, which can be done in constant time by first computing $\MMax^D$ and $\KMax^D$ from $\HMax^D$ and then performing arithmetic operations on $\kOut$ to recover the integer representing $\phi$.
\end{proof}

\section{Proof of Claims \ref{claim:hmax} and \ref{claim:output_size}}
\label{sec:proof-output-size}

In this section, we first prove \cref{claim:output_size}, and then use it to prove \cref{claim:hmax}.

\OutputSizeClaim*

\begin{proof}
Recall that $V \defeq V(\tilde{a}, \tilde{b}, \hmax) = (\tilde{b} - \tilde{a}) \cdot (\hmax + 1) / W$ is the number of valid keys in the current subproblem. We round $V$ up to the nearest integer $\tilde{V}$ of the form $\floor[\big]{(1 + 1 / \nmax^3)^t}$ for $t \in \N$. Define $\MLarge(n, \tilde{V})$ as the largest integer such that
\[
\log \binom{\tilde{V}}{n} + \frac{n - 1}{\nmax} + \frac{1}{\nmax^{3/2}} - w \cdot \MLarge(n, \tilde{V}) \ge 2 \log \nmax.
\]
Note that $\MLarge(n, \tilde{V}) \ge 0$, since in the large-universe case, $V \ge W^3 = \poly(\nmax)$ for a sufficiently large polynomial factor, so $\log \binom{\tilde{V}}{n} \ge \log V \gg 2 \log \nmax$. Next, let $\KLarge(n, \tilde{V})$ be the smallest integer such that
\[
w \cdot \MLarge(n, \tilde{V}) + \log \KLarge(n, \tilde{V}) \ge \log \binom{\tilde{V}}{n} + \frac{n - 1}{\nmax} + \frac{1}{\nmax^{3/2}},
\numberthis \label{eq:output_size_lb_large_universe}
\]
or equivalently,
\[
\KLarge(n, \tilde{V}) \defeq \ceil*{\binom{\tilde{V}}{n} \cdot 2^{(n - 1) / \nmax + 1 / \nmax^{3/2}} \bigg/ 2^{w \cdot \MLarge(n, \tilde{V})}}.
\]
By the choice of $\MLarge(n, \tilde{V})$, we have $\KLarge(n, \tilde{V}) \in \bigl[\nmax^2, \, \nmax^2 \cdot 2^w\bigr]$. Therefore,
\[
\log \KLarge(n, \tilde{V}) - \allowbreak \log \bk[\big]{\KLarge(n, \tilde{V}) - 1} = O(1 / \nmax^2),
\]
which, together with the minimality of $\KLarge(n, \tilde{V})$, yields
\[
w \cdot \MLarge(n, \tilde{V}) + \log \KLarge(n, \tilde{V}) \le \log \binom{\tilde{V}}{n} + \frac{n - 1}{\nmax} + \frac{1}{\nmax^{3/2}} + O(1 / \nmax^2).
\numberthis \label{eq:output_size_ub_large_universe}
\]

Furthermore, by the rounding rule for $\tilde{V}$, we have $\tilde{V} \le V \cdot (1 + 1 / \nmax^3)$, so
\begin{align*}
\log \binom{\tilde{V}}{n} & \le \log \binom{V}{n} + n \log (1 + 1 / \nmax^3) = \log \binom{V}{n} + O(1 / \nmax^2) \\
&= \log \mathcal{N}(n, \tilde{a}, \tilde{b}, \hmax) + O(1 / \nmax^2) \\
&= \log \mathcal{N}(n, a, b, \hmax) + O(1 / \nmax^2). \tag{by \cref{clm:rounding_loss}}
\end{align*}
Combining this with \eqref{eq:output_size_lb_large_universe} and \eqref{eq:output_size_ub_large_universe} gives the desired property \eqref{eq:output_size_prop}.

To support efficient computation, we precompute a table of all valid values of $\tilde{V}$ and their corresponding $\MLarge(n, \tilde{V})$ and $\KLarge(n, \tilde{V})$. This table occupies $\poly(\nmax, \log \Umax)$ words as there are $\poly(\nmax, \log \Umax)$ valid values of $\tilde{V}$. Given $n, a, b, \hmax$, we first compute the index of $\tilde{V}$ using logarithms, and then use the precomputed table to retrieve $\MLarge(n, \tilde{V})$ and $\KLarge(n, \tilde{V})$ in constant time.
\end{proof}

\HMaxClaim*

\begin{proof}
We use \cref{claim:output_size} to compute \[w \cdot M(n, a, b, \hmax) + \log K(n, a, b, \hmax) - \frac{n-1}{\nmax} - \frac{1}{\nmax^{3/2}} = \log \mathcal{N}(n, a, b, \hmax) + O(1/\nmax^2),\]
and then compute
\begin{align*}
\HMax(n, \tilde{a}, \tilde{b}, \hmax, h(p)) &= w \cdot M(n, a, b, \hmax) + \log K(n, a, b, \hmax) - \frac{n-1}{\nmax} - \frac{1}{\nmax^{3/2}} \\
& \hspace{2em} - \log \frac{1}{\DistWeightTd(h(p))} - \log \frac{\tilde{b}-\tilde{a}}{W}  + \Theta(1/\nmax^2)
\end{align*}
for a sufficiently large $\Theta(1/\nmax^2)$ term.
This satisfies the desired condition \eqref{eq:hmax_defn}. The computation takes constant time using lookup tables from \cref{claim:output_size}.
\end{proof}

\end{document}